\documentclass[namedreferences]{solarphysics}

\usepackage{amsmath}
\usepackage{graphicx}
\usepackage{wrapfig}
\usepackage{lscape}
\usepackage{rotating}
\usepackage{epstopdf}
\usepackage{tabularx}
\usepackage[utf8]{inputenc}
\usepackage{longtable}
\usepackage{array}

\usepackage[hyperref,optionalrh,showbiblabels]{spr-sola-addons} 
\usepackage{graphicx}        
\usepackage{color}           
\usepackage{breakurl}        

\chardef\us=`\_

\begin{document}

\begin{article}
\begin{opening}

\title{Role of Compressive Viscosity and Thermal Conductivity on the Damping of Slow Waves in the Coronal Loops With and Without Heating-Cooling Imbalance}

\author[addressref=aff1]{\inits{Abhinav}\fnm{Abhinav}~\lnm{Prasad}}
\author[addressref={aff1},corref,email={asrivastava.app@itbhu.ac.in}]{\inits{A.K.}\fnm{A.K.}~\lnm{Srivastava}}
\author[addressref=aff2]{\inits{T. J.}\fnm{T.J.}~\lnm{Wang}}

\address[id=aff1]{Department of Physics, Indian Institute of Technology (BHU), Varanasi-221005, UP, India.}
\address[id=aff2]{The Catholic University of American and NASA Goddard Space Flight Center, Code 671, Greenbelt, MD, 20771, USA.}

\runningauthor{Prasad et al.}
\runningtitle{Damping of Slow Waves in Coronal Loops}

\begin{abstract}
In the present article, we derive a new dispersion relation for slow magnetoacoustic waves invoking the effect of thermal conductivity, compressive viscosity, radiation, and unknown heating term along with the consideration of heating-cooling imbalance from linearized MHD equations. We solve the general dispersion relation to understand the role of compressive viscosity and thermal conductivity in the damping of slow waves in coronal loops with and without heating-cooling imbalance. We have analyzed the wave damping for the range of loop length $L$=50\,--\,500 Mm, temperature $T$=5\,--\,30 MK, and density $\rho$=10$^{-11}$\,--\,10$^{-9}$ kg m$^{-3}$. It was found that the inclusion of compressive viscosity along with thermal conductivity significantly enhances the damping of the fundamental mode oscillations in shorter (e.g. $L$=50 Mm) and super-hot ($T>$10 MK) loops. However, the role of viscosity in the damping is insignificant in longer (e.g. $L$=500 Mm) and hot loops (T$\leq$10 MK) where, instead, thermal conductivity along with the presence of heating-cooling imbalance plays a dominant role. For shorter loops at a super-hot regime of temperature, the increment in the loop density substantially enhances the damping of the fundamental modes due to thermal conductivity when viscosity is absent, however, when the compressive viscosity is added the increase in density substantially weakens the damping. Thermal conductivity alone is found to play a dominant role in longer loops at lower temperatures (T$\leq$10 MK), while compressive viscosity dominates the damping at super-hot temperatures ($T>$10 MK) in shorter loops. The predicted scaling law between damping time ($\tau$) and wave period ($P$) is found to better match the observed SUMER (Solar Ultraviolet Measurements of Emitted Radiation) oscillations when the heating-cooling imbalance is taken into account in addition to thermal conductivity and compressive viscosity for the damping of the fundamental slow mode oscillations. 
\end{abstract}
\end{opening}

\section{Introduction}
     \label{S-Introduction} 
Doppler shift oscillations in hot coronal loops were first directly observed by the  SUMER (Solar Ultraviolet Measurements of Emitted Radiation) spectrograph onboard the Solar and Heliospheric Observatory (SoHO)  \citep{2002ApJ...574L.101W,2003A&A...402L..17W}. These observations are found to be associated with the fundamental mode of the slow magnetoacoustic oscillations exhibiting an efficient damping \citep{2002ApJ...580L..85O}. \citet{2003A&A...406.1105W} have detected such slow-mode oscillations in numerous hot coronal loops, and using a large enough statistics established their physical properties consistently, e.g., the phase speed derived from the observed period and loop length matches approximately the local sound speed; and the intensity fluctuation lags the Doppler shift by 1/4 period. It was also found that the observed scaling of the damping time with the wave period matches the predicted scaling for slow waves when the damping effects due to thermal conduction and compressive viscosity are considered.
The numerical modeling by \citet{2008A&A...481..247T} showed that such oscillations are also expected to be detected in normal coronal loops maintained at 1\,--\,2 MK temperature, and later this supposition was confirmed by observations from the  EUV Imaging Spectrometer (EIS) onboard Hinode \citep{2008ApJ...681L..41M, 2008A&A...489L..49E,2010NewA...15....8S}. Damped slow-mode oscillations are also suggested to be produced in stellar flaring loops because some
quasi-periodic pulsations (QPPs) detected in stellar flares show many features similar to those observed in solar flares \citep{2005A&A...436.1041M,2013ApJ...778L..28S,2016ApJ...830..110C}.

The SUMER and Yohkoh/SXT (Soft X-Ray Telescope) observations suggested that such loop oscillations may be associated with an impulsive deposition of the heat at its footpoint due to localized transient events (e.g. microflares) that may lead to perturbations both in velocity and density within the loop \citep{2005A&A...435..753W}. While impulsive heating is proposed as a primary exciter of slow magnetoacoustic oscillations in solar loops \citep{2005A&A...435..753W, 2006ApJ...647.1452P,2007ApJ...659L.173T}, there are several other mechanisms that may also trigger them, e.g., pressure or velocity pulses,  kink instability, etc. \citep[e.g.][and references therein]{2005A&A...436..701S,2007ApJ...668L..83S,2008A&A...479..235H,2005A&A...438..713T}. Additionally, the pressure pulse and flows may also inevitably be associated with a response to impulsive heating. Since such loops may cool after the transient energy release and heating, thermal conduction is termed initially as a viable dissipation mechanism for interpreting the observed damping of slow magnetoacoustic oscillations \citep{2002ApJ...580L..85O}. The effect of dissipative agents, e.g., thermal conduction, compressive viscosity, and radiation have been studied on slow magnetoacoustic oscillations in a greater detail by \cite{2006SoPh..236..127P}. They have found that by varying the density from 10$^{8}$ to 10$^{10}$ cm$^{-3}$ at a fixed temperature in the range 6\,--\,10 MK as observed by SUMER, strong damping occurs at lower density and weak damping occurs at higher density. It was also noted that the effect of optically thin radiation provides some additional dissipation apart from thermal conductivity and viscosity in weak-damped oscillations. \citet{2004ApJ...605..493M} have pointed out the effect of stratification on the wave damping, and concluded that the dissipation rates of slow waves by thermal conduction and compressive viscosity are enhanced by the nonlinear effect caused by gravitational stratification. Later, \citet{2007SoPh..246..187S} argued that thermal conduction  alone cannot produce strong damping as observed in SUMER oscillations, while the inclusion of compressive viscosity is required. 

\citet{2008A&A...483..301B} have reported the effect of the radiative emission arising from a non-equilibrium ionization on damping of the slow magnetoacoustic oscillations in the loops,
and inferred that this loss  may reduce the damping timescale by up to 10\% than that typically observed by SUMER.  \citet{2008A&A...479..235H} have studied the fact that even in the absence of thermal conduction, large-amplitude slow-mode oscillations are getting damped strongly by the dissipation of the shocks. \citet{2008ApJ...685.1286V} have further found that in the presence of thermal conduction, shock dissipation at large amplitudes enhances the damping rate by 50\% higher than the rate achieved during the presence of thermal conduction enabled dissipation alone. 
\citet{2008SoPh..252..305E} have reported  the damping scenario of standing slow waves in nonisothermal, hot, gravitationally stratified coronal loops, and established the physical fact that the decay time of waves decreases with the increase of the initial temperature. \citet{2013SoPh..283..413A} have found that although the background plasma is cooling, thermal conduction is still found to cause a strong damping for the slow magnetoacoustic oscillations in hot coronal loops. However, they consider only the effect of thermal conduction that has a weak damping effect at very high temperature. Therefore, when the loop cools from super-hot to the hot regime close to a maximum damping temperature, then damping due to conduction increases with the cooling \citep{2003A&A...408..755D}. \citet{2016ApJ...824....8K} have presented a theoretical model of the standing slow magnetoacoustic mode and found that these modes are  highly sensitive to the radiative cooling and choice of the heating function as well.

It was found that the thermal conductivity is nearly suppressed and compressive viscosity is enhanced by more than an order of magnitude in very hot loops, affecting the dissipation of the various harmonics of the slow magnetoacoustic oscillations \citep{2015ApJ...811L..13W,2018ApJ...860..107W}.
Many previous studies on the dynamics of compressive MHD wave modes in coronal loops have described the significance of thermal equilibrium along with  mechanical equilibrium in the wave guiding magnetic structures \citep[e.g.][]{2016ApJ...824....8K,2017ApJ...849...62N}. Various representative theories, on the physical processes that balance the internal energy losses to maintain thermal equilibrium, have been given to explain the unknown mechanism underlying the observed coronal heating. For a long time, it has been termed as the coronal heating problem \citep[e.g.,][]{2012RSPTA.370.3217P}. Although it is established that many efforts have been made to understand the different heating and cooling mechanisms in coronal loops, no explanation has proven to be accepted preferably. Since the specific heating mechanism is unknown, many previous theoretical models of MHD waves in coronal loops have taken into account the heating function to depend upon the loop parameters, e.g., temperature, density, and magnetic field, as well as a static or time-dependent heating function, more specifically in the form of power law functions \citep[e.g.][and references  therein]{1978ApJ...220..643R,2014LRSP...11....4R,2019A&A...628A.133K}. Compressive and longitudinal MHD wave modes change the local thermal equilibrium by locally perturbing the background equilibrium quantities such as density, temperature, pressure which ultimately leads to a heating-cooling imbalance. This physical scenario is inferred as the imbalance between the equilibrium balance of radiative cooling losses and unknown coronal heating. This affects the slow wave by modifying the dispersion relation and can be attributed to the suppressed or enhanced damping of the waves. Many previous theoretical studies have concentrated on the effects of this imbalance in the limit of weak non-adiabacity. However, a more recent work by \citet{2019A&A...628A.133K} has removed this approximation by taking into account the non-adiabatic terms to be arbitrarily large, and we have followed a similar procedure in our present model.

As we mentioned above, the significance of heating-cooling imbalance and its effect on the properties of slow magnetoacoustic oscillations has been recently studied by \citet{2019A&A...628A.133K}. They have analyzed the damping of standing slow magnetoacoustic oscillations in solar coronal loops by taking into account the field-aligned thermal conductivity and a wave-induced misbalance between radiative cooling and some unspecified heating rates, and found that the slow wave dynamics is  highly sensitive to the characteristic timescales of the thermal misbalance. \citet{2019ApJ...886....2W} have recently shown the significance of suppressed thermal conductivity and enhanced effect of the compressive viscosity that determine damping properties of the magneto-acoustic oscillations in hot loops at approximately 10 MK. In the present article, we have derived a new dispersion relation taking into account the compressive viscosity, thermal conduction, and radiative cooling as dissipative mechanisms, and an appropriate heating function in order to heat the coronal loop plasma. We consider the coronal loops for a wide range of length (50\,--\,500 Mm), temperature (5\,--\,30 MK), and density (10$^{-11}$\,--\,10$^{-9}$ kg m$^{-3}$) with and without heating-cooling misbalance to understand the evolutionary and dissipative properties of the slow magnetoacoustic oscillations. We refer to hot loops as those hosting the SUMER oscillations with T=5\,--\,15 MK (including the Doppler shift oscillations observed with Yohkoh/BCS (Bragg Crystal Spectrometer) and longitudinal oscillations observed with SDO/AIA, Atmospheric Imaging Assembly) and super-hot loops with T=20\,--\,30 MK are referred to the RHESSI (Reuven Ramaty High Energy Solar Spectroscopic Imager) detected oscillations or QPPs in flares by \citet{2016ApJ...830..110C}. Moreover, we have chosen the damped oscillatory regime of the magnetoacoustic oscillations of loops both with and without heating-cooling imbalance, and also have performed a detailed analytical study of the effect of compressive viscosity and thermal conductivity on the damping of slow waves. We have also compared their individual roles in this damping under the consideration of heating-cooling imbalance.
We also have obtained new scaling relations between the damping time ($\tau$) and wave period ($P$) from the results of these studies, and have compared the theoretical results and various scaling laws with the observed damped SUMER oscillations.
In Section 2, we present the basic model and dispersion relation. The numerical solution and related results are described in Section 3. The last section presents the discussion and conclusions.

\section{Analytical Model and New Dispersion Relation}
                \label{S-general}      
The description of the model of slow magnetoacoustic oscillations in hot and dense coronal loop is given below. It depicts the properties of slow waves in the viscous, thermal conductive, and radiative plasma with a certain heating, 
as well as with and without considering the effect of heating-cooling imbalance.

\subsection{Basic MHD Equations} 

We consider the effects of thermal conductivity, imbalance of radiative cooling and unknown coronal heating, dissipative viscous forces, and heating in our model. For the generalization of the results we compare the model results with and without heating-cooling imbalance as well. To exclude the heating-cooling imbalance, we consider the constant heating term in our model. To the best of our knowledge, this is the first effort to perform such a detailed analytical calculation. Apart from these effects, we have the infinite magnetic field approximation under which the perturbations are confined along the rigid magnetic field lines. We derive a new more general dispersion relation for the slow magnetoacoustic waves in the non-ideal coronal loop plasma. The governing magnetohydrodynamic equations, \citet{2014masu.book.....P}, in 1D are given as follows:\newline
Mass Equation, \newline
\begin{equation}
    \frac{\partial \rho}{\partial t} + \frac{\partial(\rho V)}{\partial z} = 0.
\end{equation}
Here $V$ is the velocity field.\newline
Momentum Equation,
\begin{equation}
    \rho\left(\frac{\partial V}{\partial t} + V\frac{ \partial V}{\partial z}\right) + \frac{\partial p}{\partial z} - \frac{4}{3}\frac{\partial}{\partial z}\left(\eta_0\frac{\partial V}{\partial z}\right) = 0.
\end{equation}
Here \(\eta_0\) is the coefficient of compressive viscosity.\newline
Energy Equation,
\begin{multline}
    C_v\left(\frac{\partial T}{\partial t} + V\frac{\partial T}{\partial z}\right) - \left(\frac{k_BT}{\rho m}\right)\left(\frac{\partial \rho}{ \partial t} + V \frac{\partial \rho}{\partial z}\right) =\\ -Q(\rho,T) + \frac{1}{\rho}\frac{\partial }{\partial z}\left(\kappa\frac{\partial T}{\partial z}\right) + \left(\frac{4\eta_0}{3}\right)\left(\frac{\partial V}{\partial z}\right)^2.
\end{multline}
Here \(C_v\) is defined as 
\begin{equation}
    C_v = \frac{k_B}{m(\gamma-1)}
\end{equation}
and \(Q(\rho,T)\) is composed of two functions 
\begin{equation}
    Q(\rho,T) = \mathcal{L}(\rho,T) - H(\rho,T),
\end{equation}
\begin{equation}
    \mathcal{L}(\rho,T) = \chi \rho T^{\alpha}\,\,\,\,\,\text{(radiative cooling function)},
\end{equation}
\begin{equation}
    H(\rho,T) = h\rho^a T^b \,\,\,\,\text{(unknown heating function)},
\end{equation}
where $a$, $b$ are power index factors and $h$, \(\chi,\alpha\) are the unknown heating coefficient and radiative cooling coefficients, while \(\kappa\) is the thermal conductivity.\newline
Gas Equation,
\begin{equation}
    p = \frac{\rho k_B T}{m},
\end{equation}
where \(k_B\) is the Boltzmann constant and $m$ is the mean particle mass equal to \(0.6 m_p\), where \(m_p\) is the proton mass.\newline
Further we take the temperature dependence of viscosity as \citep{1965RvPP....1..205B}  
\begin{equation}
    \eta_0 = 10^{-17}T^{5/2}\,\, {\text k g}\,{\text m}^{-1}\,{\text s}^{-1}.
\end{equation}
We also take the temperature dependence of thermal conductivity as \citep{1965RvPP....1..205B}  
\begin{equation}
    \kappa = 9\times10^{-12}T^{5/2}\,\, {\text W}\,{\text m}^{-1}\,{\text K}^{-1}.
\end{equation}

\subsection{Linearised MHD Equations} 
{
In order to study the dynamics of slow magnetoacoustic waves, we consider linear perturbations in the basic plasma state. In equilibrium, the plasma is isothermal with a constant uniform density and pressure throughout. It is also assumed to be stationary having zero velocity field and the gravitational effects are entirely ignored in the analysis.
\begin{equation}
    p = p_0 + p_1\,\,\,\,\,\text{(pressure)},
\end{equation}
\begin{equation}
    \rho = \rho_0 + \rho_1\,\,\,\,\,\text{(density)},
\end{equation}
\begin{equation}
    T = T_0 + T_1\,\,\,\,\text{(temperature)},
\end{equation}
\begin{equation}
    V = V_1\,\,\,\,\,\,\text{(velocity field)}.
\end{equation}
Further using these linear perturbations, we linearize the MHD equations as: \newline
 Linearized Mass equation,
\begin{equation}
    \frac{\partial \rho_1}{\partial t} + \rho_0 \left(\frac{\partial V_1}{\partial z}\right) = 0.   
\end{equation}
 Linearized  Momentum equation,
\begin{equation}
    \rho_0 \left( \frac{\partial V_1}{\partial t} \right) + \frac{\partial p_1}{\partial z} = \left(\frac{4\eta_0} {3}\right)\left(\frac{\partial^2V_1}{\partial z^2}\right).
\end{equation}
 Linearized Energy equation,
\begin{multline}
    \frac{\partial T_1}{\partial t} - \left(\frac{(\gamma-1)T_0 }{\rho_0}\right)\left(\frac{\partial \rho_1}{\partial t}\right) = \left(\frac{\kappa}{\rho_0 C_v}\right)\left( \frac{\partial^2 T}{\partial z^2}\right) \\- \left(\frac{T_1}{\tau_2}\right) - \left(\frac{1}{\tau_2} - \frac{\gamma}{\tau_1}\right)\left(\frac{T_0}{\rho_0}\right)\rho_1.
\end{multline}
Here $\gamma = \frac{5}{3}$ is the adiabatic index and \(\tau_1 \), \(\tau_2\) are the corresponding characteristic time scales of the heating-cooling mechanisms involved. They are defined below according to \citet{2019A&A...628A.133K}:\newline
\begin{equation*}
    \tau_2 = \frac{C_v}{(\partial Q/\partial T)},
\end{equation*}
\begin{equation*}
    \tau_1 = \frac{\gamma C_v}{(\partial Q/\partial T - (\rho_0/T_0)(\partial Q/\partial \rho))},
\end{equation*}
 Ideal gas equation,
\begin{equation}
    \frac{p_1}{p_0} = \frac{T_1}{T_0} + \frac{\rho_1}{\rho_0}.
\end{equation}
}
\subsection{Fourier Analysis and New Dispersion Relation}
{
We propose Fourier solutions of the form
\begin{equation}
    F = \hat{F}e^{i(kz-\omega t)}
\end{equation}
to obtain our dispersion relation 
\begin{equation}
    \omega^3 + A\omega^2 + B\omega + C = 0,
\end{equation}
where 
\begin{equation*}
    A = i\left (\frac{4\eta_0k^2}{3\rho_0} + \frac{\kappa k^2}{\rho_0 C_v} + \frac{1}{\tau_2}\right),
\end{equation*}
\begin{equation*}
    B = -\left(\frac{\gamma p_0 k^2}{\rho_0} + \frac{4\eta_0 k^2}{3\rho_0}\left(\frac{\kappa k^2}{\rho_0 C_v} + \frac{1}{\tau_2}  \right) \right),
\end{equation*}
\begin{equation*}
    C = -i\left(\frac{k^2p_0}{\rho_0}\left( \frac{\kappa k^2}{\rho_0 C_v} + \frac{\gamma}{\tau_1}\right)\right).
\end{equation*}

}

\section{Numerical Solutions of the Dispersion Relation} 
      \label{S-features}      
{
In order to solve our dispersion relation (Equation 20), we consider the following form of the unknown heating function 
 \begin{equation}
     H(\rho,T) = h\rho^{-1/2}T^{-3},
 \end{equation}
  where the power index factors $a$, $b$ are chosen as -0.5 and -3 respectively, which is associated with determining a good quality factor of the damped oscillatory slow magnetoacoustic oscillations during heating-cooling imbalance  \citep{2019A&A...628A.133K} and $h$ is determined by the initial equilibrium condition, i.e. \(Q(\rho_0,T_0) = 0\). Throughout the article we shall use this form of heating function whenever heating-cooling imbalance is to be considered, otherwise $H(\rho,T)=$ constant is assumed.
  
    Many previous studies have investigated the damping of slow waves by
assuming the unspecific heating term as constant ($H$=constant), which infers that the heating only plays a role in balancing the optically thin radiation loss in the equilibrium but does not contribute to the evolution of waves \citep{2006SoPh..236..127P,2007SoPh..246..187S}. These studies have demonstrated that the effect of radiation on slow wave damping in the hot plasma is weak or negligible. However, when considering heating as a function of density and temperature, the waves will, in turn, cause variations of heating, leading to the so-called heating-cooling imbalance that can significantly change the behavior of the wave evolution \citep{2016ApJ...824....8K,2017ApJ...849...62N,2019A&A...628A.133K}.\newline

We have used the CHIANTI atomic database v. 9.0.1 in determining the specific values of \(\chi\) and \(\alpha\) at different temperature and density for the radiative cooling.
Using this we obtain the values of \(\tau_1\) and \(\tau_2\) for different temperatures at a density of \( 10^{-11}\)  kg \( {\text m}^{-3}\). In the hot regime of temperature ($T\leq$10 MK), for $T$=5.0, 6.3, 8.9, 10 MK, $\tau_{1}$ is estimated as 36, 37, 65, 100 mins, while $\tau_{2}$ as 10, 12, 19, 22 mins respectively.
In the super-hot regime of temperature ($T>$10 MK), for $T$=20, 30 MK, $\tau_{1}$ is estimated as 308, 566 mins, while $\tau_{2}$ as 123, 226 mins, respectively.
The values of \(\tau_1\) and \(\tau_2\) are reduced by 
10, and 100 times respectively at each given temperature for the loops with density \( 10^{-10}\) kg \({\text m}^{-3}\) and \( 10^{-9}\) kg \({\text m}^{-3}\).

We solve our dispersion relation numerically using the Wolfram Mathematica environment from 2016 for solution of standing wave form, i.e. assuming the cyclic frequency \(\omega\) to be complex \(\omega_R + i\omega_I\) while the wavenumber $k$ is real.
Further we solve our dispersion relation for a range of temperature, density (normal and over-dense postflare loops), and loop lengths which also include the values for observed temperature and loop length of SUMER oscillations \citep[e.g.][ and references therein]{2003A&A...406.1105W,2011SSRv..158..397W}.\newline

\section{Theoretical Results} 
In Section 4.1, we analyze the effect of viscosity and thermal conductivity in the damping of slow waves in coronal loops with and without considering heating-cooling imbalance. Thermal conductivity and radiative cooling is always present as the damping mechanism in these analyses. However, we switch on and off the effect of compressive viscosity both in the case of heating-cooling imbalance and without it in order to understand its effects with respect to thermal conductivity on the damping of slow magnetoacoustic oscillations. 
The role of loop density in the damping of slow waves in coronal loops of length $L$=500 Mm within the hot regime of temperature (T$\leq$10 MK) is discussed in Section 4.2. In Section 4.3, we study the role of loop density in the damping of slow waves in coronal loops of length $L$=50 Mm within the super-hot regime of temperature (T$>$10 MK). In Section 4.4, we compare the individual role of viscosity and thermal conductivity on the damping of slow waves in coronal loops in the presence of heating-cooling imbalance. Thereafter, in the light of detailed analytical results, we made new scaling laws between the damping time ($\tau$) and the wave period ($P$) of the fundamental mode of slow magnetoacoustic oscillations and compare them with the SUMER oscillations in Section 4.5. 

\subsection{Effect of Viscosity and Thermal Conductivity on the Damping of Slow Waves in Coronal Loops with and without Heating-Cooling Imbalance}

We have solved a new dispersion relation (Equation 18) in order to understand the effect of compressive viscosity and thermal conductivity at different sets of physical parameters, e.g. loop length, temperature, and density.\newline
The main objective of the present work is to understand the evolution and damping of the fundamental mode of the slow magnetoacoustic oscillations, i.e. \(k = \pi/L\) ,where $L$ is the loop length and $k$ is the wave number. However, the most general solution of the dispersion relation depends on the dimensionless wave number $K$, where it is defined as $K=kL/\pi$ ($K$=1 corresponds to the
fundamental mode (or first harmonic), $K$=2, 3, 4 correspond to the second, third, fourth harmonics respectively).
We study the damping of slow magnetoacoustic oscillations with and without heating-cooling imbalance in our analysis due to the linear perturbation of the plasma. This is a very detailed parametric study that considers the variation of the parameters in a wide range, e.g. loop length ($L$=50\,--\,500 Mm), temperature ($T$=5\,--\,30 MK), and density ($\rho$=10$^{-11}$\,--\,10$^{-9}$ kg m$^{-3}$) in order to understand the evolution and damping of slow magnetoacoustic oscillations. We define two regimes of loops based on their temperature, i.e. (i) hot loops with $T\leq$10 MK and (ii) super-hot loops with $T>$10 MK. Later, for the loop lengths of $L$=50 Mm (shortest) and 500 Mm (longest), we perform similar parametric studies at three different densities, e.g. normal ($\rho$=10$^{-11}$ kg m$^{-3}$), and over-dense ($\rho$=10$^{-10}$, 10$^{-9}$ kg m$^{-3}$) hot loops that are also maintained at a wide range of the temperature (5\,--\,30 MK). 


 \begin{figure}   
   \centerline{\hspace*{0.015\textwidth}
               \includegraphics[width=0.615\textwidth,clip=]{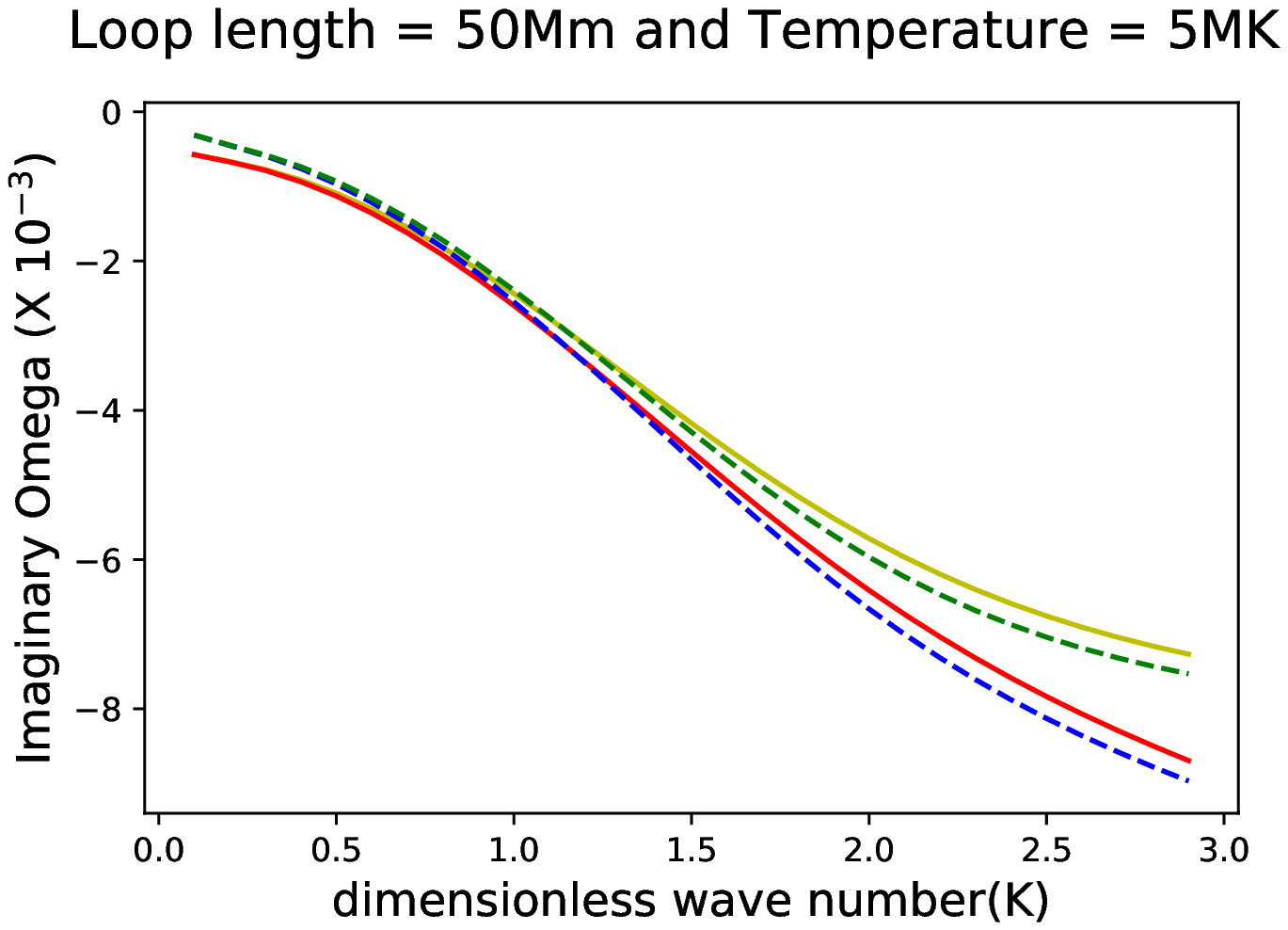}
               \hspace*{-0.03\textwidth}
               \includegraphics[width=0.615\textwidth,clip=]{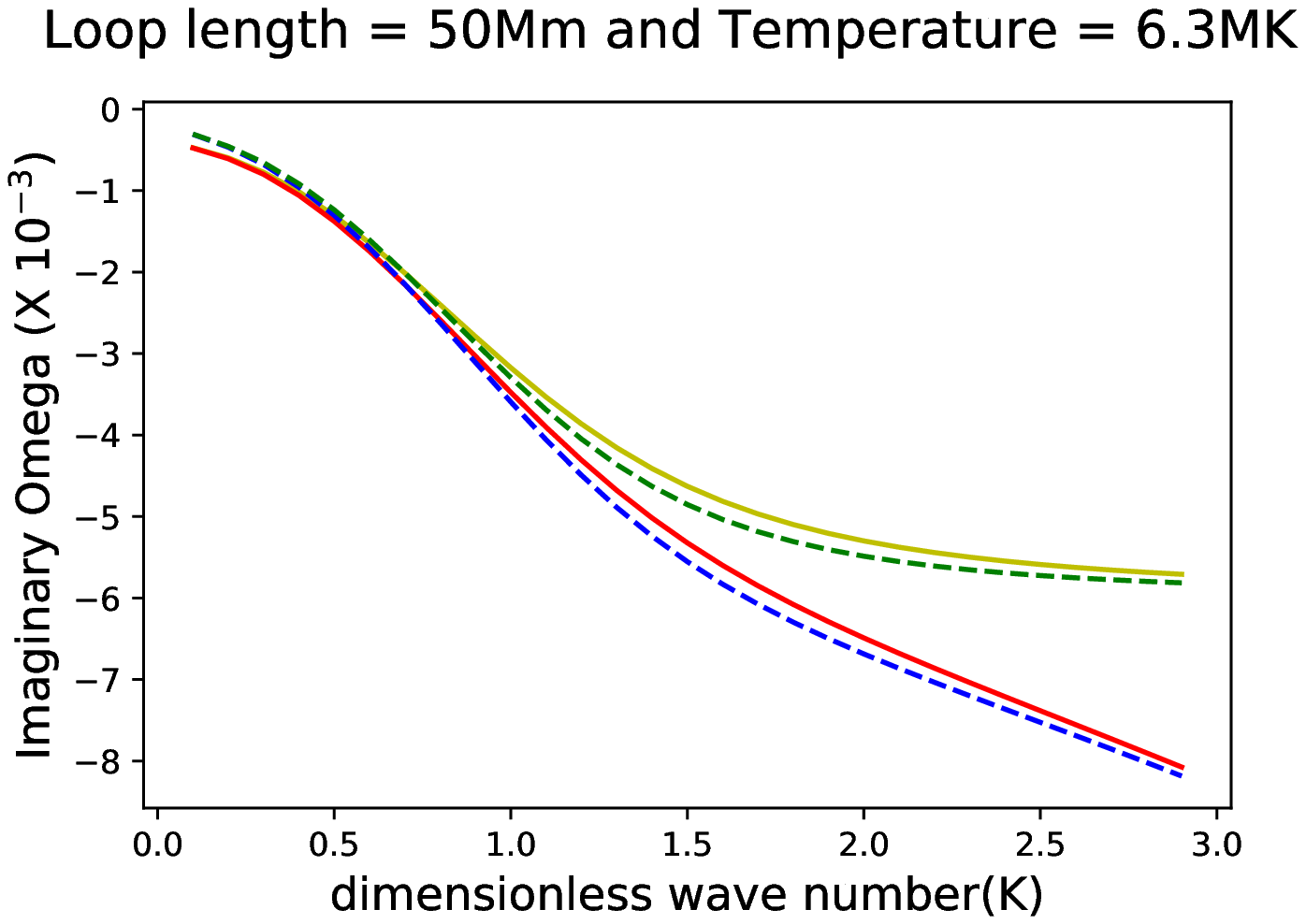}
              }
     \vspace{-0.35\textwidth}   
     \centerline{\Large \bf     
      \hspace{0.0 \textwidth}  \color{white}{(a)}
      \hspace{0.415\textwidth}  \color{white}{(b)}
         \hfill}
     \vspace{0.31\textwidth}    
   \centerline{\hspace*{0.015\textwidth}
               \includegraphics[width=0.615\textwidth,clip=]{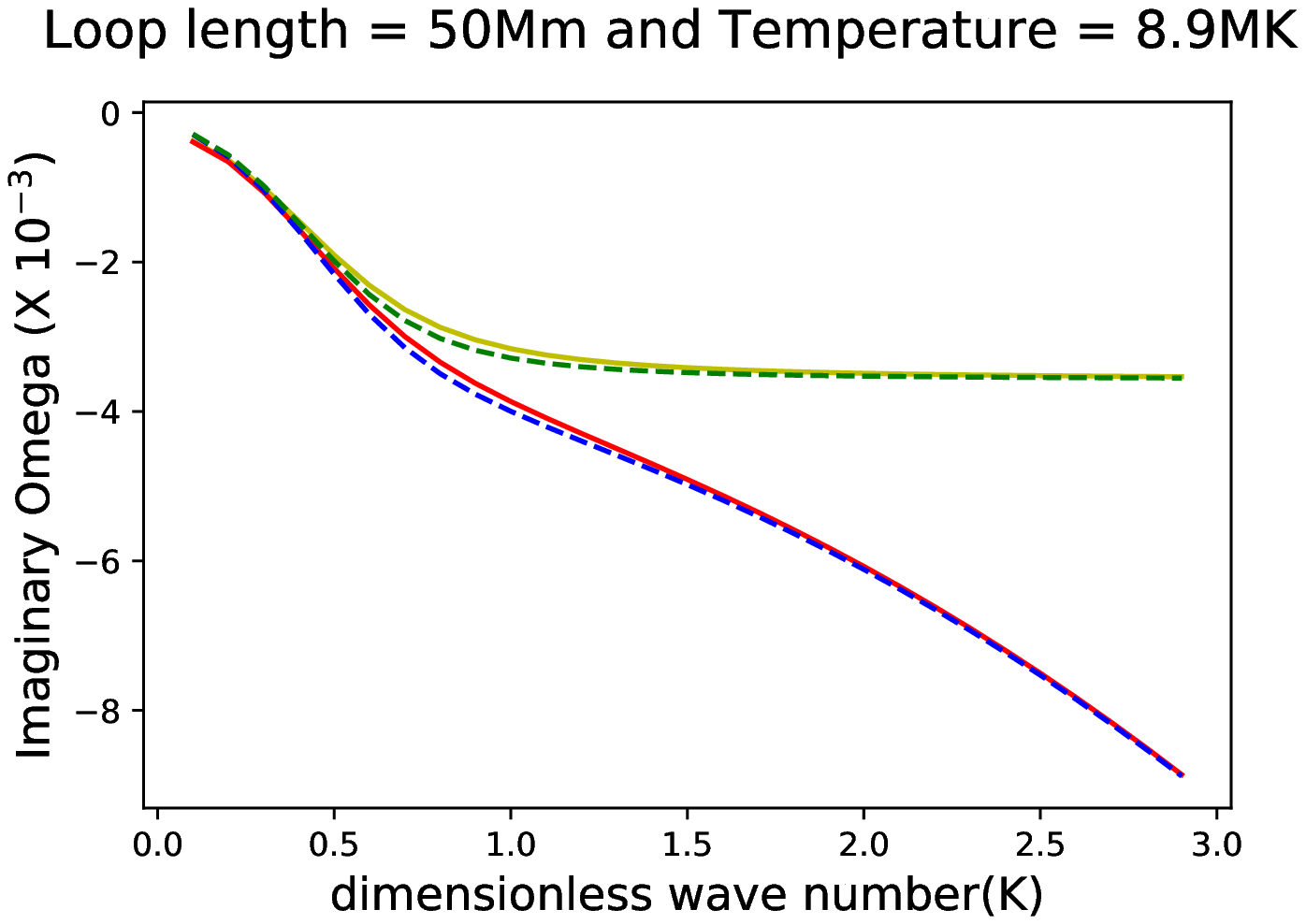}
               \hspace*{-0.03\textwidth}
               \includegraphics[width=0.615\textwidth,clip=]{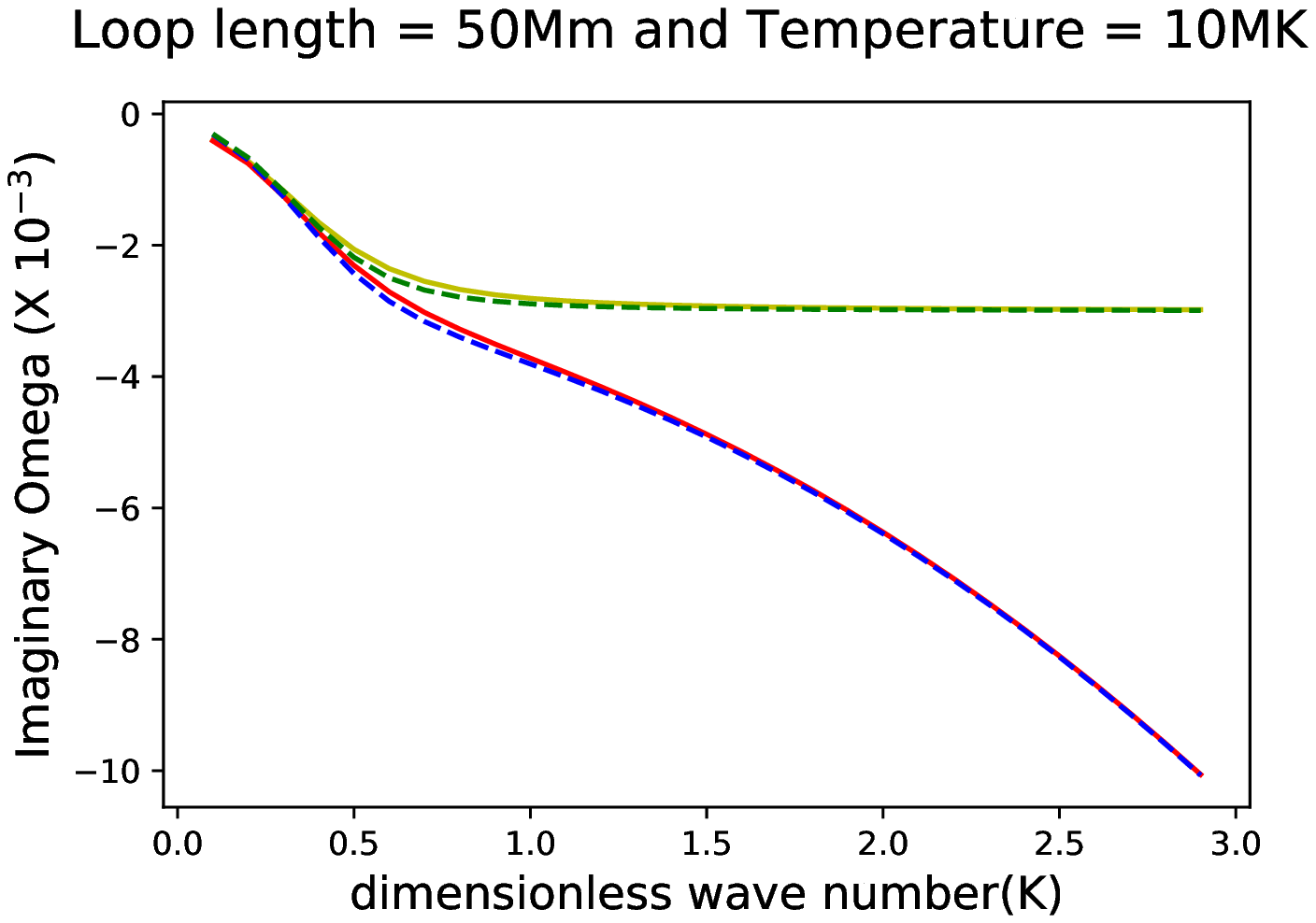}
              }
     \vspace{-0.35\textwidth}   
     \centerline{\Large \bf     
      \hspace{0.0 \textwidth} \color{white}{(c)}
      \hspace{0.415\textwidth}  \color{white}{(d)}
         \hfill}
     \vspace{0.31\textwidth}    
     
     \centerline{\hspace*{0.015\textwidth}
               \includegraphics[width=0.615\textwidth,clip=]{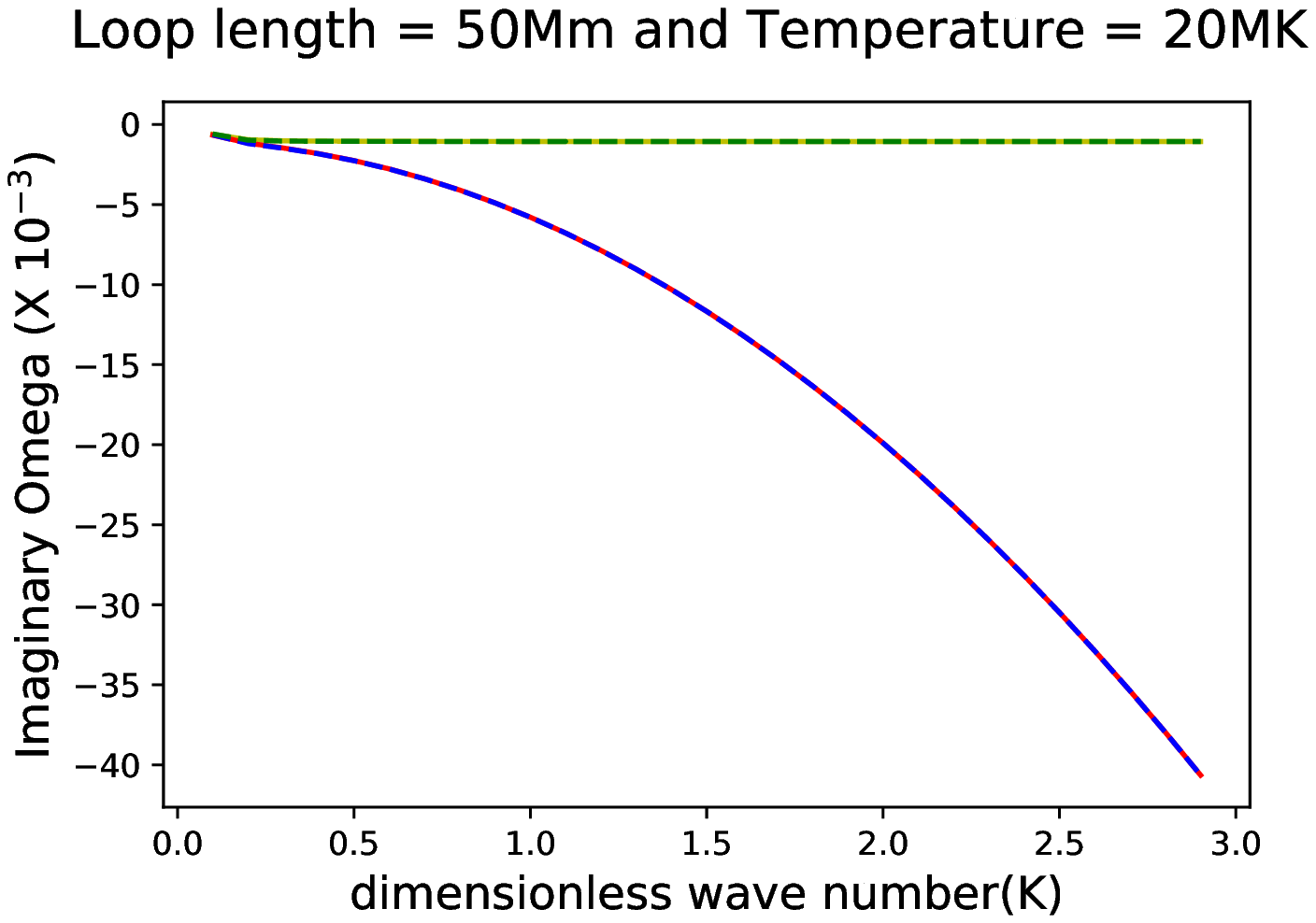}
               \hspace*{-0.03\textwidth}
               \includegraphics[width=0.615\textwidth,clip=]{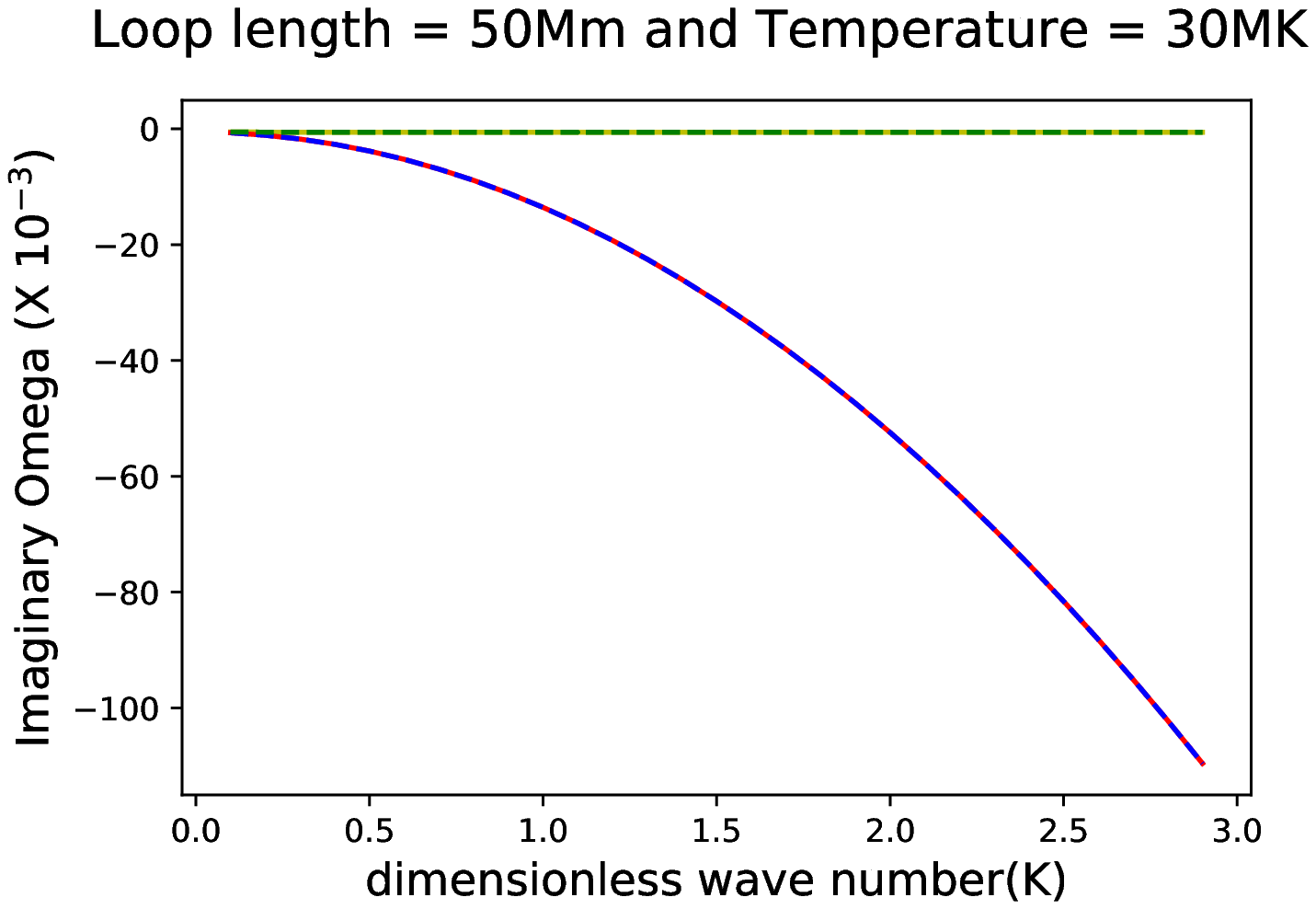}
              }
     \vspace{-0.35\textwidth}   
     \centerline{\Large \bf     
      \hspace{0.0 \textwidth} \color{white}{(c)}
      \hspace{0.415\textwidth}  \color{white}{(d)}
         \hfill}
     \vspace{0.31\textwidth}    
              
\caption{The panels show the variation of \(\omega_I\) with  dimensionless wave number $K$ at a fixed loop-length of 50 Mm for different temperatures from 5 MK to 30 MK. In each panel, red and yellow curves correspond to the solution of the dispersion relation with and without the effect of compressive viscosity respectively when the heating-cooling imbalance is present. The blue-dotted and green-dotted curves represent the solution of the dispersion relation with and without the effect of compressive viscosity respectively when the heating-cooling imbalance is not present. Thermal conductivity is always present as a damping mechanism in these analyses.}
 \label{F-4panels}
 \end{figure}
  \begin{figure}    
   \centerline{\hspace*{0.015\textwidth}
               \includegraphics[width=0.615\textwidth,clip=]{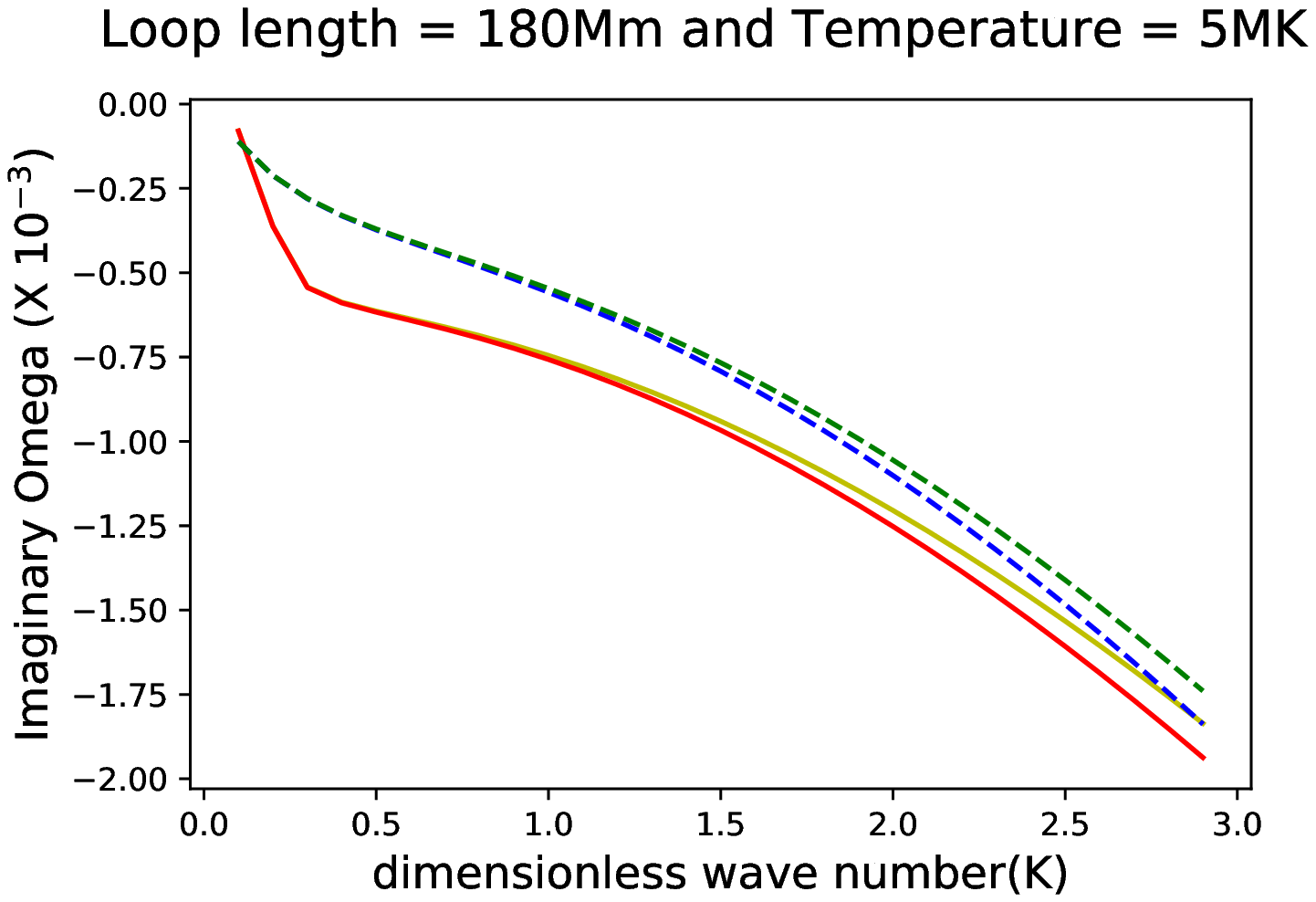}
               \hspace*{-0.03\textwidth}
               \includegraphics[width=0.615\textwidth,clip=]{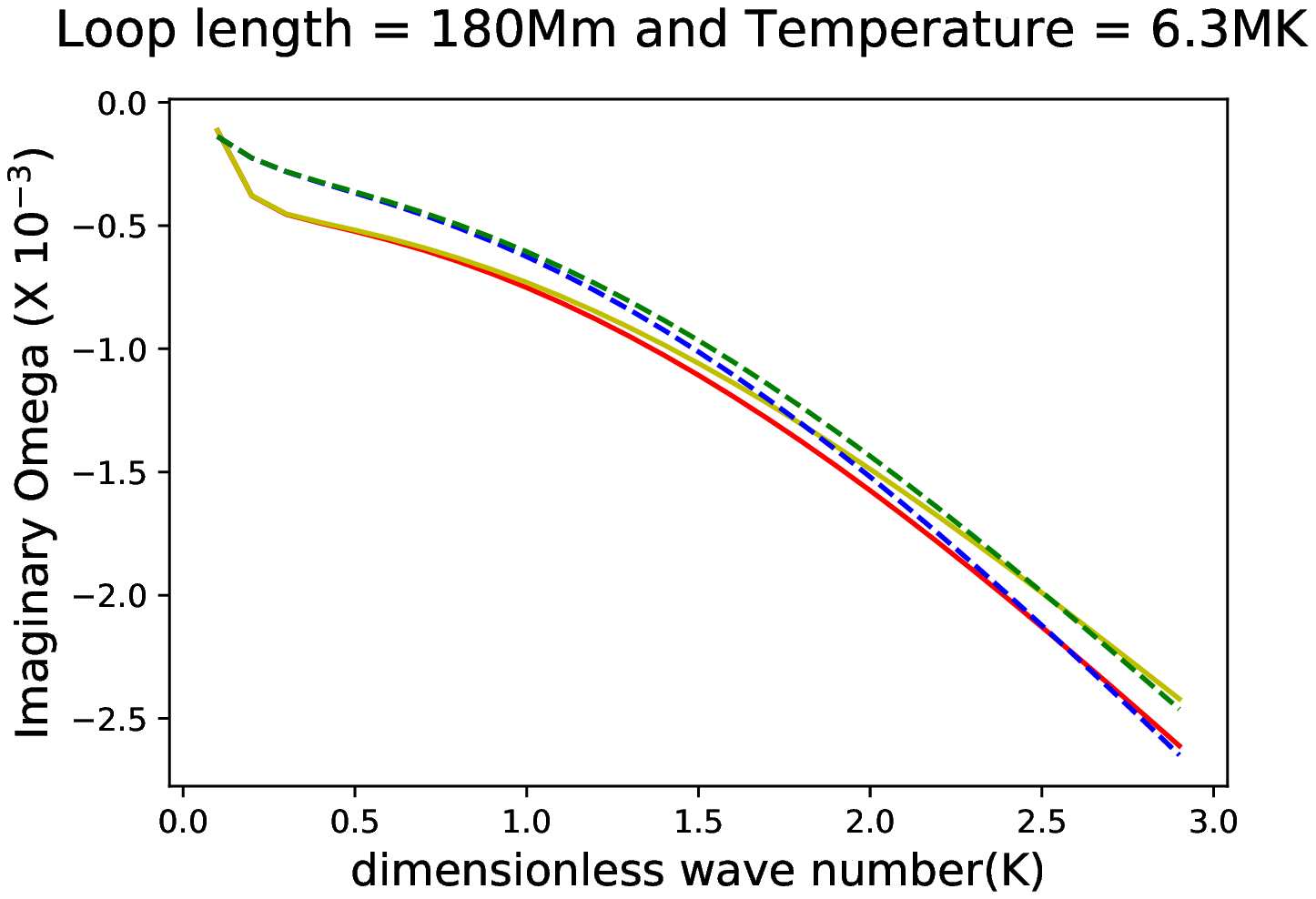}
              }
     \vspace{-0.35\textwidth}   
     \centerline{\Large \bf     
      \hspace{0.0 \textwidth}  \color{white}{(a)}
      \hspace{0.415\textwidth}  \color{white}{(b)}
         \hfill}
     \vspace{0.31\textwidth}    
   \centerline{\hspace*{0.015\textwidth}
               \includegraphics[width=0.615\textwidth,clip=]{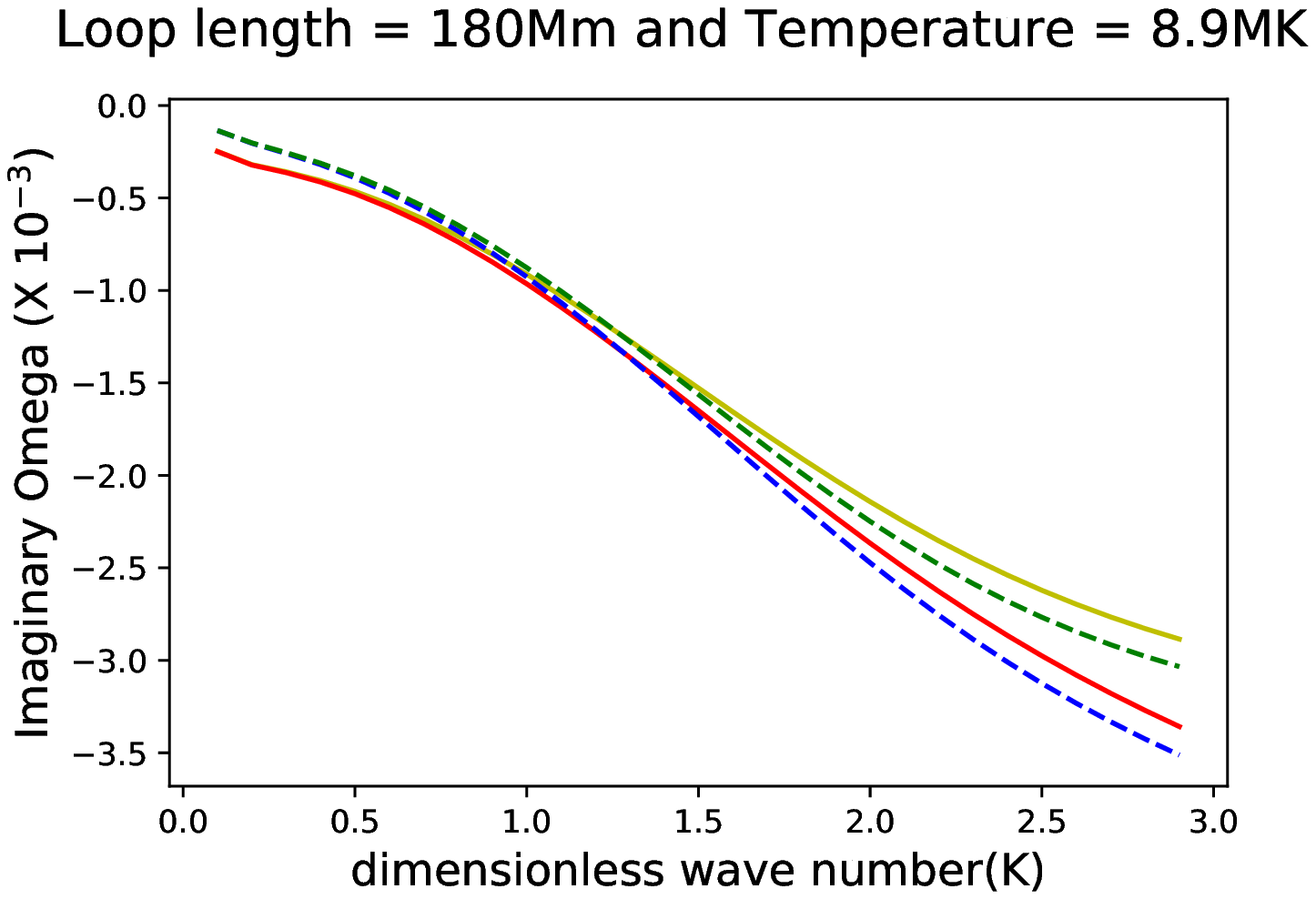}
               \hspace*{-0.03\textwidth}
               \includegraphics[width=0.615\textwidth,clip=]{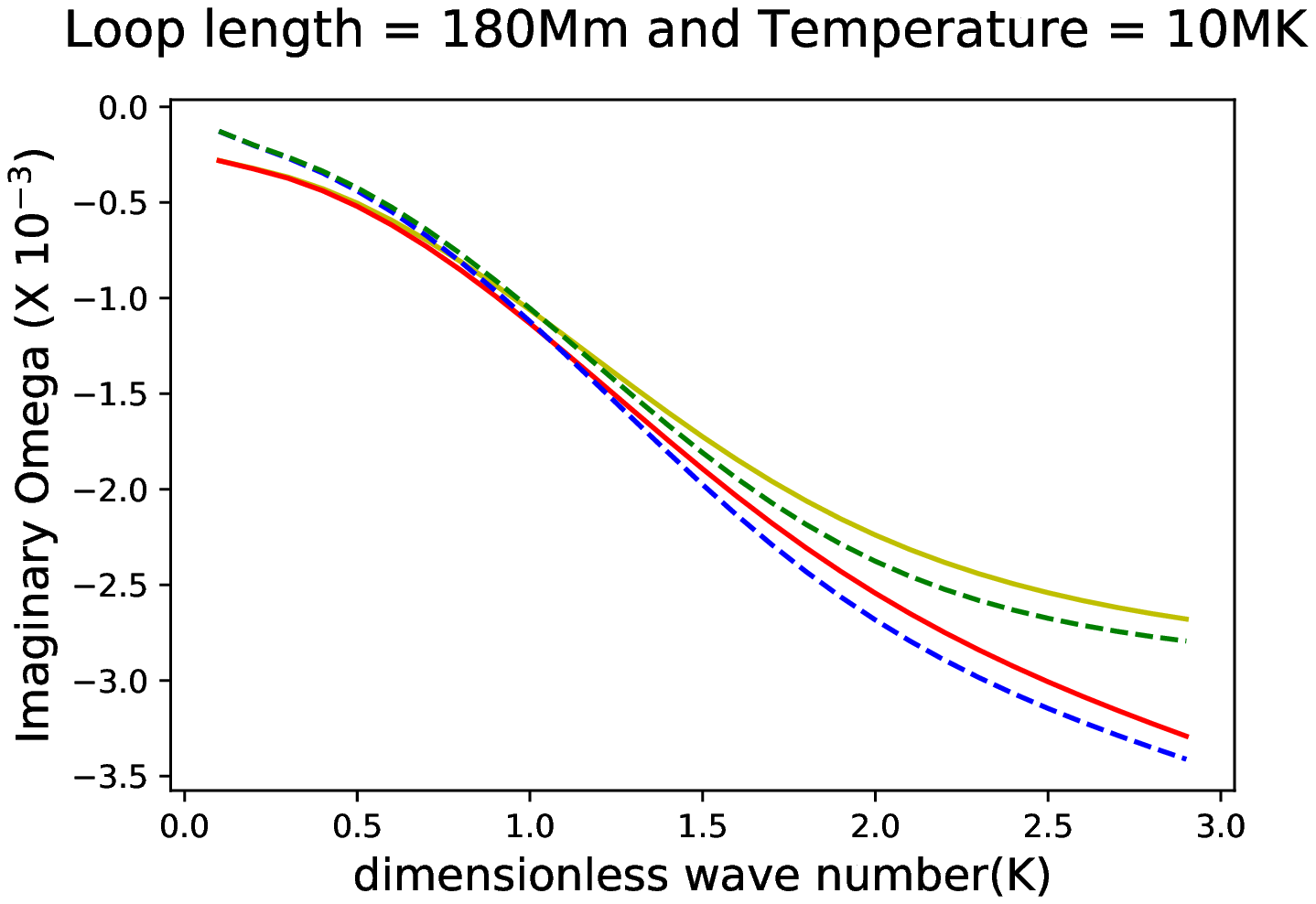}
              }
     \vspace{-0.35\textwidth}   
     \centerline{\Large \bf     
      \hspace{0.0 \textwidth} \color{white}{(c)}
      \hspace{0.415\textwidth}  \color{white}{(d)}
         \hfill}
     \vspace{0.31\textwidth}    
              
     \centerline{\hspace*{0.015\textwidth}
               \includegraphics[width=0.615\textwidth,clip=]{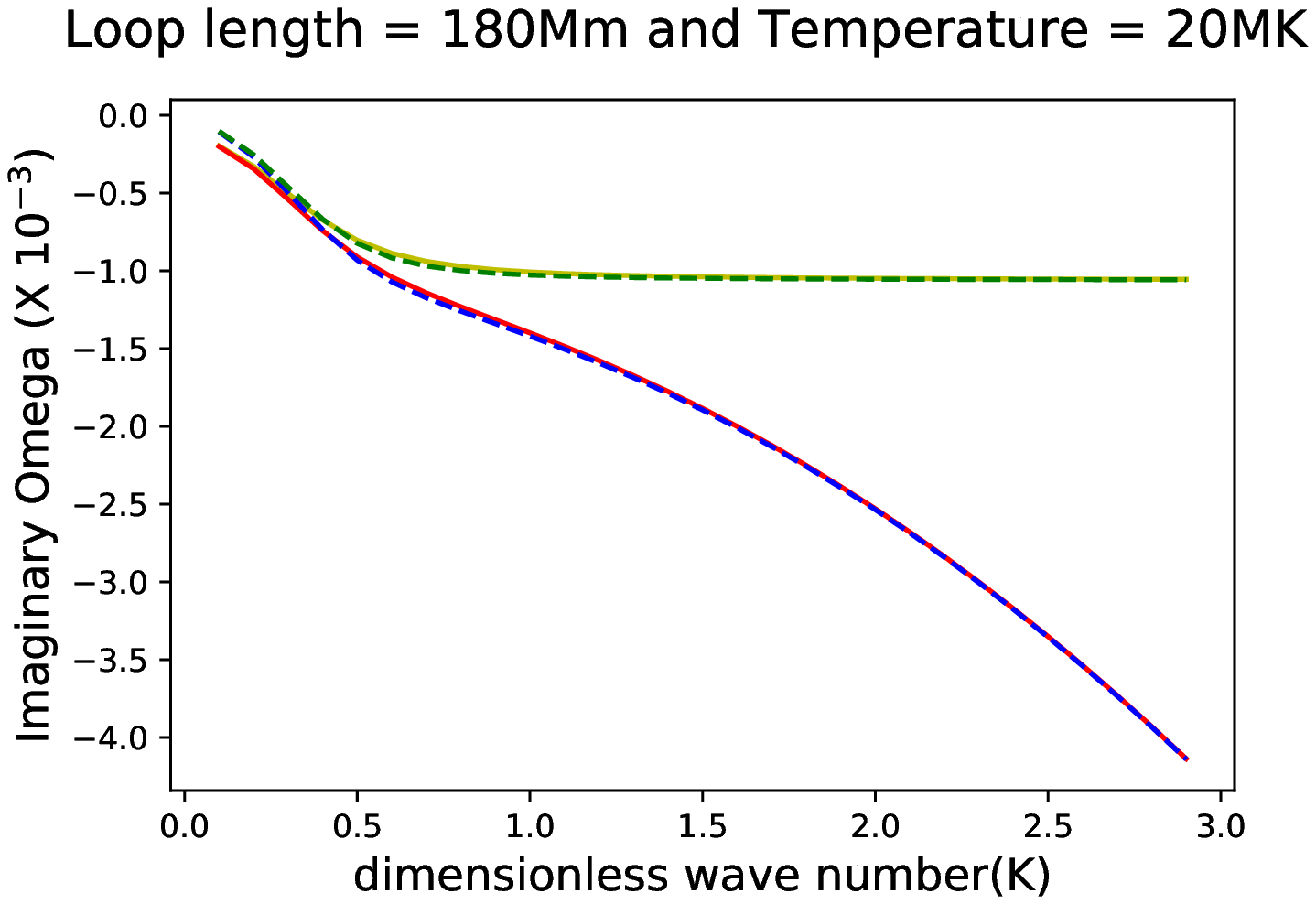}
               \hspace*{-0.03\textwidth}
               \includegraphics[width=0.615\textwidth,clip=]{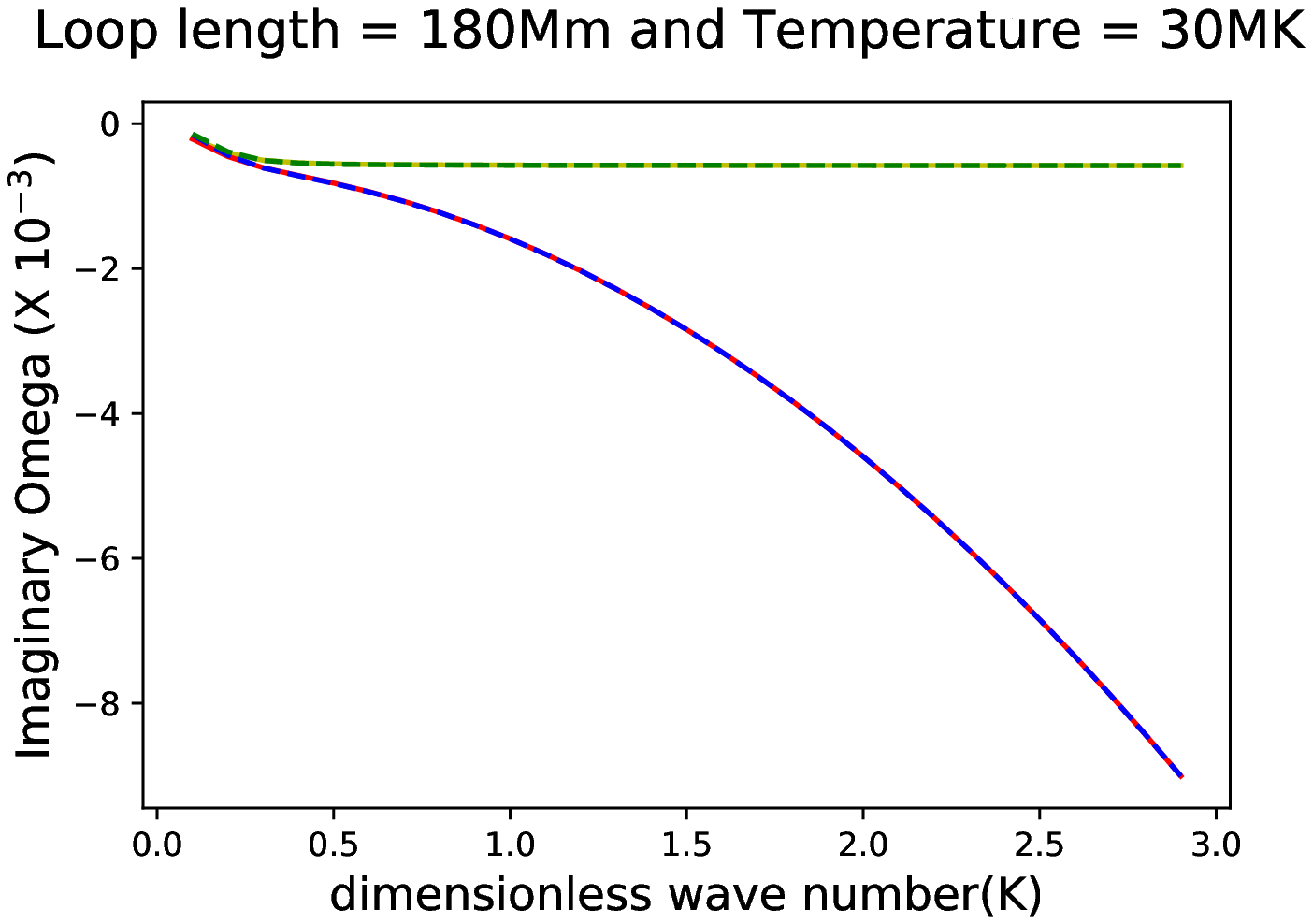}
              }
     \vspace{-0.35\textwidth}   
     \centerline{\Large \bf     
      \hspace{0.0 \textwidth} \color{white}{(c)}
      \hspace{0.415\textwidth}  \color{white}{(d)}
         \hfill}
     \vspace{0.31\textwidth}    
              
\caption{The panels show the variation of \(\omega_I\) with  dimensionless wave number $K$ at a fixed loop-length of 180 Mm for different temperatures from 5 MK to 30 MK. In each panel, the red and yellow curves correspond to the solution of the dispersion relation with and without the effect of compressive viscosity respectively when the heating-cooling imbalance is present. The blue-dotted and green-dotted curves  represent the solution of the dispersion relation with and without the effect of compressive viscosity respectively when the heating-cooling imbalance is not present.Thermal conductivity is always present as a damping mechanism in these analyses.}
   \label{F-4panels}
   \end{figure}
From the following analysis of the results shown in Figures 1-5, we find that the effect of heating-cooling imbalance on damping is more remarkable for the loops of lower temperatures but greater lengths. Panels in Figure 1-3 show the variation of \(\omega_I\) with  dimensionless wave number $K$ at fixed loop lengths (50, 180, 500 Mm) for different temperatures from 5 MK to 30 MK. In each panel, the red and yellow curves are the solution of the dispersion relation respectively with and without the effect of compressive viscosity when the heating-cooling imbalance is present. The blue-dotted and green-dotted curves are the solution of the dispersion relation respectively with and without the effect of compressive viscosity when the heating-cooling imbalance is not present. It should be noted that thermal conductivity is by default present in these analytical calculations. Therefore, when we consider an absence of compressive viscosity, this means that in the  background thermal conductivity is present as a natural damping mechanism. When we include compressive viscosity, its effect will be added to the presence of thermal conductivity. These estimations and related analyses are performed both with and without heating-cooling imbalance. Therefore, in principle, the red curve shows the joint effect of both thermal conductivity and compressive viscosity during heating-cooling imbalance, while the yellow one exhibits only the presence of thermal conductivity with heating-cooling imbalance. Similarly the blue-dotted curve shows the joint effect of both thermal conductivity and compressive viscosity during the absence of heating-cooling imbalance, while the green-dotted curve includes only the effect of thermal conductivity without heating-cooling imbalance. 

In the next two sections, we will provide the detailed results and their analyses/interpretations regarding the effect of thermal conductivity and viscosity on the 
fundamental modes, and higher order harmonics of the slow waves respectively. 


\subsubsection{Effect of Thermal Conductivity and Viscosity on the Damping of the Fundamental Modes}

In Figure 1, we present the variation of \(\omega_I\)  w.r.t. $K$ for the fundamental modes in a loop of length $L$=50 Mm. We show this variation at $T$=5.0, 6.3, 8.9, 10, 20, and 30 MK. In the super-hot regime at $T$=20, and 30 MK, for the fundamental mode ($K$=1.0), the value of \(\omega_I\) represented by red and blue-dotted curves drops down significantly compared to the same depicted by the yellow and green-dotted curve (cf. bottom-left and right panels in Figure 1). This shows that the compressive viscosity causes an enhanced damping of the fundamental mode in a super-hot regime of shorter loops, when it is added in the presence of thermal conductivity. 
The red and blue-dotted (or yellow and green-dotted) curves coincide with each other (cf. bottom-left and right panels in Figure 1), which depicts that heating-cooling imbalance has no significant effects on the damping of the fundamental modes in the super-hot regime of shorter loops ($T>$10 MK).
In conclusion, Figure 1 demonstrates that the compressive viscosity  clearly enhances the damping of the fundamental modes in shorter loops ($L$=50 Mm) at a super-hot regime of $T>$10 MK in the presence of thermal conductivity.
At a hot regime of $T\leq$10 MK at $T$= 8.9 and 10 MK (cf. middle-left and right panels in Figure 1), the compressive viscosity still plays certain roles in the damping of the fundamental mode of slow waves along with thermal conductivity, but not as effective as we have already seen for the super-hot regime.
Again, the effect of heating-cooling imbalance in this condition is still negligible.
As temperature goes down to $T$=6.3 MK and then further to $T$=5.0 MK (cf. top-left and right panels in Figure 1), for the fundamental mode ($K$=1.0), the value of \(\omega_I\) shown by red, blue-dotted, yellow, and green-dotted curves is almost the same. This indicates that the effects of both heating-cooling imbalance and viscosity  on the damping of the fundamental mode are insignificant compared to the thermal conductivity.

In Figure 2, we present the variation of \(\omega_I\)  w.r.t. $K$ in the case of $L$=180 Mm and $T$=5.0, 6.3, 8.9, 10, 20, and 30 MK. In the super-hot regime at $T$=20, and 30 MK, for the fundamental mode ($K$=1.0), the value of \(\omega_I\) shown by red and blue-dotted curves drops down compared to that depicted by yellow and green-dotted curves (cf. bottom-left and right panels in Figure 2). This shows that compressive viscosity along with thermal conductivity cause an enhanced damping of the fundamental mode in a super-hot regime for the loops of intermediate length ($L$=180 Mm).
It is also seen that the red and blue-dotted (or yellow and green-dotted) curves coincide with each other, which means that heating-cooling imbalance has little effects on the damping of the fundamental mode oscillations in the super-hot regime loops even with an intermediate length.
In conclusion, compressive viscosity  has a definite role along with the presence of thermal conductivity in the damping of the fundamental oscillations for intermediate loops (e.g. $L$=180 Mm) at the super-hot regime of $T>$10 MK. However, this physical effect is comparatively weaker compared to that in the case of shorter loops (cf. compare the bottom panels of Figures 1 and 2). In the hot regime at $T$= 8.9 and 10 MK (cf. middle-left and right panels in Figure 2), for the fundamental mode ($K$=1.0), the value of \(\omega_I\) on red/blue-dotted curves has a smaller drop in its value compared to the yellow/green-dotted curves. Moreover, red and blue-dotted curves coincide with each other, and the same is true for the yellow and green-dotted ones. These results infer that the heating-cooling imbalance affects little the damping of the fundamental mode oscillations. Additionally it indicates that the damping caused by the viscosity is insignificant compared to that caused by thermal conductivity. The effect of the compressive viscosity is nearly minimal at these temperatures. At $T$=5.0 and 6.3 MK (cf. top-left and right panels in Figure 2), for the fundamental mode, the value of \(\omega_I\) represented by red/yellow curves is lower compared to that shown by blue/green-dotted curve. Moreover, the red curve coincides with the yellow one and the same is valid for the green and blue-dotted curves. The physical scenario for the damping of the fundamental mode in the loops of intermediate length (e.g. $L$=180 Mm) at these lower temperatures now demonstrates that (i) the heating-cooling imbalance enhances the damping of the fundamental mode, (ii) the compressive viscosity has almost no effect on the damping, therefore, thermal conductivity dominates.

In Figure 3, we present the variation of \(\omega_I\) with $K$ in the case of $L$=500 Mm and $T$=5.0, 6.3, 8.9, 10, 20, and 30 MK. At the highest temperature of 30 MK in the super-hot regime, for the fundamental mode ($K$=1.0), the value of \(\omega_I\) shown by red and blue-dotted curves drops down compared to that represented by yellow and green-dotted curves (cf. bottom-right panel in Figure 3). This shows that the compressive viscosity along with thermal conductivity causes an enhanced damping of the fundamental mode at the highest temperature in the longest loop considered in our analysis ($L$=500 Mm).
It is also seen that the red and blue-dotted (or yellow and green-dotted) curves coincide with each other at 30 MK, which means that heating-cooling imbalance has no distinct effects on the damping of the fundamental mode  compared to the case when it is absent.
At $T$=20 MK (cf. bottom-left panel in Figure 3), for the fundamental mode ($K$=1.0), the value of \(\omega_I\) shown by red, blue-dotted, yellow, and green-dotted curves is almost the same. The heating-cooling imbalance, therefore, does not cause any enhanced damping of the fundamental mode compared to the case when it is absent. Moreover, the damping effect of the compressive viscosity is not important compared to that of thermal conductivity even at this high temperature.
In the hot regime at $T$=5\,--\,10 MK, for the fundamental mode ($K$=1.0), \(\omega_I\) shown by red/yellow curves has lower values compared to the case of the blue/green-dotted curves (cf. top-left and right, middle-left and right in Figure 3). Further, the red curve coincides with the yellow one, and the same is true for the green and blue-dotted curves. This suggests that the heating-cooling imbalance causes more damping of the fundamental mode compared to the case when it is not present. In addition, this indicates that the compressive viscosity has insignificant effect on the damping of the fundamental mode, therefore, thermal conductivity dominates. Moreover, as the temperature goes towards the lowest value the heating-cooling imbalance causes larger damping compared to the case without it. 
 
\subsubsection{Effect of Thermal Conductivity and Viscosity on the Damping of the Higher Order Harmonics}

Figure 1 also shows that in the case of shorter loops ($L$=50 Mm) at the super-hot regime ($T>10$ MK) the compressive viscosity dominates in the damping and its effect increases with the harmonic number, whereas heating-cooling imbalance has little effects on the damping and is nearly independent of the harmonic number.


Figure 2 shows that in the case of intermediate loops (e.g. $L$=180 Mm), the behavior of the compressive viscosity and heating-cooling imbalance on the damping of higher harmonics is similar to that for shorter loops in the super-hot regime. While at the lowest temperature of $T$=5 MK,  the thermal conduction dominates in the damping and its effect increases with the harmonic number, the heating-cooling imbalance plays a little role and its dependence on the harmonic number is weak.


Figure 3 shows that in the super-hot regime, the longest chosen loop ($L$= 500 Mm) has similar damping properties for higher harmonics as the intermediate loops. While at the lowest temperature of  $T$=5 MK, thermal conduction still dominates in the damping and its effect increases with the harmonic number. Compared to the intermediate loops, the longest loop shows a significant effect of the heating-cooling imbalance on the damping, which appears to be independent of the harmonic number.

In the next section, we present a brief summary of the results obtained in Sections 4.1.1 and 4.1.2.

\subsubsection{Brief Summary}

Note that in Sections 4.1.1 and 4.1.2, we have not compared the individual roles of thermal conduction and viscosity in the damping of the fundamental modes and higher order harmonics of slow waves, which will be done next in Section 4.2. In summary, from Figures 1-3, we find that (i) viscosity causes a significant and dominant damping of slow modes in the super-hot regime for shorter loops with a damping rate even higher for the higher harmonics (cf. bottom panels of Figure 1), while the heating-cooling imbalance has little effect on the damping of slow modes in this condition; (ii) the heating-cooling imbalance causes a significantly enhanced damping of slow modes in the less hot and longer loops with effects that weakly depend on the frequency (or harmonic number; cf. top panels of Figure 3), while the viscosity plays nearly no role or a very small one in the damping for this condition.
It is worth noting that the y-axes on Figures 1-3 have varying scales of \(\omega_I\), therefore, describing significantly different decay rates for the different cases.

Many previous studies on the damping of standing slow MHD waves in coronal loops have defined dimensionless parameters in order to quantify the damping effect of thermal conductivity, viscosity, and radiative losses/gains \citep{2003A&A...408..755D,2004A&A...415..705D,2007SoPh..246..187S}. These parameters are defined as
\begin{equation}
    \epsilon = \frac{\eta_0}{\rho_0Lc_s}\,\,\,\,{\text {(viscous ratio)}},
\end{equation}
\begin{equation}
    d = \frac{(\gamma-1)\kappa\rho_0T_0c_s}{\gamma^2 p_0^2 L } \,\,\,\,\, {\text{(thermal ratio)}},
\end{equation}
\begin{equation}
    r = \frac{(\gamma-1)L\rho_0\mathcal{L}(\rho_0,T_0)}{\gamma p_0 c_s}\,\,\,\,{\text{(radiative ratio)}}.
\end{equation}
Here $c_s = \left (\frac{\gamma k_B T_0}{m}\right )^{1/2}$.\newline
The damping rate of the standing waves has been analytically studied in terms of these dimensionless ratios and it was found that it is a monotonically increasing function of $\epsilon$ \citep{2007SoPh..246..187S} whereas it increases upto a peak maximum value and then reduces with increasing $d$ ratio \citep{2003A&A...408..755D}. In the regime for hot (5\,--\,10 MK) and longer loops (500 Mm) with normal density, both $d$ and $\epsilon$ are small ($\ll1$), this is the condition for the so called weak dissipation approximation, in this case, $\frac{d}{\epsilon}$ is a constant and $\frac{d}{\epsilon} \gg1$, so thermal conduction dominates in the damping. In the regime of super-hot (20\,--\,30MK) and short loops (50 Mm) with normal density both thermal ratio ($d \propto \frac{T_0^2}{\rho_0L}$) and viscous ratio ($\epsilon \propto \frac{T_0^2}{\rho_0 L}$) are large ($\gg1$). The different behaviour of the damping rate depending on $d$ and $\epsilon$ implies that when both $d$ and $\epsilon$ are $\gg1$, the viscous damping is dominant.

In the next section, we analyze the velocity oscillations of the fundamental mode in the hot regime of temperature $T\leq$10 MK.

\begin{figure}   
   \centerline{\hspace*{0.015\textwidth}
               \includegraphics[width=0.615\textwidth,clip=]{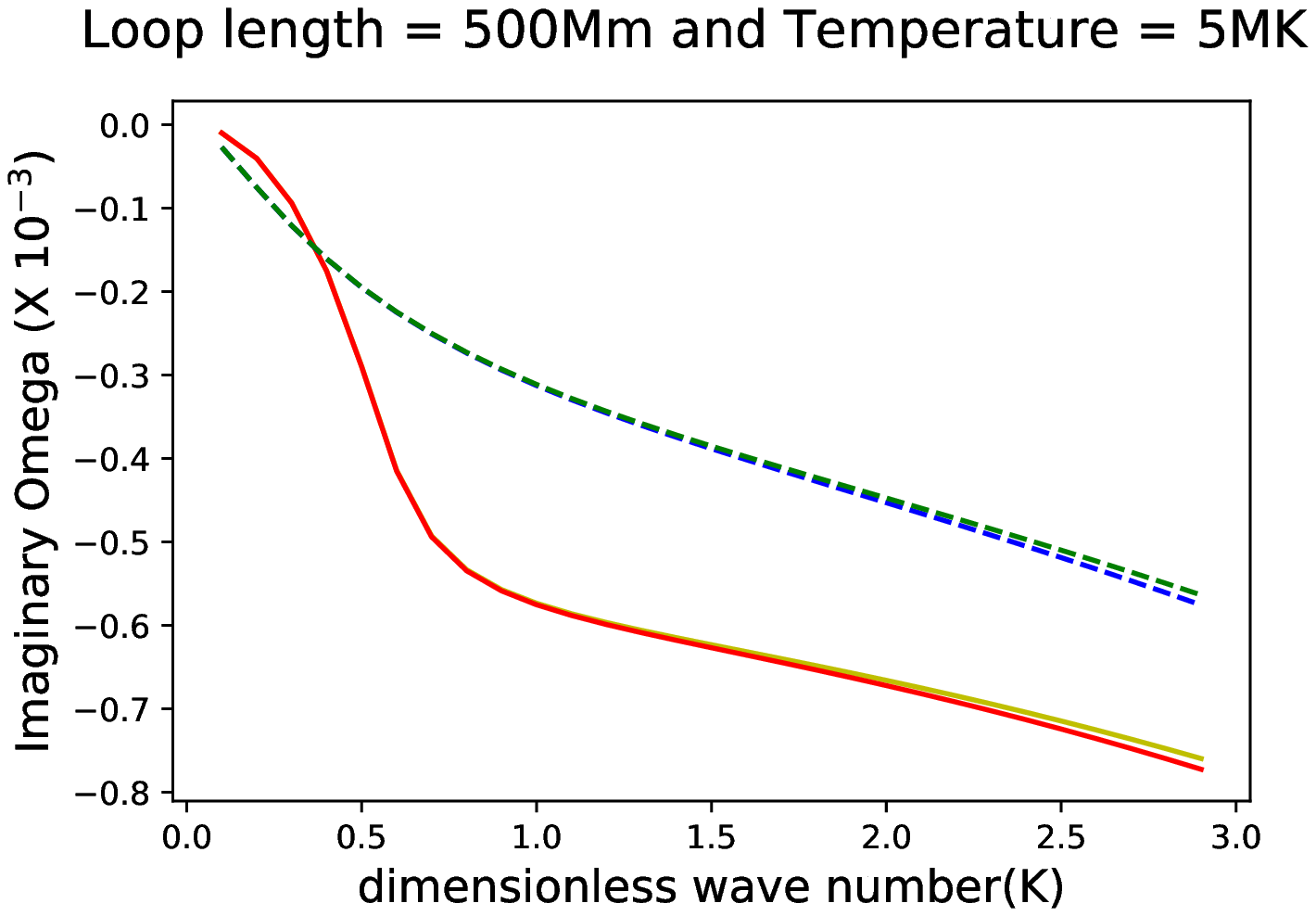}
               \hspace*{-0.03\textwidth}
               \includegraphics[width=0.615\textwidth,clip=]{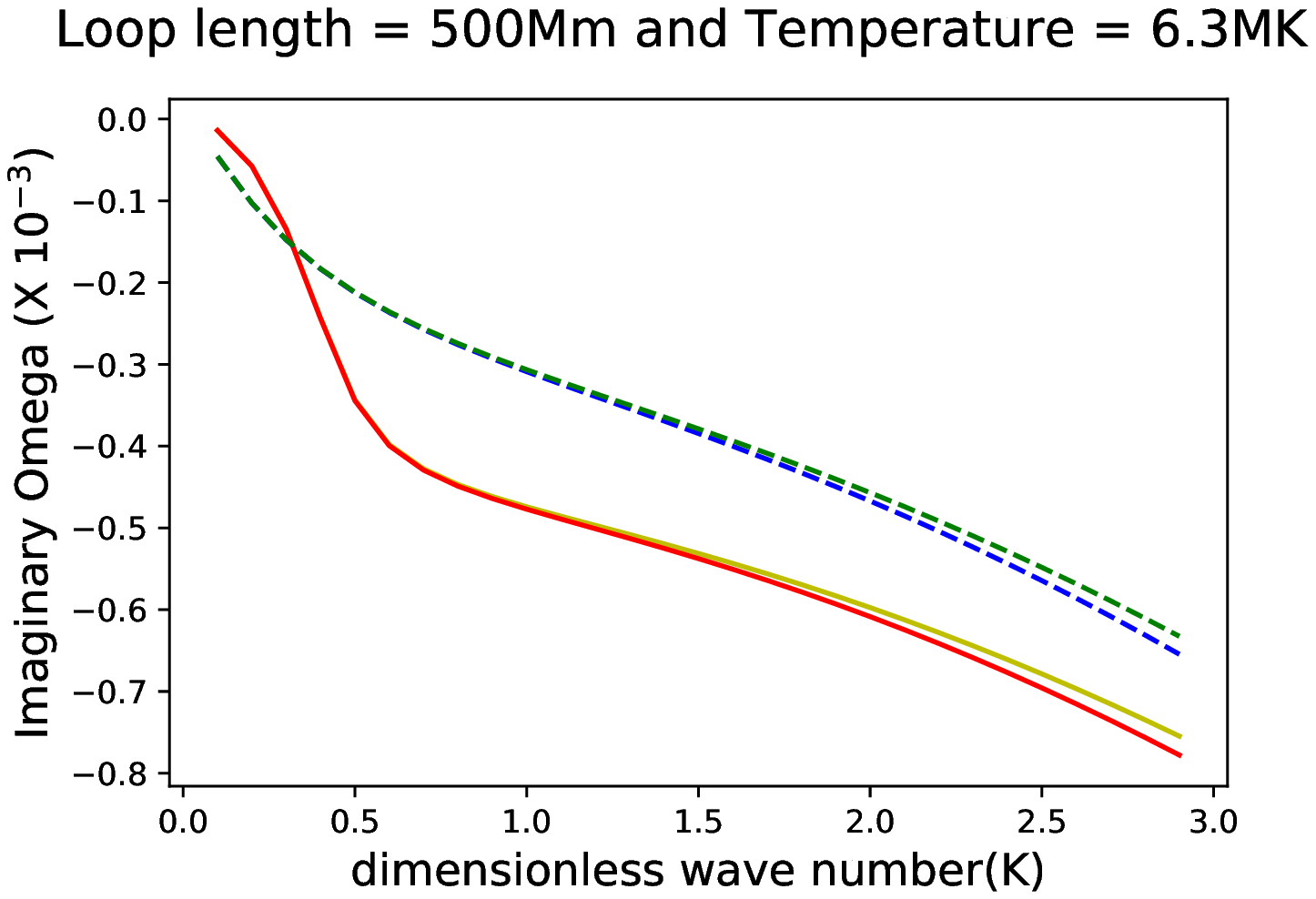}
              }
     \vspace{-0.35\textwidth}   
     \centerline{\Large \bf     
      \hspace{0.0 \textwidth}  \color{white}{(a)}
      \hspace{0.415\textwidth}  \color{white}{(b)}
         \hfill}
     \vspace{0.31\textwidth}    
   \centerline{\hspace*{0.015\textwidth}
               \includegraphics[width=0.615\textwidth,clip=]{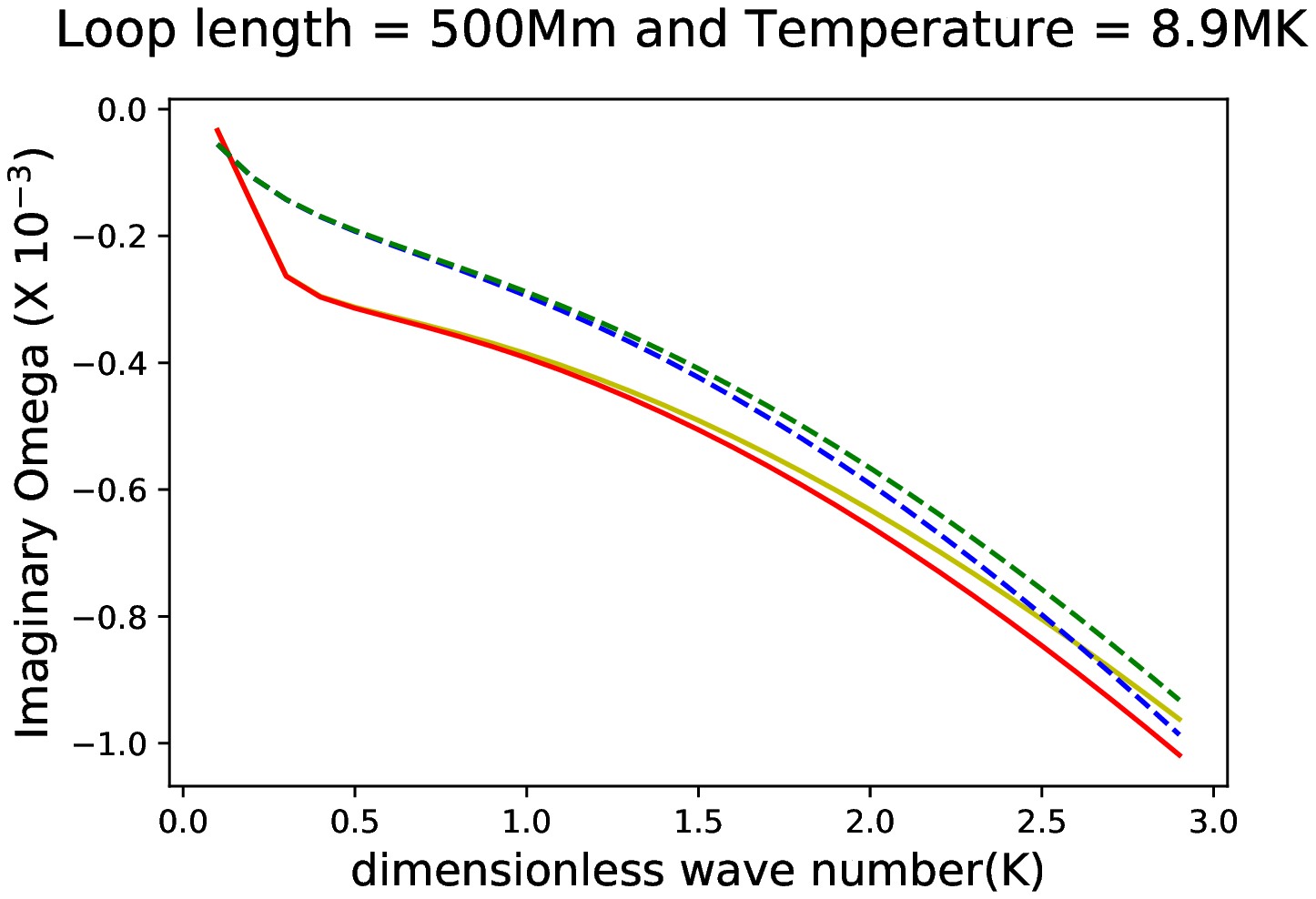}
               \hspace*{-0.03\textwidth}
               \includegraphics[width=0.615\textwidth,clip=]{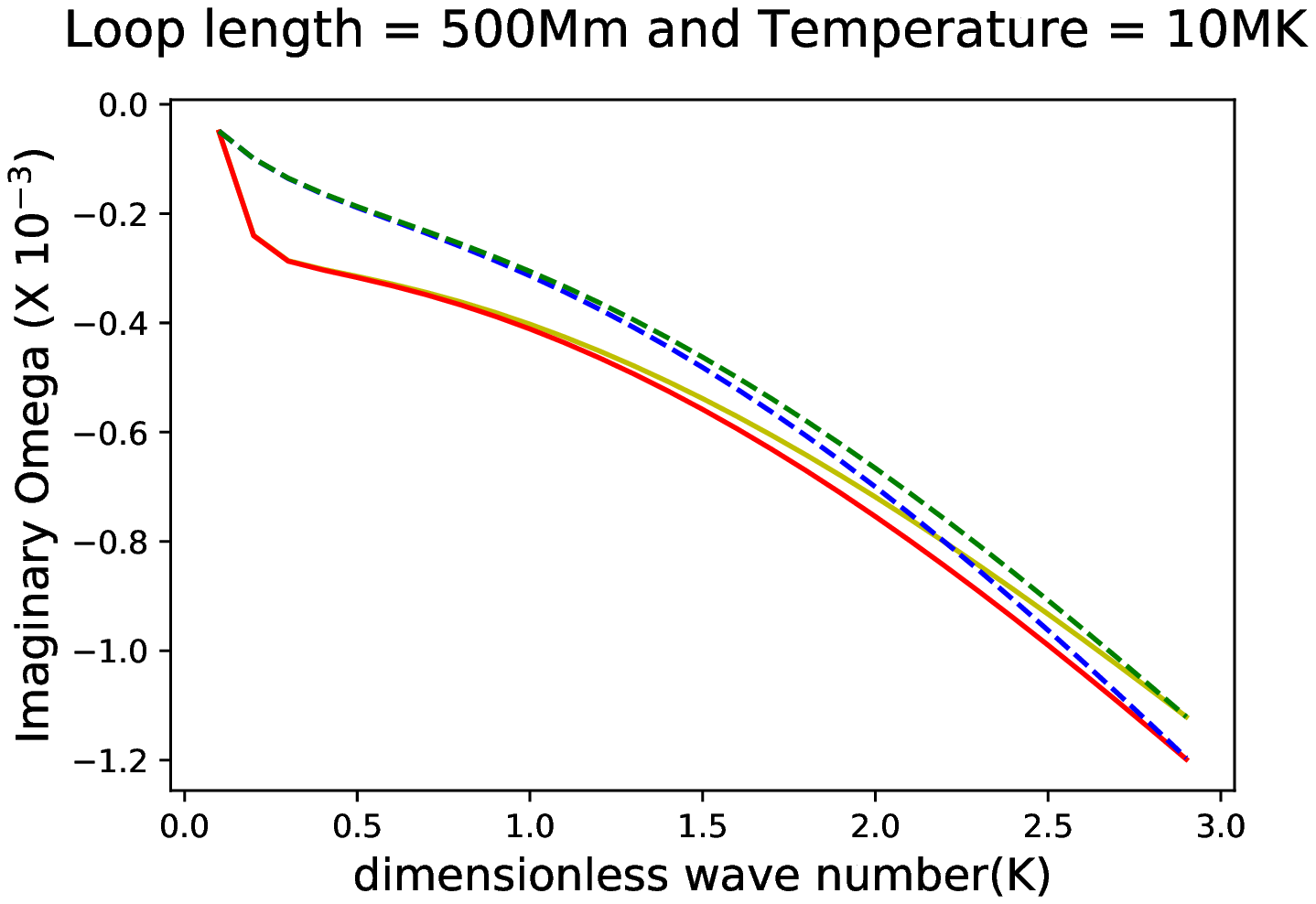}
              }
     \vspace{-0.35\textwidth}   
     \centerline{\Large \bf     
      \hspace{0.0 \textwidth} \color{white}{(c)}
      \hspace{0.415\textwidth}  \color{white}{(d)}
         \hfill}
     \vspace{0.31\textwidth}    
     
     \centerline{\hspace*{0.015\textwidth}
               \includegraphics[width=0.615\textwidth,clip=]{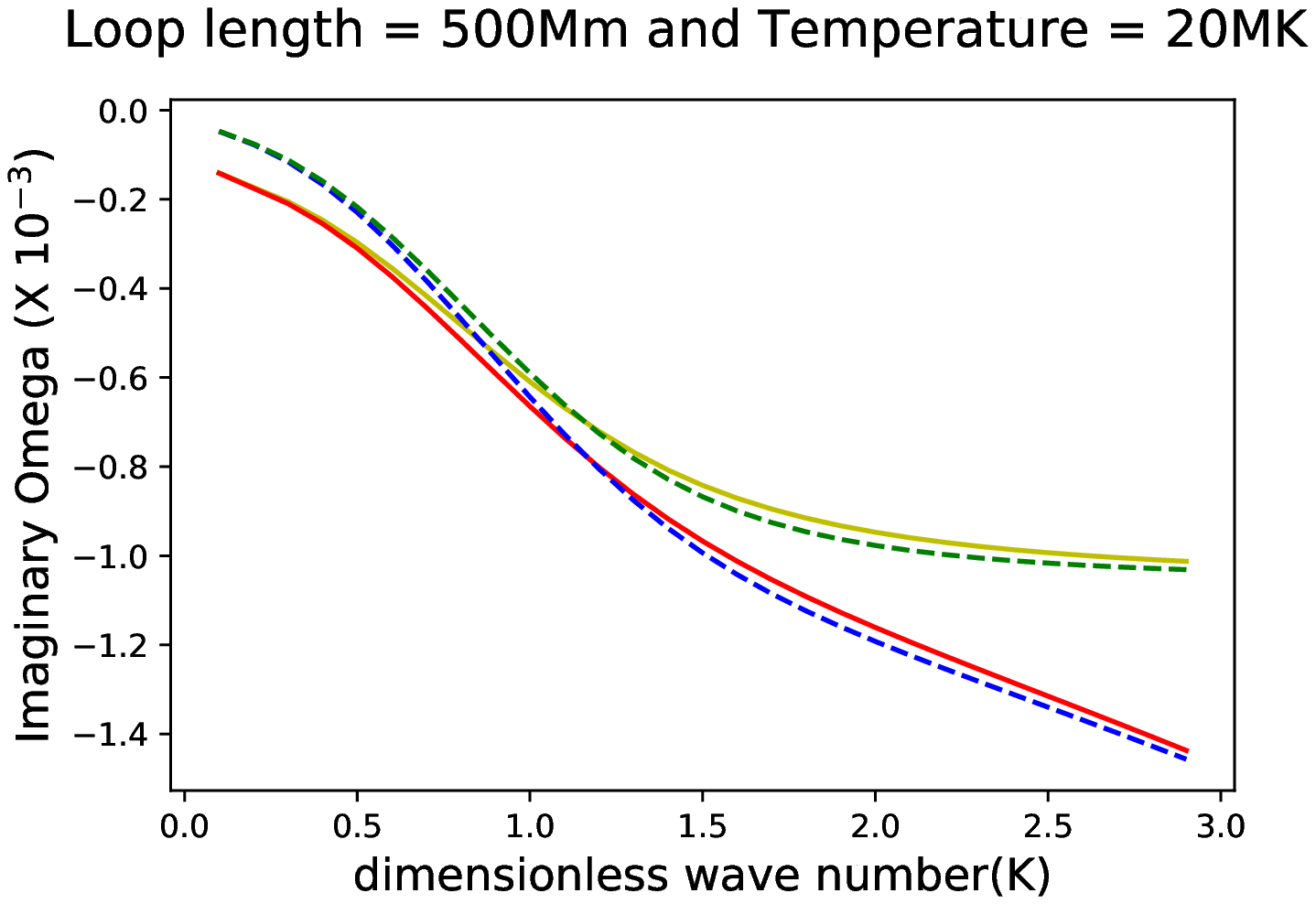}
               \hspace*{-0.03\textwidth}
               \includegraphics[width=0.615\textwidth,clip=]{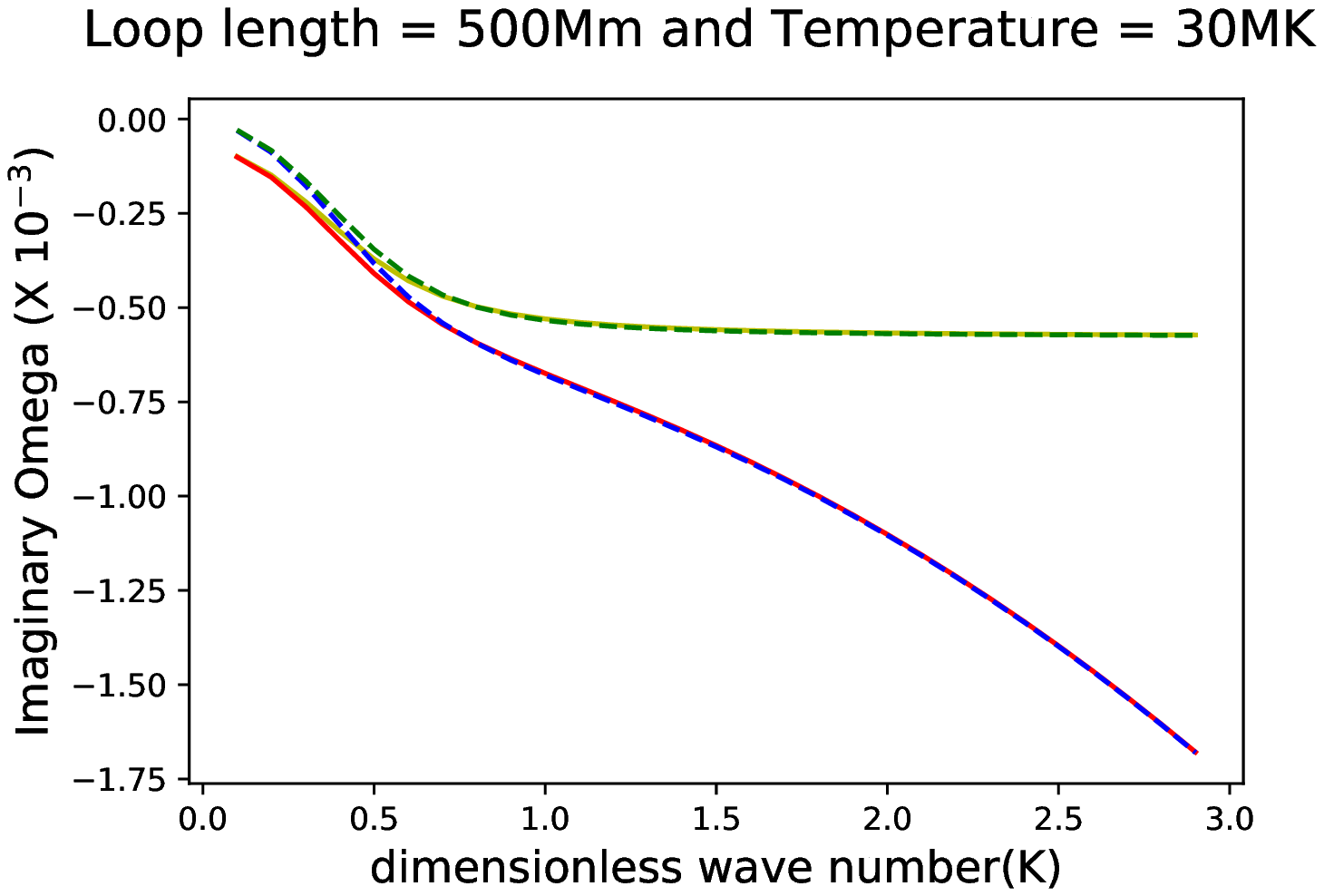}
              }
     \vspace{-0.35\textwidth}   
     \centerline{\Large \bf     
      \hspace{0.0 \textwidth} \color{white}{(c)}
      \hspace{0.415\textwidth}  \color{white}{(d)}
         \hfill}
     \vspace{0.31\textwidth}    
              
\caption{The panels show the variation of \(\omega_I\) with  dimensionless wave number K at a fixed loop-length of 500 Mm for different temperatures from 5 MK to 30 MK. In each panel, the red and yellow curves correspond to the solution of the dispersion relation with and without the effect of compressive viscosity respectively when the heating-cooling imbalance is present. The blue-dotted and green-dotted curves  represent the solution of the dispersion relation with and without the effect of compressive viscosity respectively when the heating-cooling imbalance is not present. Thermal conductivity is always present as a damping mechanism in these analyses.}
   \label{F-4panels}
   \end{figure}
\begin{figure}   
   \centerline{\hspace*{0.015\textwidth}
               \includegraphics[width=0.495\textwidth,clip=]{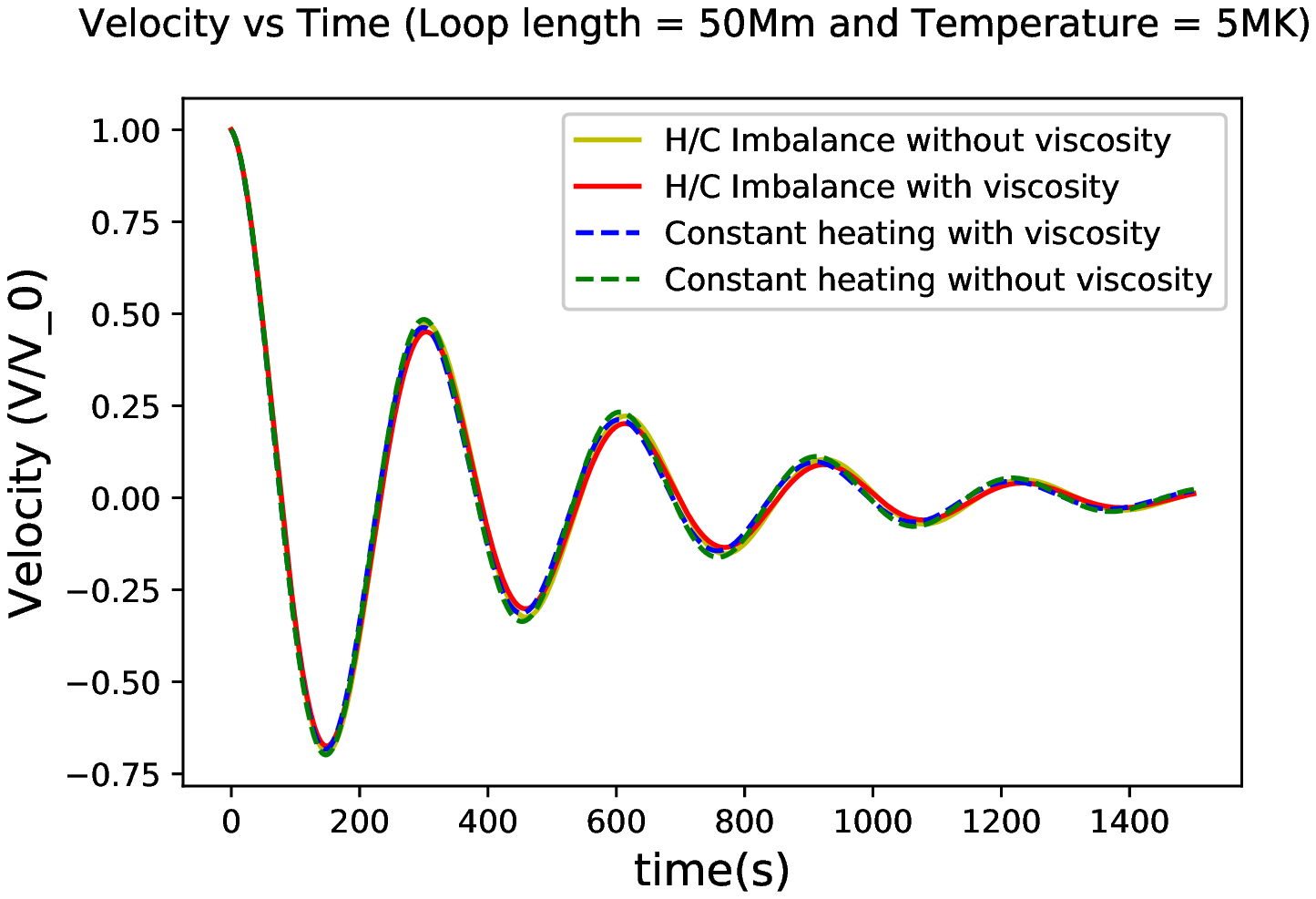}
               \hspace*{-0.015\textwidth}
               \includegraphics[width=0.495\textwidth,clip=]{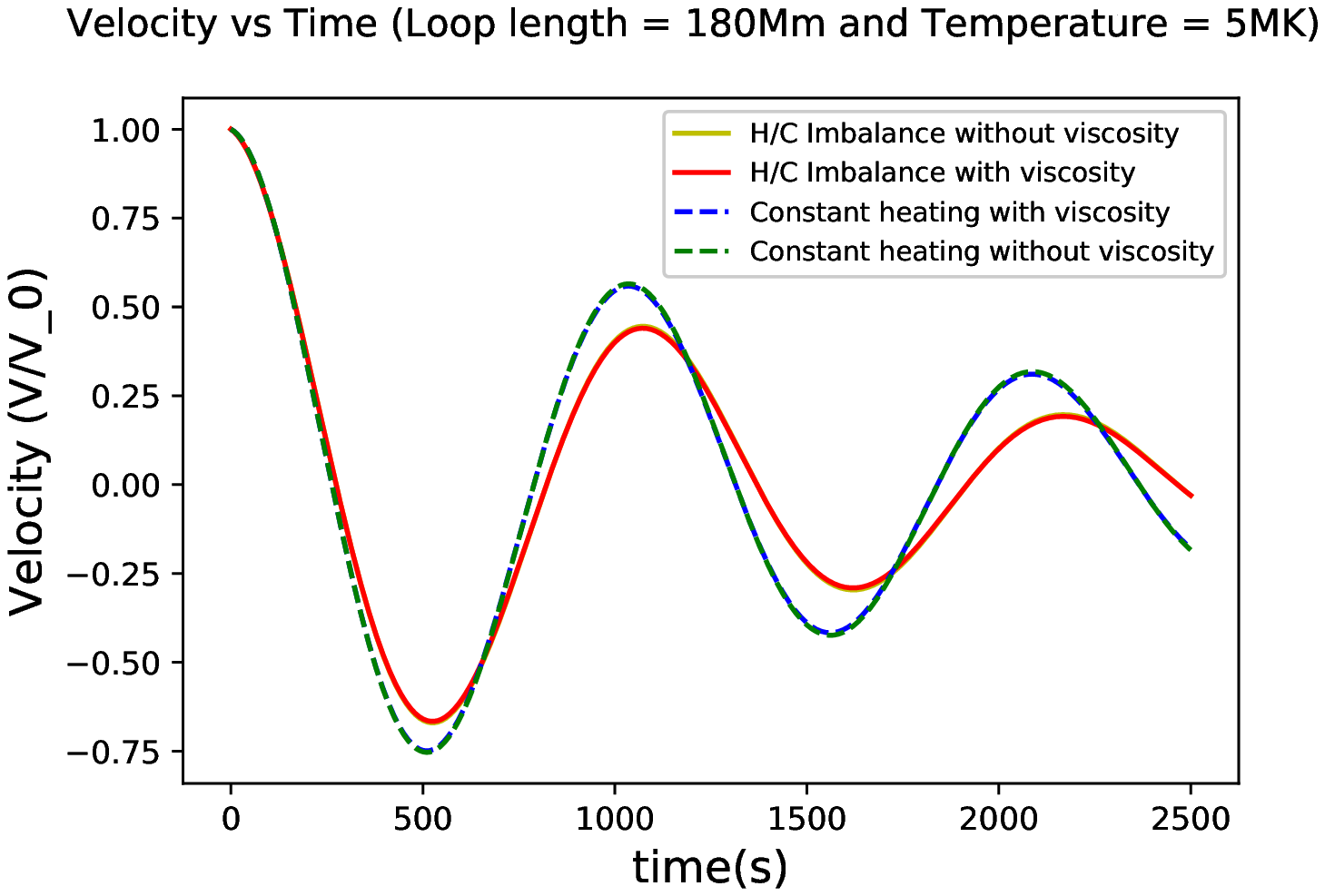}
               \hspace*{-0.015\textwidth}
               \includegraphics[width=0.495\textwidth,clip=]{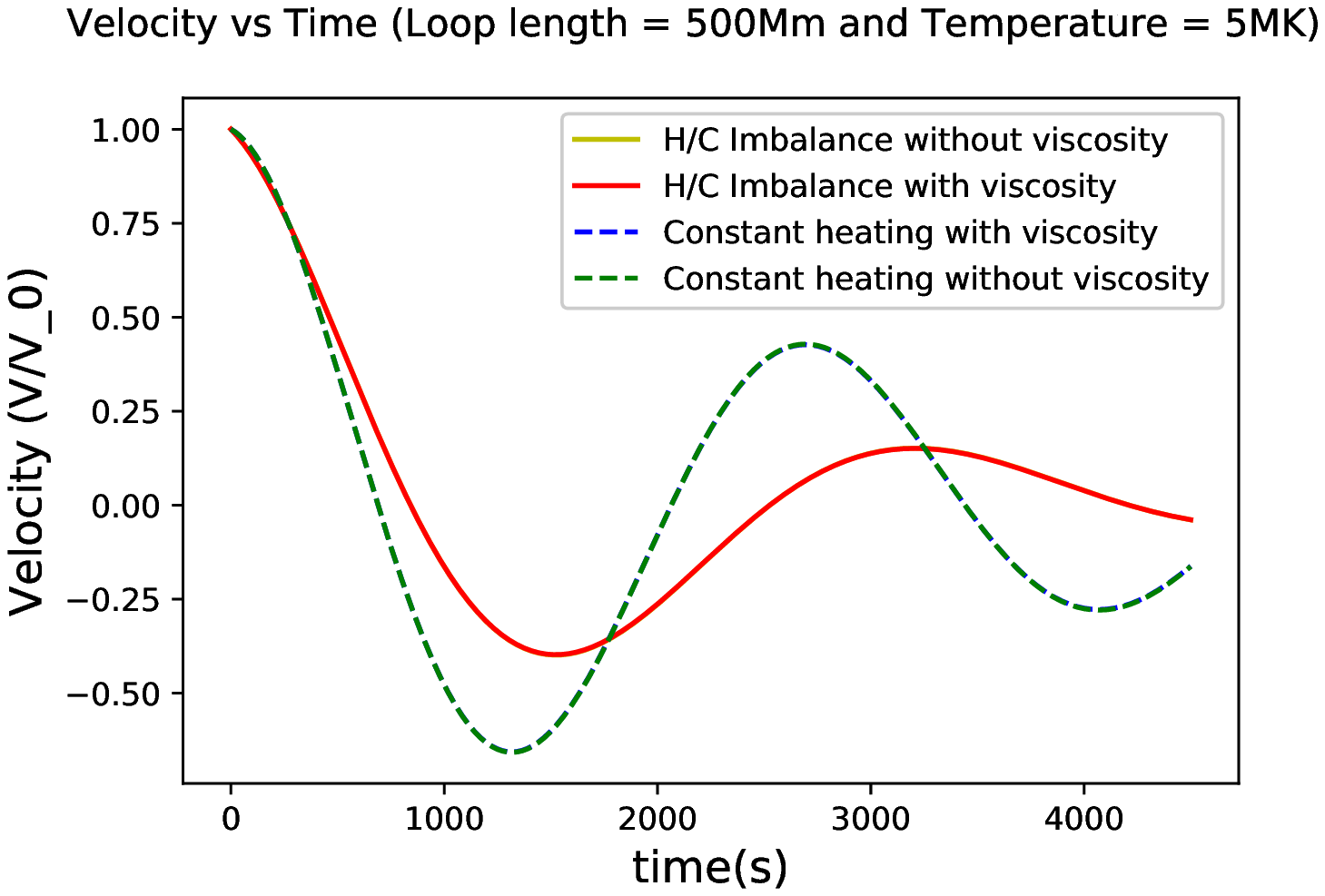}
              }
     \vspace{-0.35\textwidth}   
     \centerline{\Large \bf     
      \hspace{0.0 \textwidth}  \color{white}{(a)}
      \hspace{0.415\textwidth}  \color{white}{(b)}
         \hfill}
     \vspace{0.31\textwidth}    
   \centerline{\hspace*{0.015\textwidth}
               \includegraphics[width=0.495\textwidth,clip=]{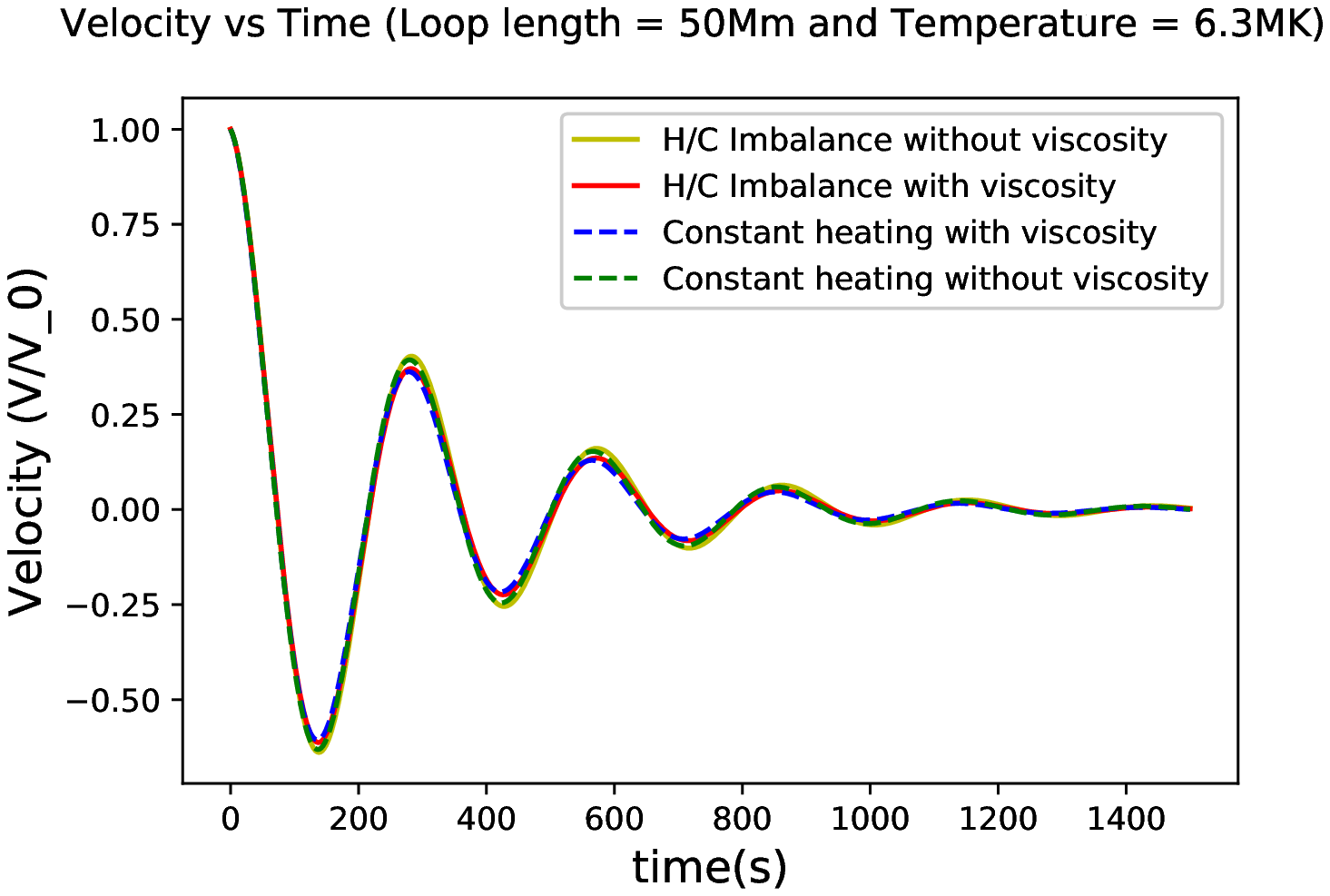}
               \hspace*{-0.015\textwidth}
               \includegraphics[width=0.495\textwidth,clip=]{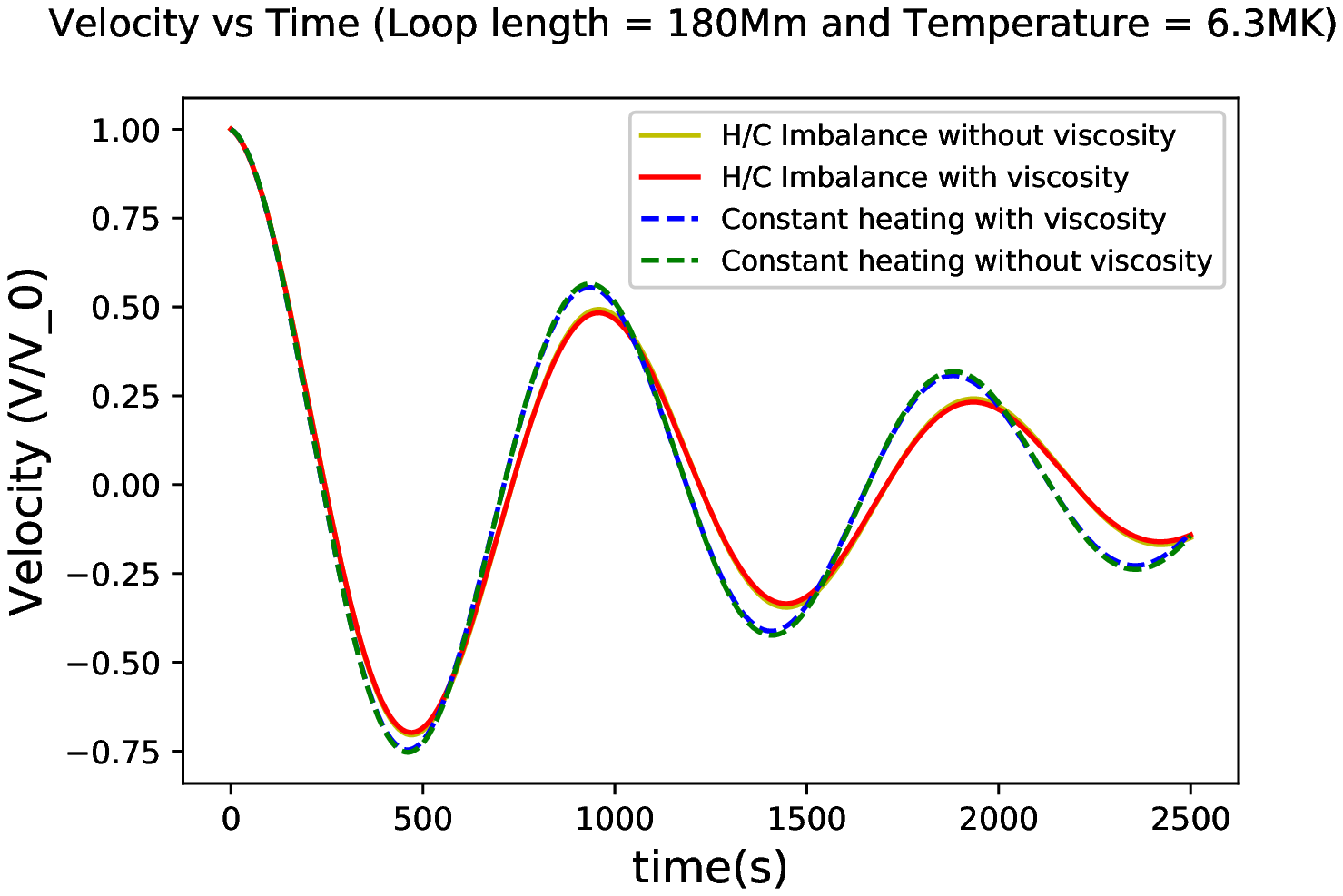}
               \hspace*{-0.015\textwidth}
               \includegraphics[width=0.495\textwidth,clip=]{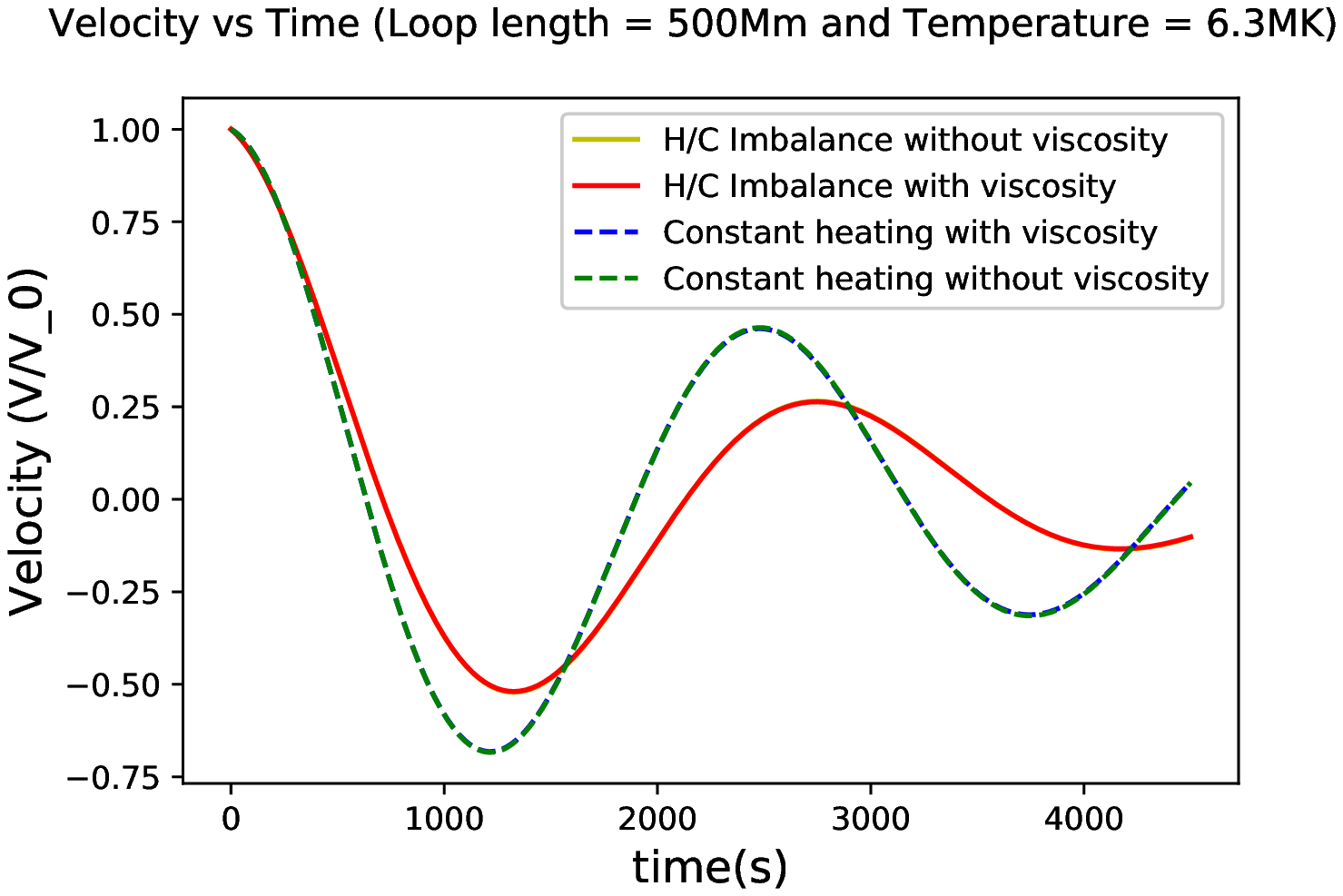}
              }
     \vspace{-0.35\textwidth}   
     \centerline{\Large \bf     
      \hspace{0.0 \textwidth} \color{white}{(c)}
      \hspace{0.415\textwidth}  \color{white}{(d)}
         \hfill}
     \vspace{0.31\textwidth}    
              
     \centerline{\hspace*{0.015\textwidth}
               \includegraphics[width=0.495\textwidth,clip=]{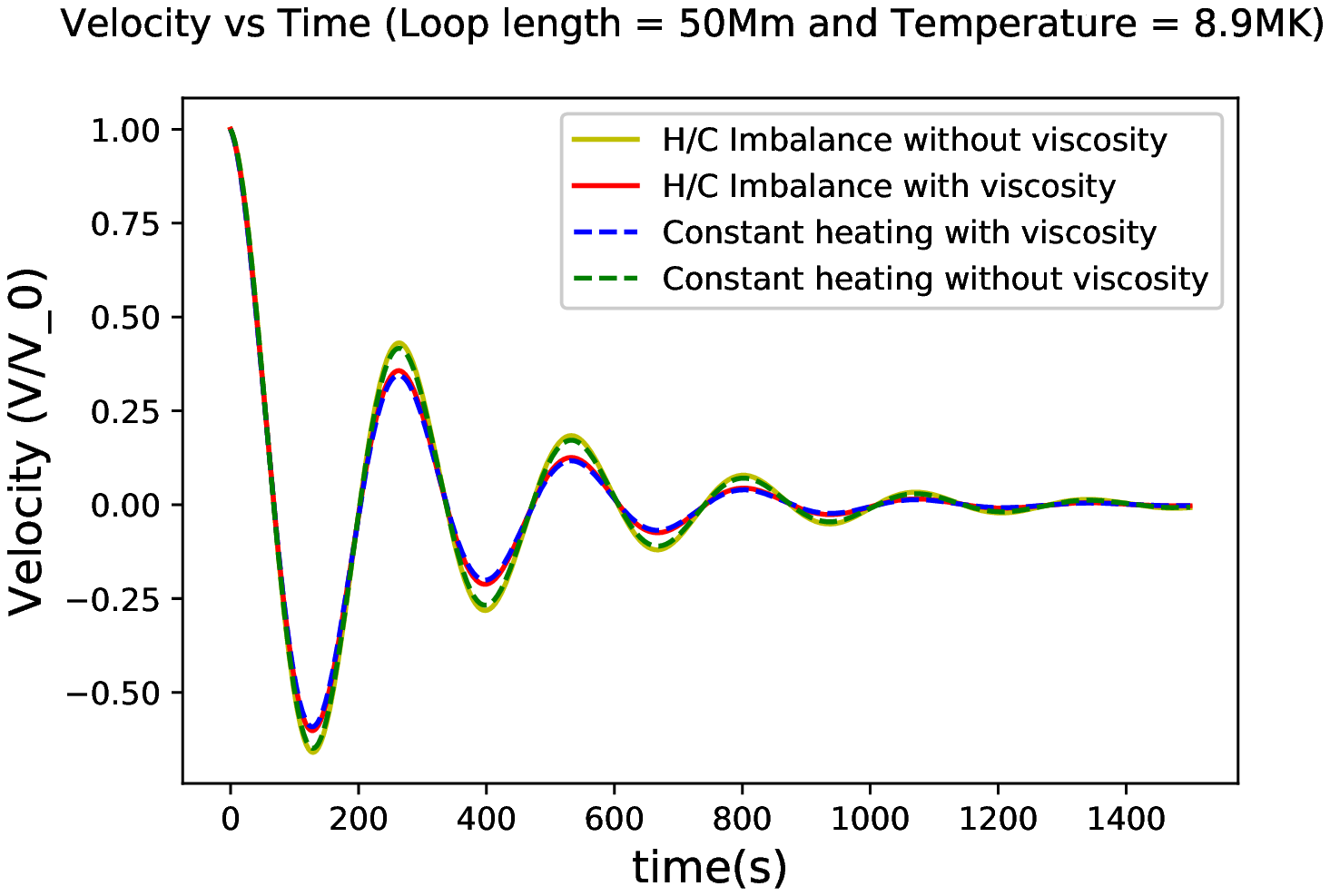}
               \hspace*{-0.015\textwidth}
               \includegraphics[width=0.495\textwidth,clip=]{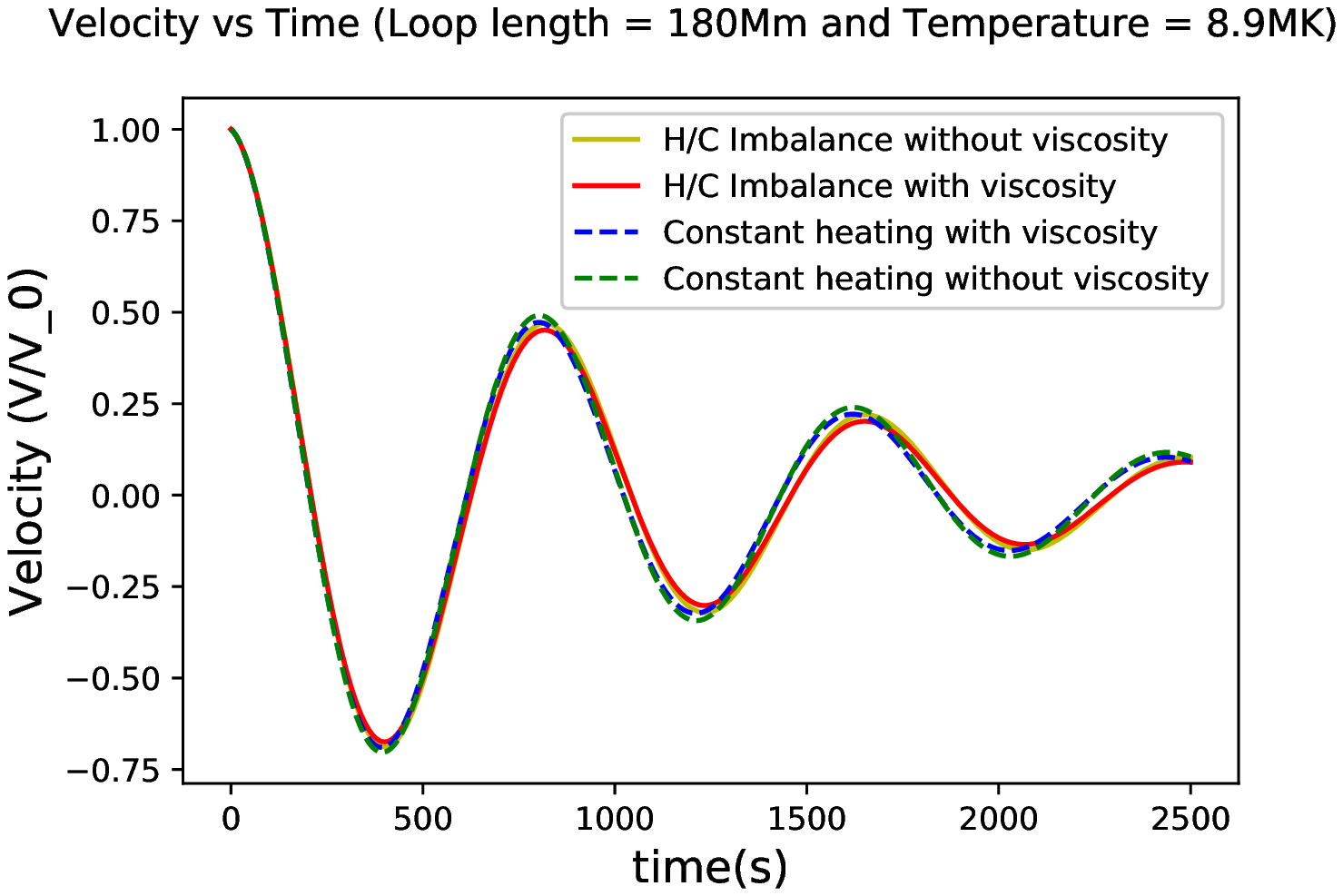}
               \hspace*{-0.015\textwidth}
               \includegraphics[width=0.495\textwidth,clip=]{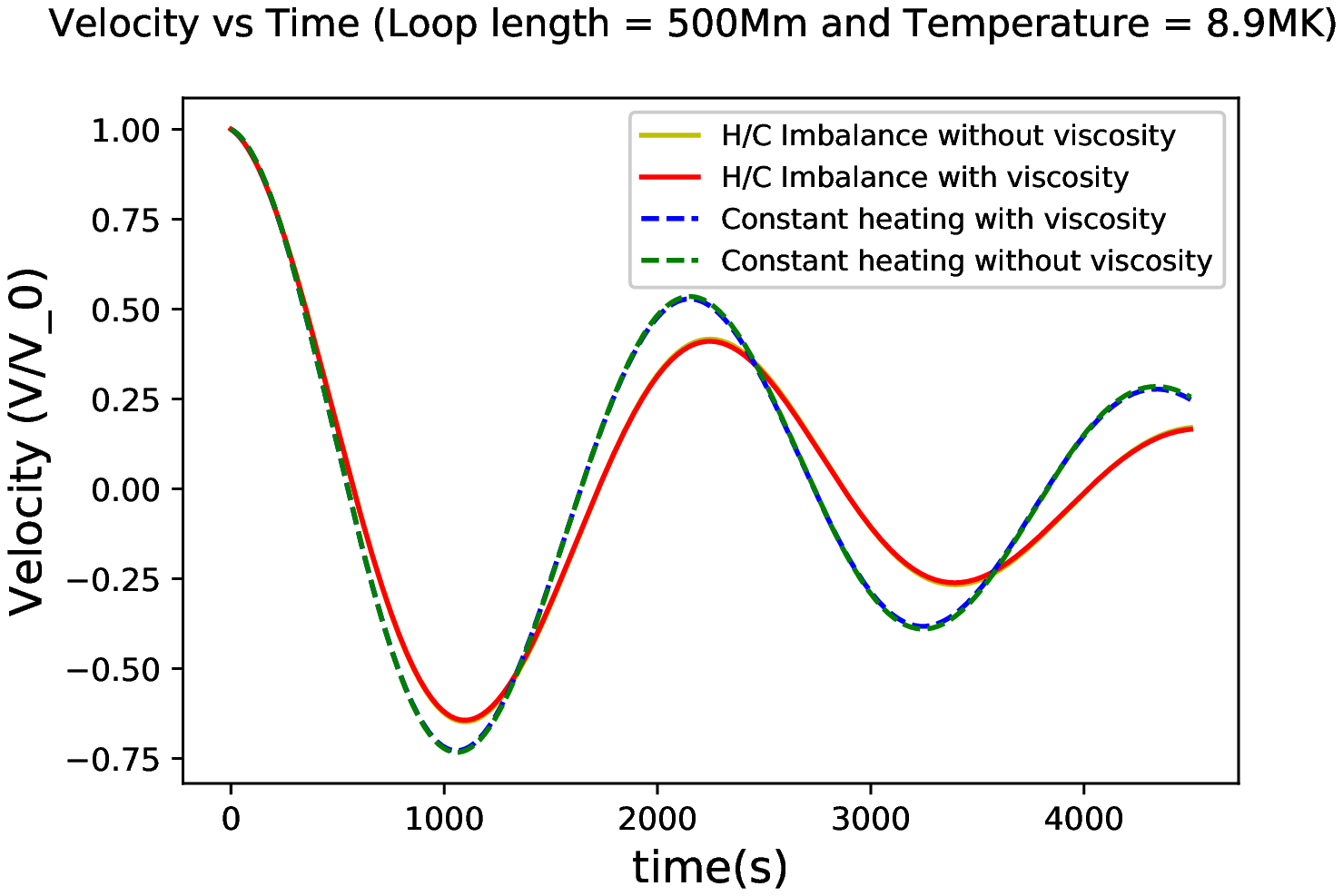}
              }
     \vspace{-0.35\textwidth}   
     \centerline{\Large \bf     
      \hspace{0.0 \textwidth} \color{white}{(c)}
      \hspace{0.415\textwidth}  \color{white}{(d)}
         \hfill}
     \vspace{0.31\textwidth}    
\caption{Variation of $V$ with time for the fundamental slow magnetoacoustic oscillations at $z=\frac{L}{2}$ in loops of length $L$=50 Mm (left column), 180 Mm (middle column), and 500 Mm (right column) at $T$=5.0, 6.3, 8.9 MK.  The yellow (red) curve represents the oscillations without (with) the effect of viscosity under heating-cooling imbalance. The dotted-green (dotted-blue) curve represents the oscillations without (with) the effect of viscosity in the absence of heating-cooling imbalance. Thermal conductivity is always present as a damping mechanism in these analyses.
        }
   \label{F-4panels}
   \end{figure}

\begin{figure}   
   \centerline{\hspace*{0.015\textwidth}
               \includegraphics[width=0.495\textwidth,clip=]{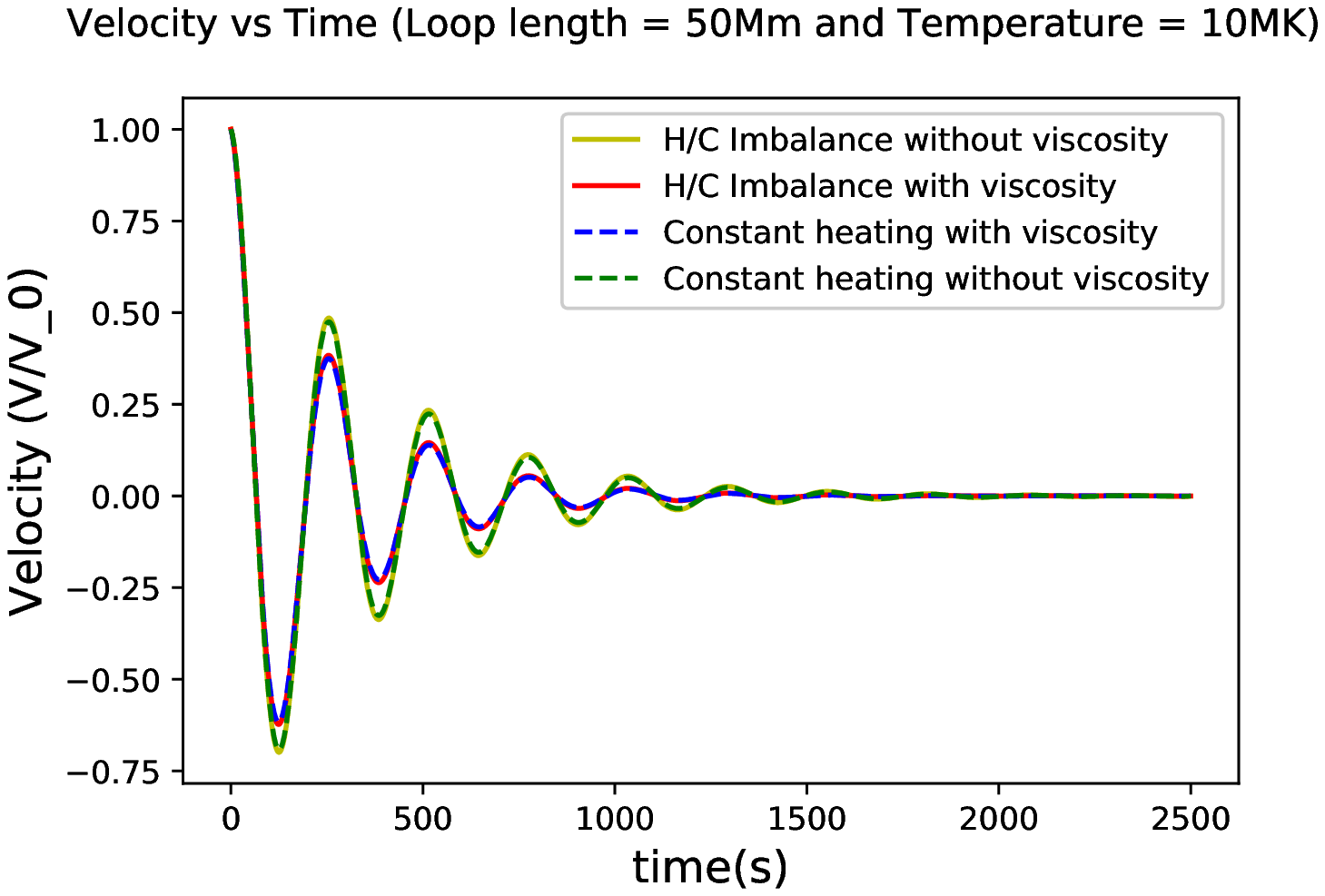}
               \hspace*{-0.015\textwidth}
               \includegraphics[width=0.495\textwidth,clip=]{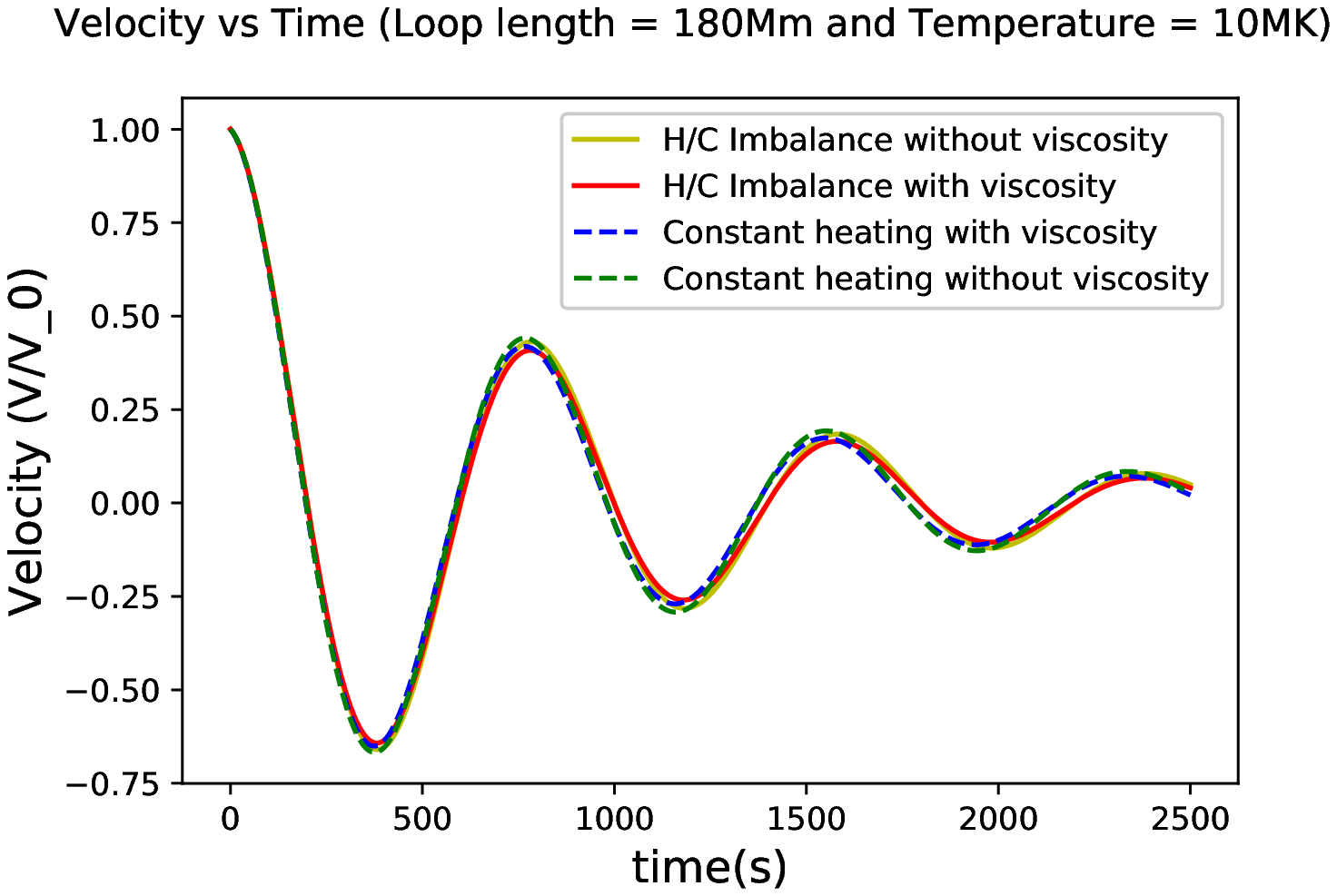}
               \hspace*{-0.015\textwidth}
               \includegraphics[width=0.495\textwidth,clip=]{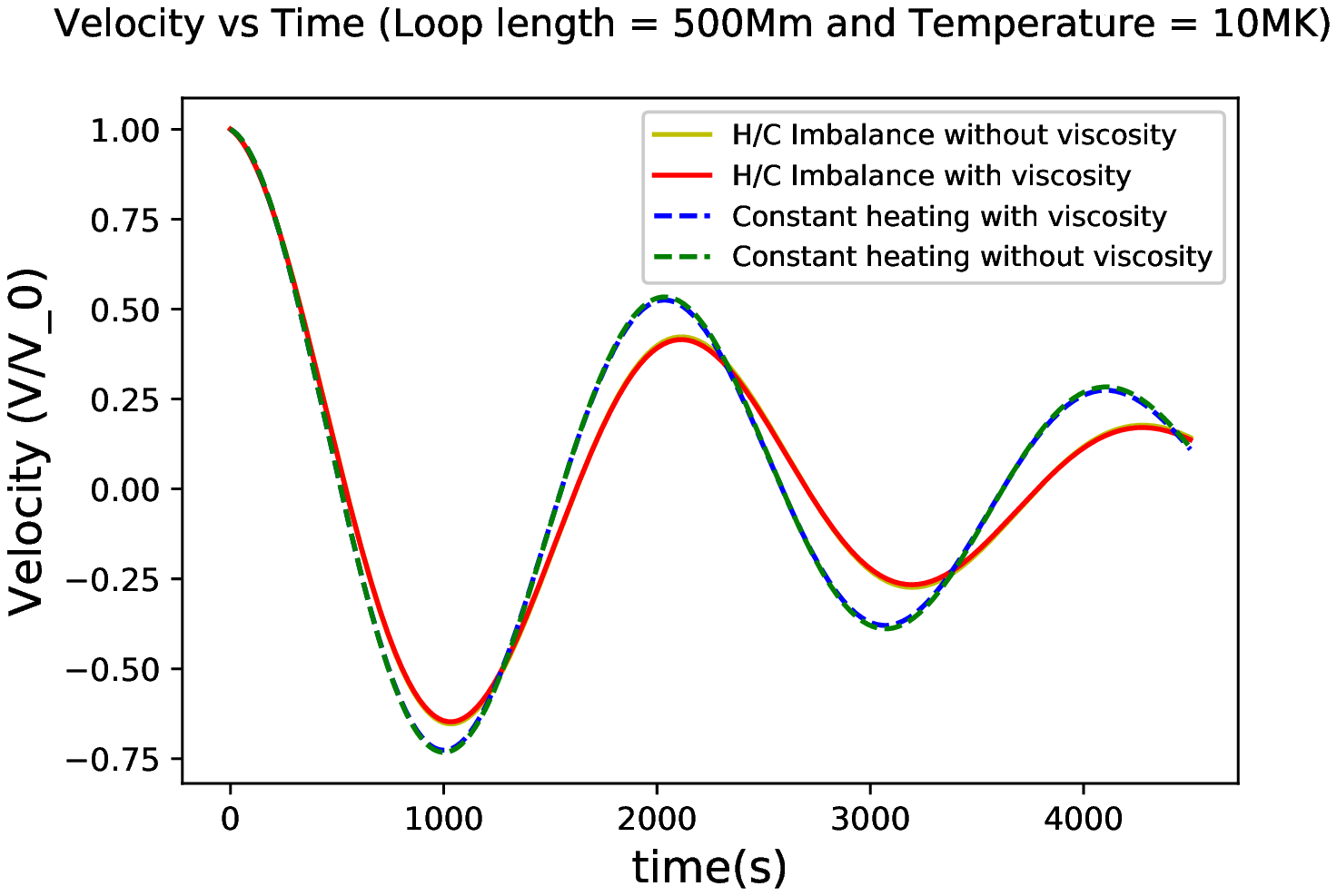}
              }
     \vspace{-0.35\textwidth}   
     \centerline{\Large \bf     
      \hspace{0.0 \textwidth}  \color{white}{(a)}
      \hspace{0.415\textwidth}  \color{white}{(b)}
         \hfill}
     \vspace{0.31\textwidth}    
   \centerline{\hspace*{0.015\textwidth}
               \includegraphics[width=0.495\textwidth,clip=]{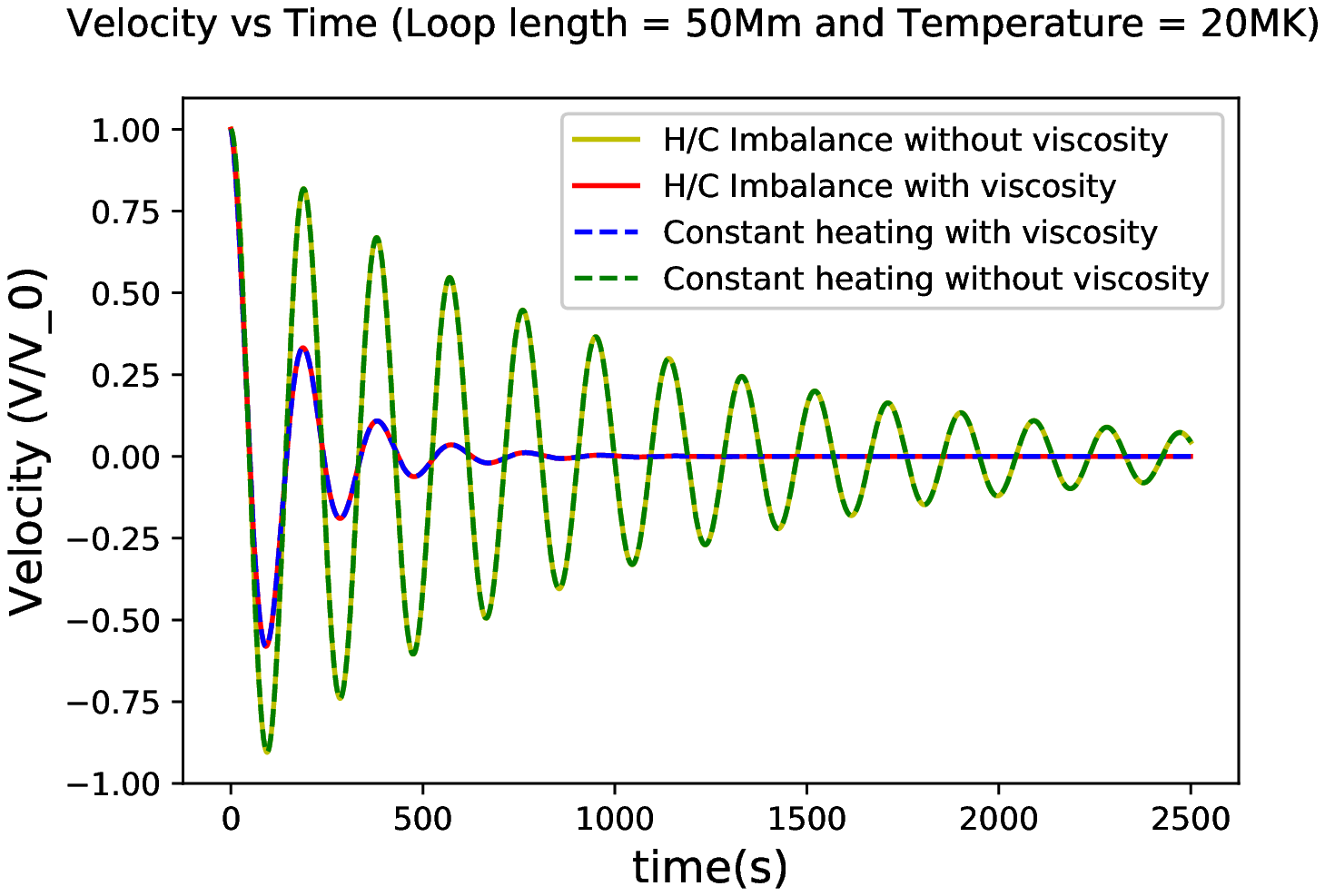}
               \hspace*{-0.015\textwidth}
               \includegraphics[width=0.495\textwidth,clip=]{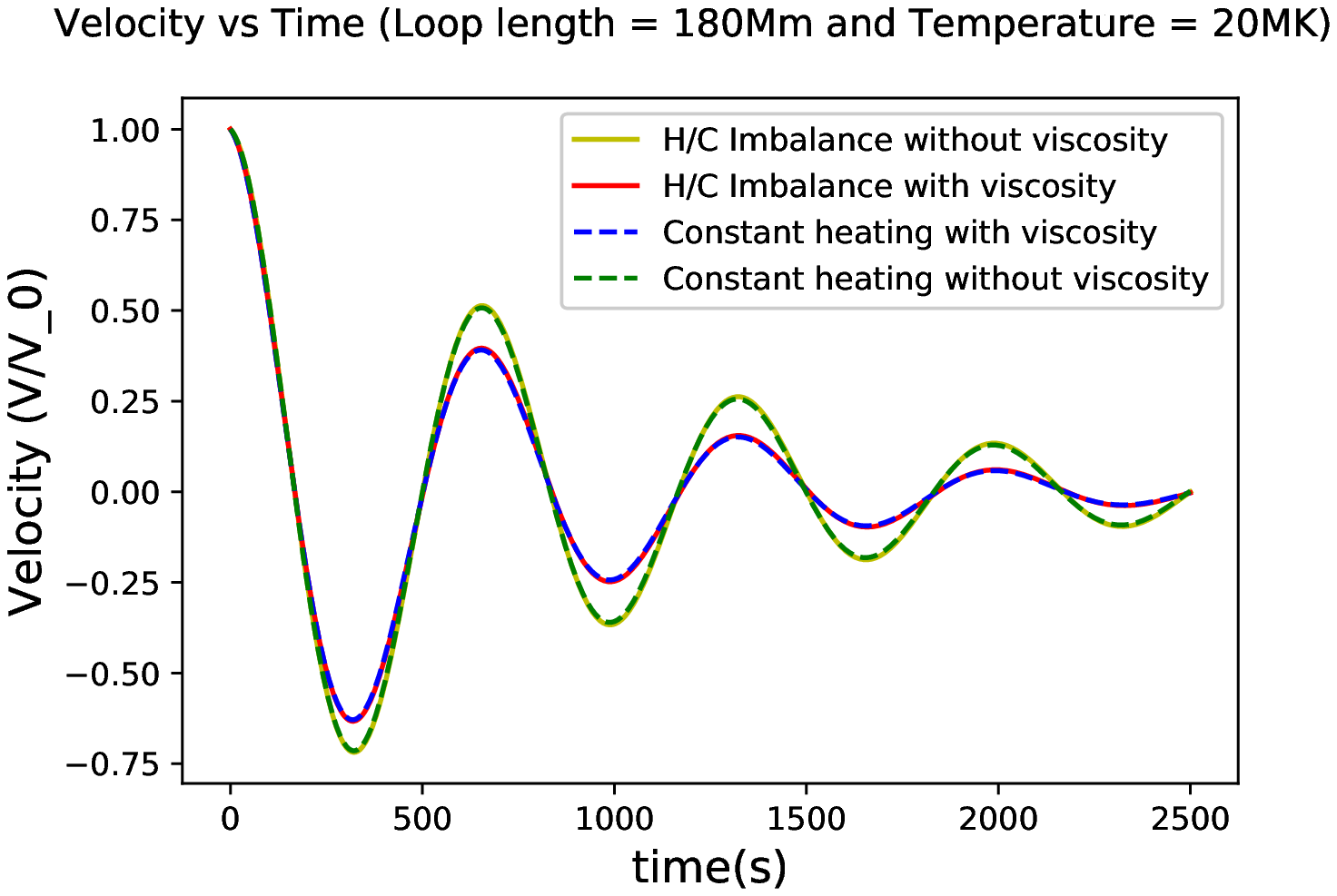}
               \hspace*{-0.015\textwidth}
               \includegraphics[width=0.495\textwidth,clip=]{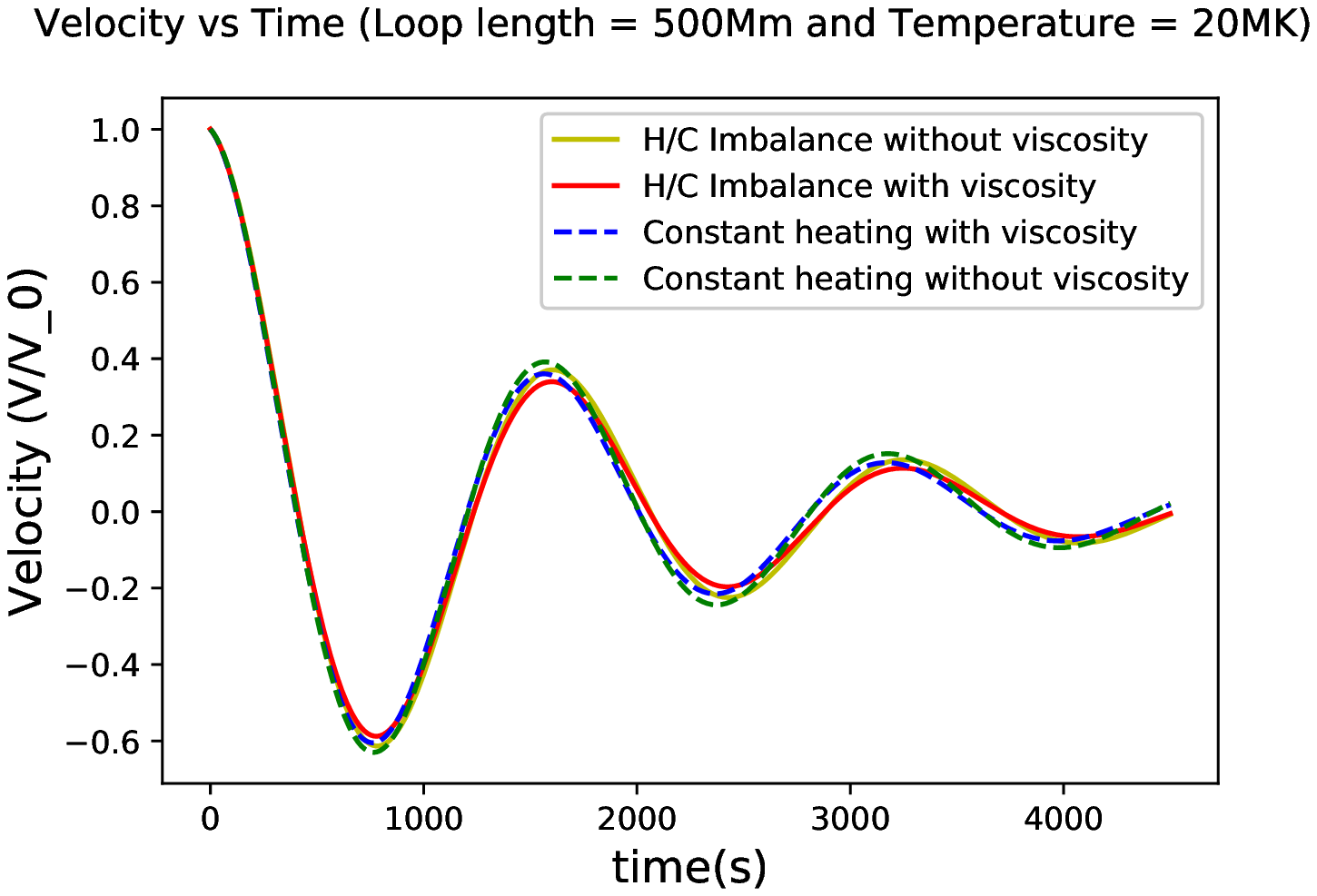}
              }
     \vspace{-0.35\textwidth}   
     \centerline{\Large \bf     
      \hspace{0.0 \textwidth} \color{white}{(c)}
      \hspace{0.415\textwidth}  \color{white}{(d)}
         \hfill}
     \vspace{0.31\textwidth}    
              
     \centerline{\hspace*{0.015\textwidth}
               \includegraphics[width=0.495\textwidth,clip=]{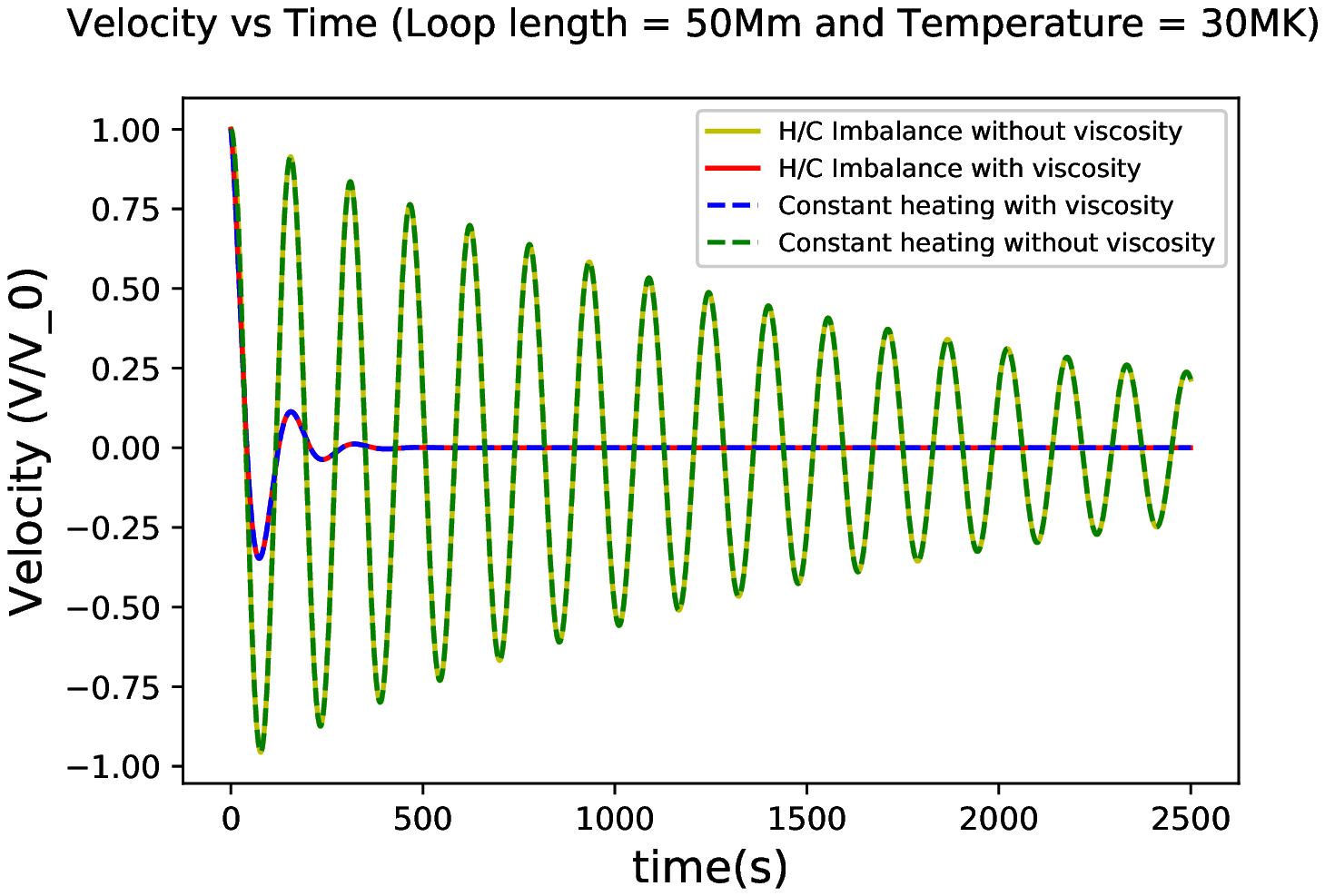}
               \hspace*{-0.015\textwidth}
               \includegraphics[width=0.495\textwidth,clip=]{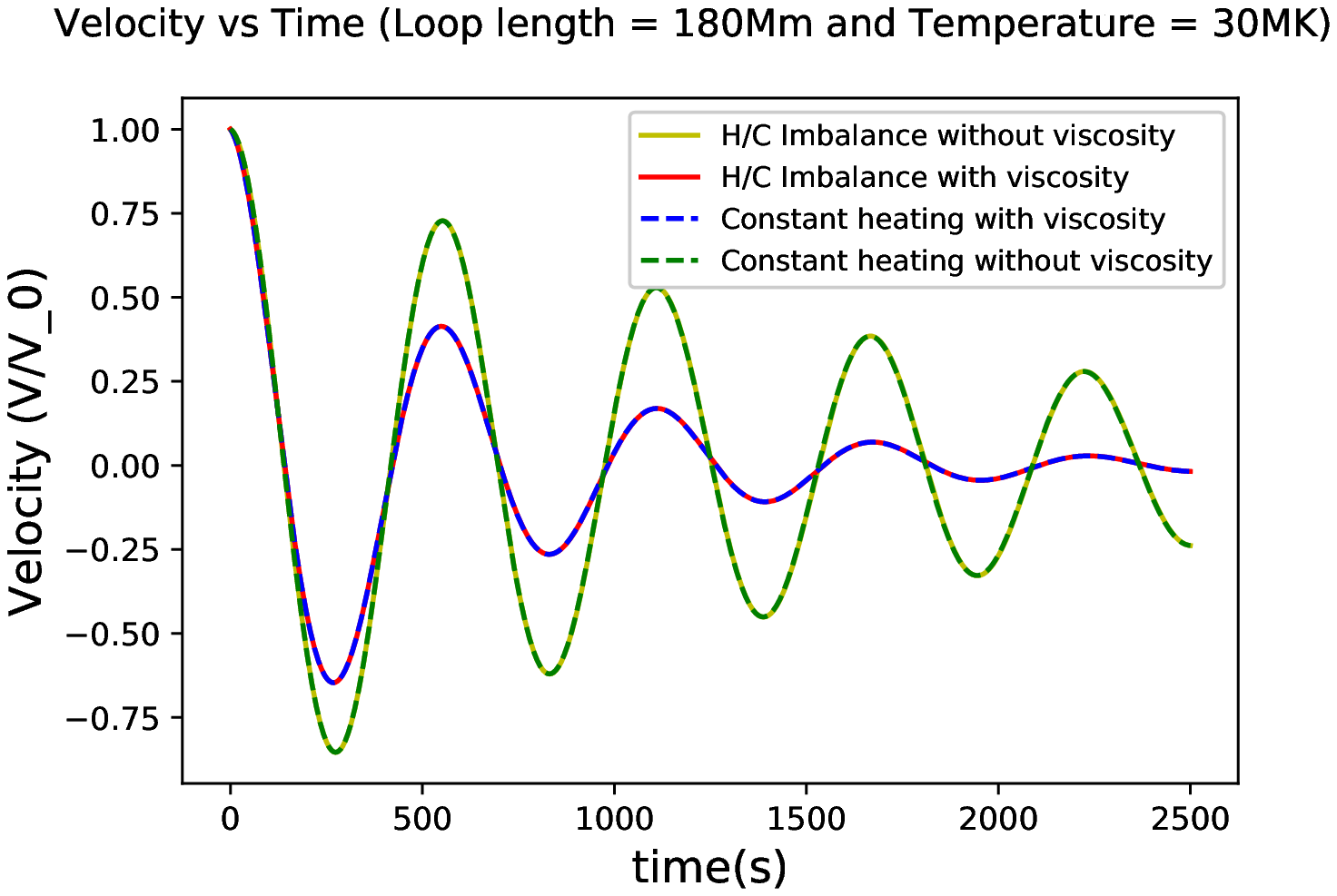}
               \hspace*{-0.015\textwidth}
               \includegraphics[width=0.495\textwidth,clip=]{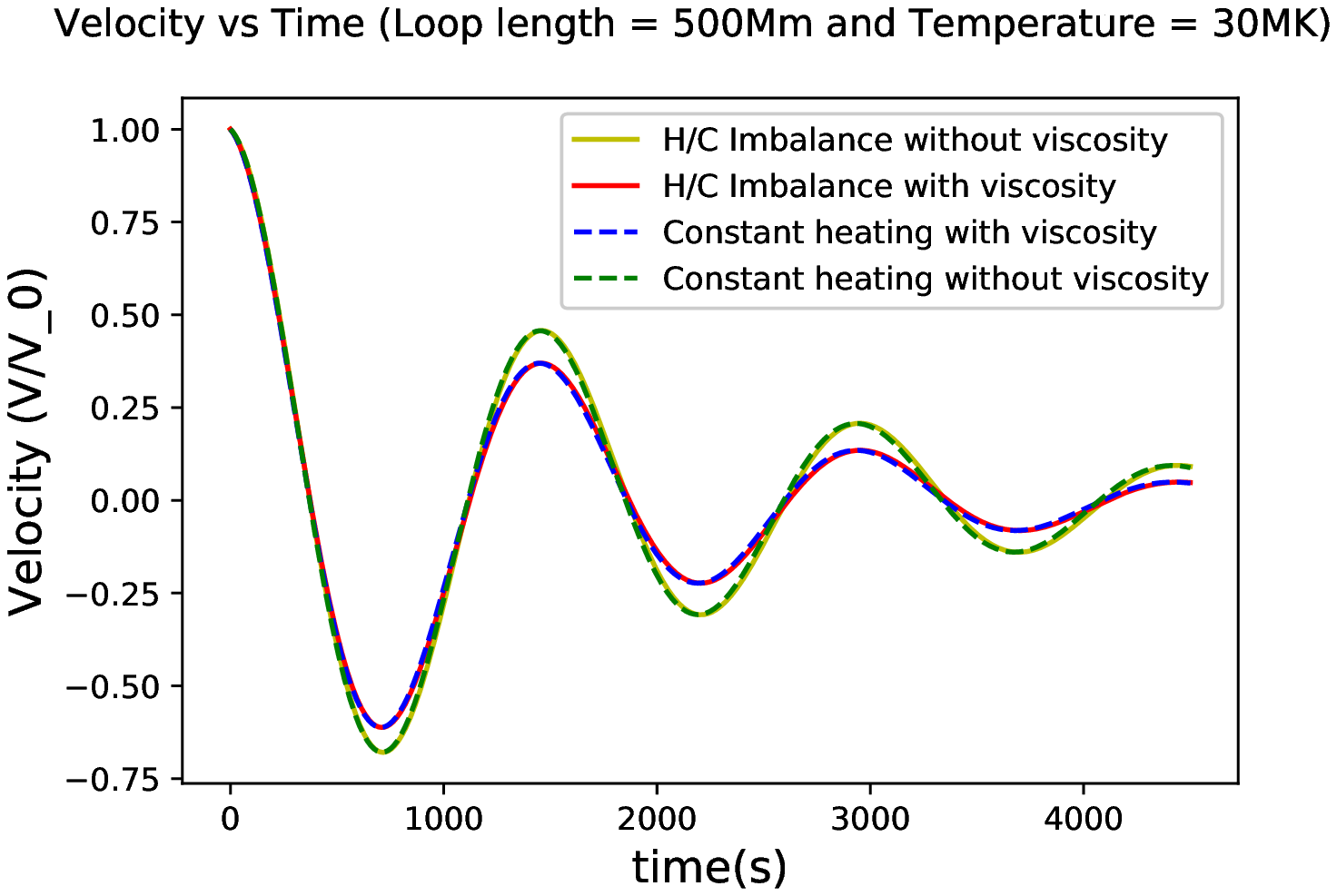}
              }
     \vspace{-0.35\textwidth}   
     \centerline{\Large \bf     
      \hspace{0.0 \textwidth} \color{white}{(c)}
      \hspace{0.415\textwidth}  \color{white}{(d)}
         \hfill}
     \vspace{0.31\textwidth}    
\caption{Variation of $V$ with time for the fundamental slow magnetoacoustic oscillations at $z=\frac{L}{2}$ in loops of length $L$=50 Mm (left column), 180 Mm (middle column), and 500 Mm (right column) at $T$=10, 20, 30 MK. The yellow (red) curve represents the oscillations without (with) the effect of viscosity under heating-cooling imbalance. The dotted-green (dotted-blue) curve represents the oscillations without (with) the effect of viscosity in the absence of heating-cooling imbalance. Thermal conductivity is always present as a damping mechanism in these analyses.
        }
   \label{F-4panels}
   \end{figure}

\subsubsection{Analyses of Velocity Oscillations of the Fundamental Mode in the Hot Regime of Temperature $T\leq$10 MK \label{sct: HV}}
 
We analyze the time evolution of damped slow-mode oscillations in a similar manner as done by \citet{2003A&A...402L..17W} and consider the velocity perturbations as 
\begin{equation}
    V(z,t) = V_0\sin{(kz)}\cos{(\omega_Rt)}e^{-\mid \omega_I \mid t},
\end{equation}
Here $\omega_R$ is the real component of cyclic frequency while \(k\) and \(V_0\) are the wave number and amplitude of velocity oscillations, respectively. We calculate the time variation of velocity oscillations using our numerical solutions for the dispersion relation in different cases.\newline
In Figure 4, the time variation of $V$ at the chosen position of $z=\frac{L}{2}$ for the damped fundamental mode oscillations in the loops of $L$=50 Mm (left column), 180 Mm (middle column), and 500 Mm (right column) are plotted at $T$=5.0, 6.3, 8.9 MK respectively. Since we estimate normalized temporal variations of $V$, and $sin (kz)$ varies between $\pm$1, any of the arbitrary choices of $z$ will not change the shape of the velocity oscillations. In each panel, the yellow (red) solid curve shows the temporal variation of the oscillations without (with) the effect of compressive viscosity under the consideration of heating-cooling imbalance and presence of thermal conductivity as a default damping mechanism. The dotted-green (dotted-blue) curve shows the temporal variation of the oscillations without (with) the effect of compressive viscosity without the consideration of heating-cooling imbalance and presence of thermal conductivity as a default damping mechanism.

It is clear that the velocity oscillations decay quickly in the shortest loop of 50 Mm length at $T$=5.0\,--\,8.9 MK. However, all the curves are merged together (cf. left column in Figure 4).
Although, the case shown in the left-bottom panel ($L$=50 MK and $T$=8.9) indicates a weak effect of viscous damping. This implies no difference in the damping between the cases with and without consideration of heating-cooling imbalance. In both cases, thermal conductivity is always switched on as a default damping mechanism. Therefore, it is clear that the fundamental mode oscillations are damped dominantly due to thermal conductivity in the case of an $L$=50 Mm loop, and both the heating-cooling imbalance and viscosity cause no enhancement of the damping.
Top and middle panels in the left-column in Figure 4 for $T$=5.0 and 6.3 MK clearly demonstrate such cases where thermal conduction dominates the damping, while both the viscosity and heating-cooling imbalance have nearly no effects on it.

Velocity oscillations also show decay in the loop of 180 Mm length at $T$=5.0\,--\,8.9 MK. At $T$=5 MK, the heating-cooling imbalance causes a reduction in the velocity amplitude (red/yellow curve) and thus an enhanced damping compared to the case in the absence of imbalance (dotted-green/dotted-blue curve; cf. top panel in the middle column). The red and yellow curves superimpose on each other, which means that the compressive viscosity has insignificant effect on enhancing the damping in the presence of heating-cooling imbalance, and its damping is mainly caused by thermal conductivity. Similarly the dotted green and blue curves also superimpose with each other indicating that the effect of viscous damping is negligible compared to the damping due to thermal conductivity in the absence of heating-cooling imbalance. As temperature reaches 6.3 MK, an enhanced damping is again observed in the presence of heating-cooling imbalance, but its effect decreases as there is a reduction in the velocity amplitude compared to the one observed at $T$=5.0 MK (cf. middle panel in the middle column). At $T$=8.9 MK, the heating-cooling imbalance has still a little higher effect on the damping of the fundamental mode oscillations compared to the case without it. However, the reduction in the velocity amplitude oscillations is even smaller compared to the one seen at lower temperatures of 6.3 and 5.0 MK (cf. bottom panel in the middle column). In conclusion, this scenario indicates that for the intermediate loop length (e.g. $L$=180 Mm here), the damping is mainly due to the presence of thermal conductivity and the effect of heating-cooling imbalance is evident mostly at the hot regime of temperature ($T\leq$10 MK). Heating-cooling imbalance causes more damping at lower temperature in the longer loops. 

For the loop of $L$=500 Mm maintained at $T$=5.0 MK,  the heating-cooling imbalance enhances the damping of the fundamental mode oscillations significantly. The velocity oscillation is reduced significantly in magnitude (cf. top panel in the right column). When temperature increases to 6.3 MK, the heating-cooling imbalance still affects strongly the damping of the fundamental mode as the velocity oscillations are still reduced significantly (cf. middle panel in the right column). In the case of $T$=8.9 MK, the same scenario holds with the reduction of the velocity oscillations in the presence of heating cooling imbalance (cf. bottom panel in the right column). In this case compressive viscosity has no appreciable role in the damping of the fundamental modes compared to thermal conductivity, and even the presence of heating-cooling imbalance enhances the dissipation caused by it.
In conclusion, in the hot regime $T\leq$10 MK, the compressive viscosity has an insignificant role in enhancing the damping of the fundamental mode oscillations. Whereas thermal conductivity plays an appreciable role and its effect is enhanced by the presence of heating-cooling imbalance especially in longer loops of lower temperature.
The case at $T$=5.0 MK in the longest loop shows the most significant effect of heating-cooling imbalance in the damping while no effect is observed due to viscosity.
In the hot temperature regime, especially in the longer loops, there is a phase-shift in the velocity oscillations (e.g. top right panel of Figure 4). A detailed study on the 
physical cause of this is out of the scope of the present article, and will be taken-up extensively in future studies.

In the next section, we analyze the velocity oscillations of the fundamental mode in the super-hot regime with $T>$10 MK.}

\subsubsection{Analyses of Velocity Oscillations of the Fundamental Mode in the Super-Hot Regime of Temperature $T>$10 MK \label{sct: SHV}}

In Figure 5, the time variation of $V$ at $z=\frac{L}{2}$ for the fundamental mode oscillations in the loops of $L$=50 Mm (left column), 180 Mm (middle column), and 500 Mm (right column) are plotted at $T$=10, 20, 30 MK respectively. In each panel, the yellow (red) solid curve shows the temporal variation of the oscillations without (with) the effect of compressive viscosity under the consideration of heating-cooling imbalance and presence of thermal conductivity as a default damping mechanism.The dotted-green (dotted-blue) curve shows the temporal variation of the oscillations without (with) the effect of compressive viscosity without the consideration of heating-cooling imbalance.

For the loop of 50 Mm length, at a temperature of 10 MK (cf. top panel in the left column), the velocity curves without viscosity in both the cases with heating-cooling imbalance (yellow) and without it (dotted-green curve) are superimposed with each other and have higher amplitude of oscillations. On the other hand,  the velocity curves with viscosity in both cases with heating-cooling imbalance (red) and without it (dotted-blue curve) are also almost superimposed and have lower amplitude of oscillations. This scenario is also true for higher temperatures of 20 and 30 MK, in general. It describes that the heating-cooling imbalance or constant heating rate without imbalance plays no role in the damping of fundamental mode oscillations in shorter loops. On the other hand in the super-hot regime, it is obvious that the inclusion of viscosity (red/blue-dotted curve) significantly enhances the damping of the fundamental modes compared to the case of thermal conductivity alone (yellow/green-dotted curve).  At $T$=10 MK, the velocity oscillations are damped (cf. Figure 5, top panel in the left column) due to the effects of both compressive viscosity and thermal conductivity. It can be seen that when temperature increases to 20 MK for the loop of $L$=50 Mm, the damping effect due to viscosity becomes even more prominent compared to the case with thermal conduction alone (cf. Figure 5, middle panel in the left column). We find that the velocity oscillations show the most serious damping (only one cycle of the oscillations is visible) at $T$=30 MK for the shortest loop of $L$=50 Mm when the viscosity is included, while the damping of oscillations in the case with thermal conduction alone become weaker with increasing temperature (cf. Figure 5, bottom panel in the left column). The cases with $L$=50 Mm and $T$=20 and 30 MK have the damping dominated by viscosity because the cases without it (or with thermal conduction alone) show only  a weak damping. Additionally the effect of heating-cooling imbalance for this condition is negligible.

For the intermediate loop length $L$=180 Mm, the red/yellow as well as blue$/$green dotted curves are almost superimposed with each other at 10 MK temperature. According to this behavior it is inferred that the damping of the fundamental mode oscillations is mostly due to the presence of thermal conductivity, and on the other hand the presence of heating-cooling imbalance and compressive viscosity do not enhance the amount of damping (cf. Figure 5, middle panel in the middle column). At $T$= 20 MK, the velocity oscillations are seen damped due to the joint effect of viscosity and conductivity while the imbalance plays no role in the damping in the case of 180 Mm  loops (cf. Figure 5, middle panel in the middle column). At $T$=30 MK, velocity oscillations are damped significantly due to the joint effect of viscosity and conductivity while the imbalance plays no role in the damping for the case of 180 Mm loops (cf. Figure 5, bottom panel in the middle column).

For the highest considered loop length $L$=500 Mm at 10 MK temperature (cf. Figure 5, top panel in the right column), heating-cooling imbalance dominates along with thermal conductivity (red/yellow curve) the damping of the fundamental modes. The effect of the compressive viscosity is not likely the cause of enhanced damping of the fundamental modes at 10 MK as indicated by the almost superimposed yellow and red curves. At $T$=20 MK, the red/yellow and blue/green-dotted curves are almost superimposed with each other, which suggests that the damping of the fundamental mode slow magnetoacoustic oscillations is mostly due to the presence of thermal conductivity while both heating-cooling imbalance and compressive viscosity do not play a role in the damping (cf. Figure 5, middle panel in the right column). At $T$=30 MK, the damping effect of the compressive viscosity only slightly takes over that of thermal conductivity, but both cases attenuate the velocity oscillations significantly in the case of 500 Mm  loop, while the effect of heating-cooling imbalance on the damping is negligible (cf. Figure 5, bottom panel in the right column). This implies that thermal conduction is dominant. In conclusion, the super-hot loops have quite appreciable velocity damping of the fundamental modes under the joint effects of compressive viscosity and thermal conductivity especially in the shorter loops, where the viscous damping dominates over thermal conduction damping, while the effect of heating-cooling imbalance is negligible. Nevertheless in the longest loop ($L$=500 Mm), once we go towards 10 MK temperature the effect of heating-cooling imbalance along with thermal conductivity causes a dominant effect on the damping. While, towards the highest temperature of 30 MK, as usual the joint effect of compressive viscosity and thermal conductivity causes an enhanced damping. It should be noted that all these estimations are made for the appropriate density values of the normal coronal loops, i.e., typically of the order of $\rho$=10$^{-11}$ kg m$^{-3}$.

In the next section, we present a brief summary of the results obtained in Sections 4.1.4 and 4.1.5. 

\subsubsection{Brief Summary}

In summary, there are mainly three different cases on the damping of slow modes by thermal conduction, viscosity and heating-cooling imbalance which show the dominant role in different conditions: (i) the viscosity dominates the damping in super-hot shorter loops (cf. bottom-left panel of Figure 5) while the heating-cooling imbalance plays nearly no role in this condition; (ii) the slow modes are damped mainly due to the joint effects of thermal conduction and  heating-cooling imbalance for the less hot loops with longer length (cf. top-right panel of Figure 4) while the viscous effect is negligible in this condition; (iii) thermal conduction dominates the damping while the effects of both viscosity and heating-cooling imbalance are negligible (cf. cases with $L$=50 and $T$=5.0 and 6.3 MK in the upper two panels of the left column of Figure 4, cases with $L$=180 and $T$=8.9 and 10 MK in bottom-middle panel of Figure 4 and top-middle panel of Figure 5, and the case with $L$=500 Mm, and $T$=20 MK in the middle-right panel of Figure 5).

Up to this point we have analyzed and established  comprehensively the damping scenario of the slow waves in the presence of thermal conductivity and compressive viscosity with and without heating-cooling imbalance for a wide range of loop lengths and temperatures. In the following sections, we will analyze the role of loop density in the damping of slow waves in two extreme cases: one for the longest loop of 500 Mm length in the hot regime $T\leq$10 MK (Section 4.2), and another for the shortest loop of 50 Mm length in the super-hot regime  $T>$10 MK (Section 4.3).

  \subsection{Role of  Loop Density on the Damping of Slow Waves in a Coronal Loop of Length $L$=500 Mm within the Hot Regime of Temperature ($T\leq$10 MK)}

\begin{figure}    
   \centerline{\hspace*{0.015\textwidth}
               \includegraphics[width=0.495\textwidth,clip=]{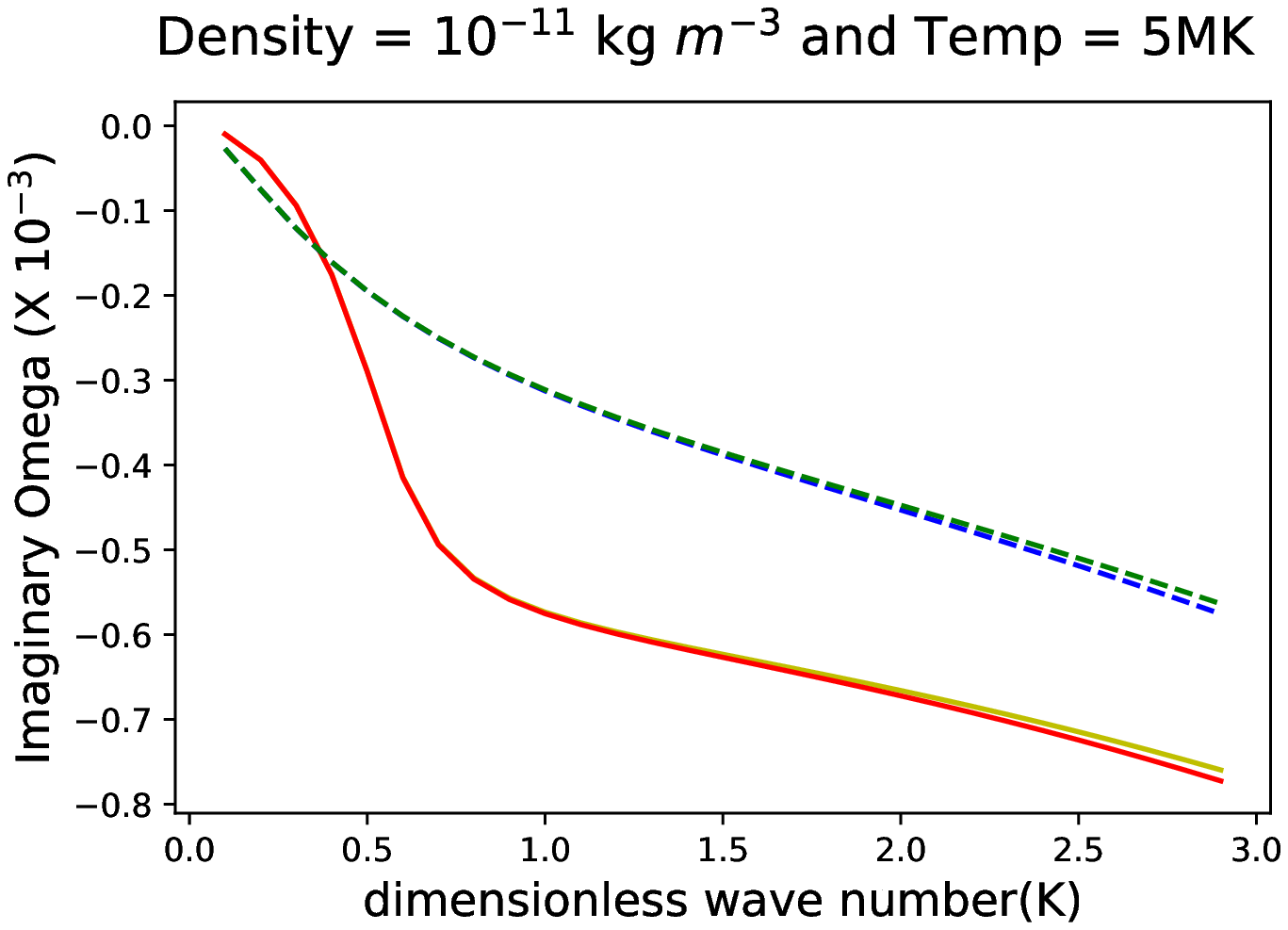}
               \hspace*{-0.015\textwidth}
               \includegraphics[width=0.495\textwidth,clip=]{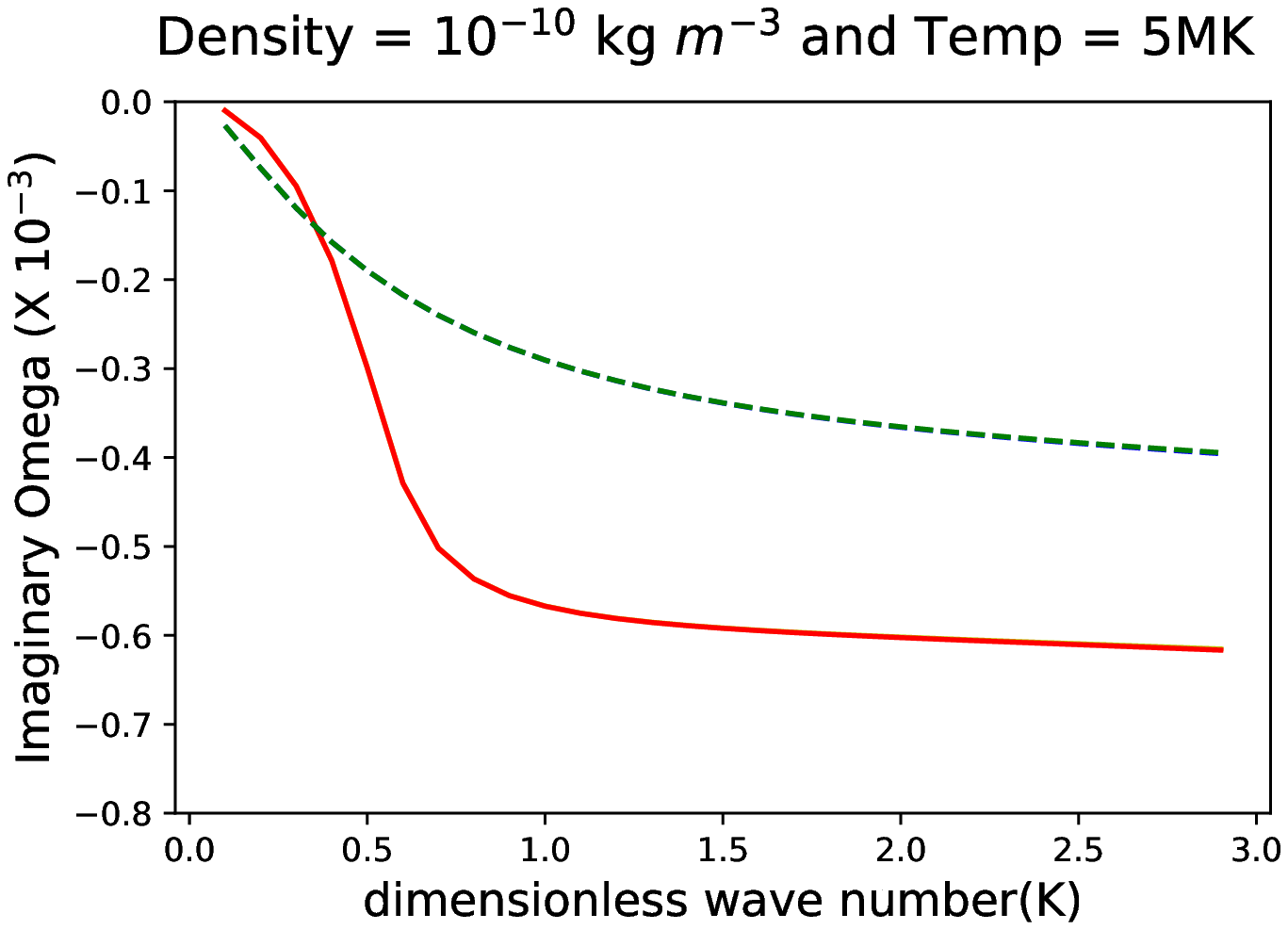}
               \hspace*{-0.015\textwidth}
               \includegraphics[width=0.495\textwidth,clip=]{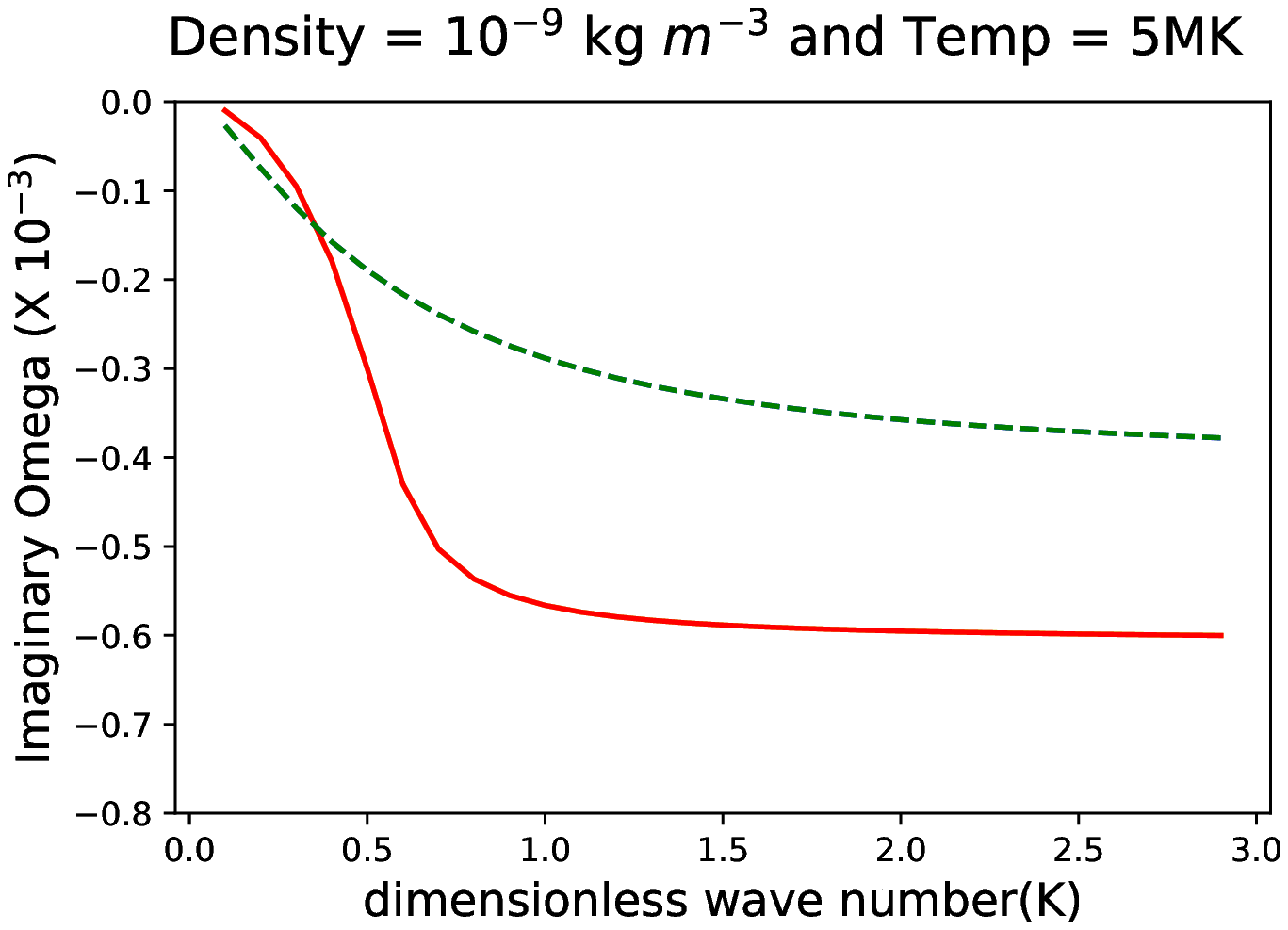}
              }
     \vspace{-0.35\textwidth}   
     \centerline{\Large \bf     
      \hspace{0.0 \textwidth}  \color{white}{(a)}
      \hspace{0.415\textwidth}  \color{white}{(b)}
         \hfill}
     \vspace{0.31\textwidth}    
   \centerline{\hspace*{0.015\textwidth}
               \includegraphics[width=0.495\textwidth,clip=]{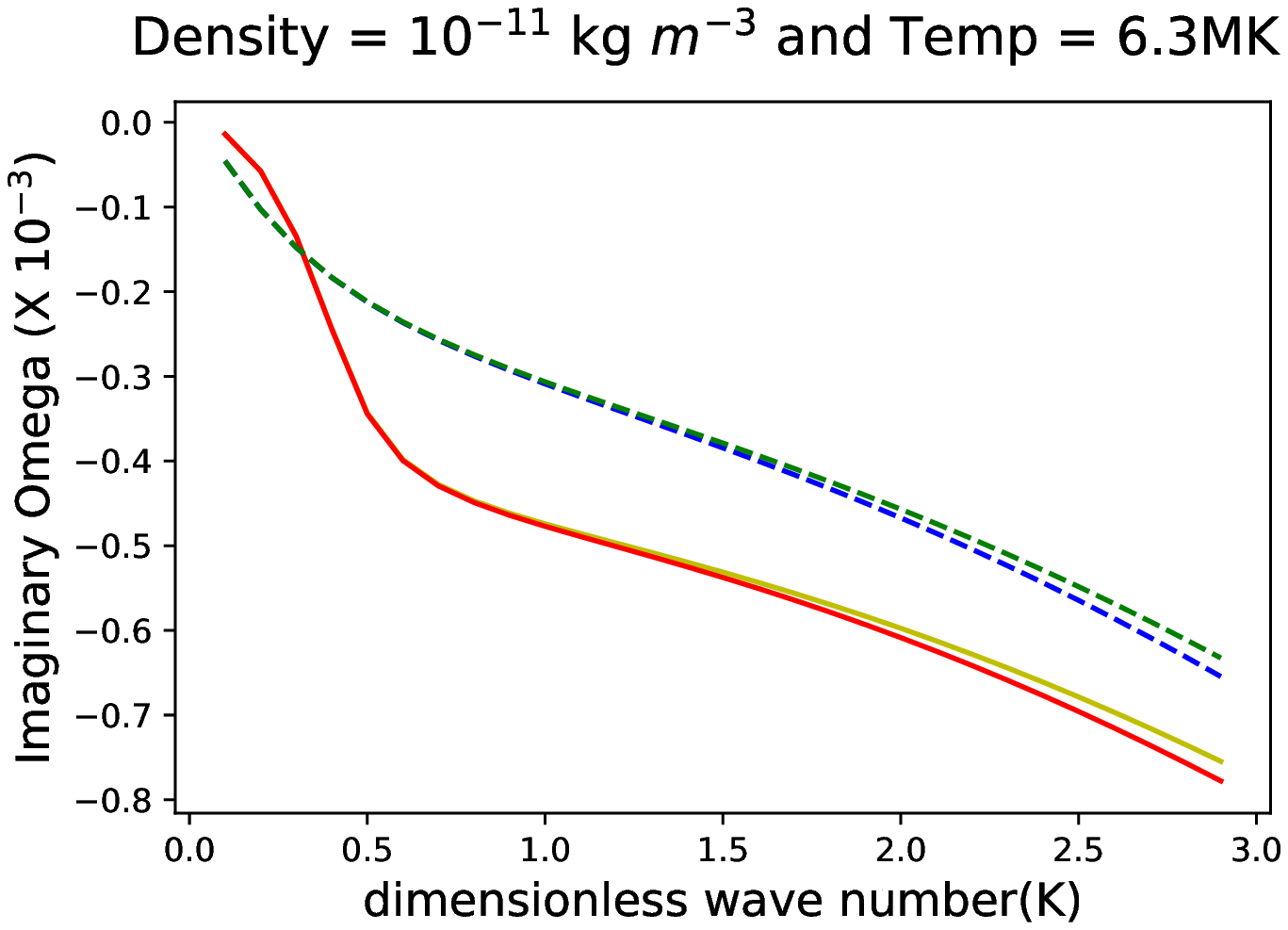}
               \hspace*{-0.015\textwidth}
               \includegraphics[width=0.495\textwidth,clip=]{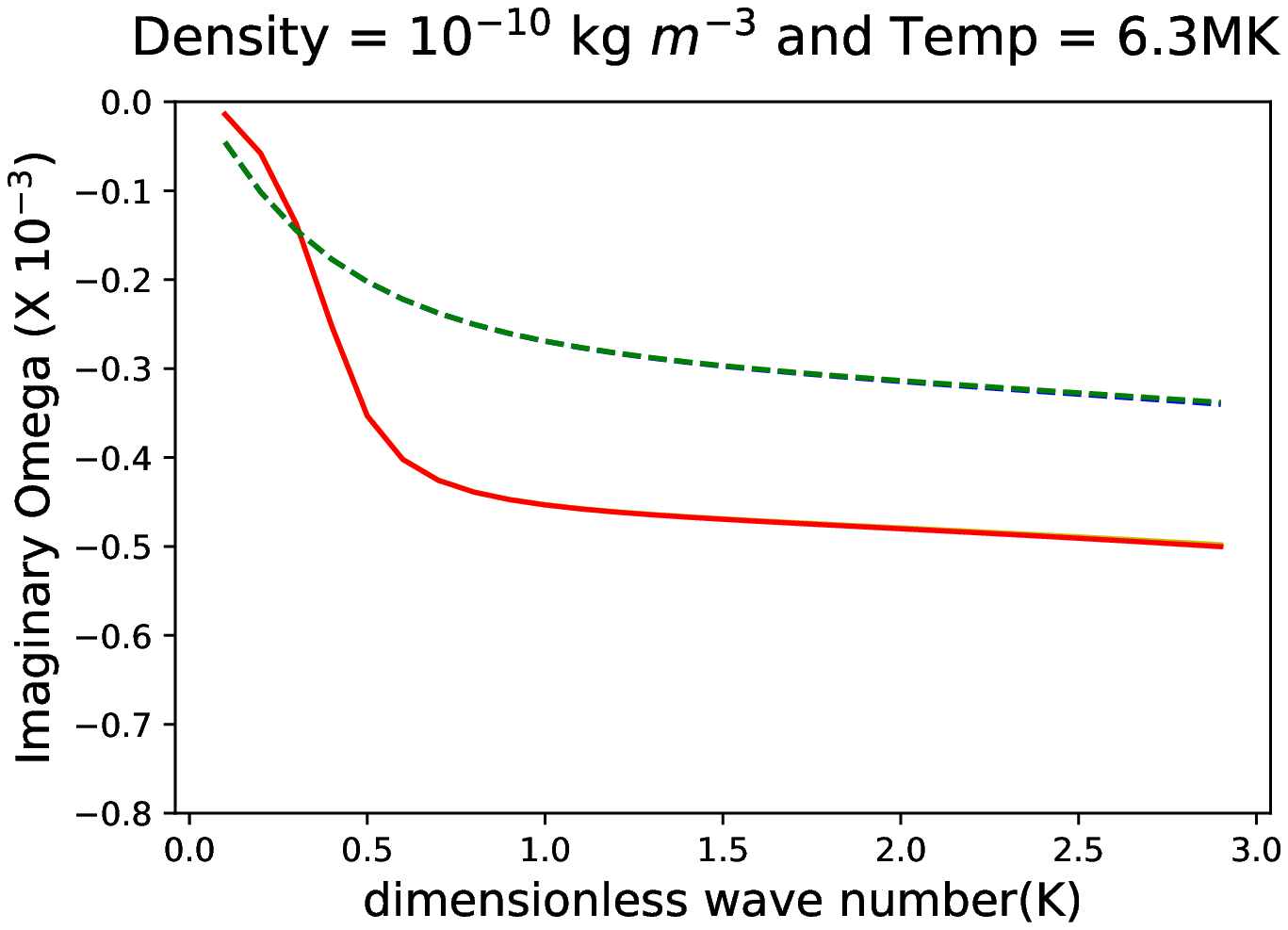}
               \hspace*{-0.015\textwidth}
               \includegraphics[width=0.495\textwidth,clip=]{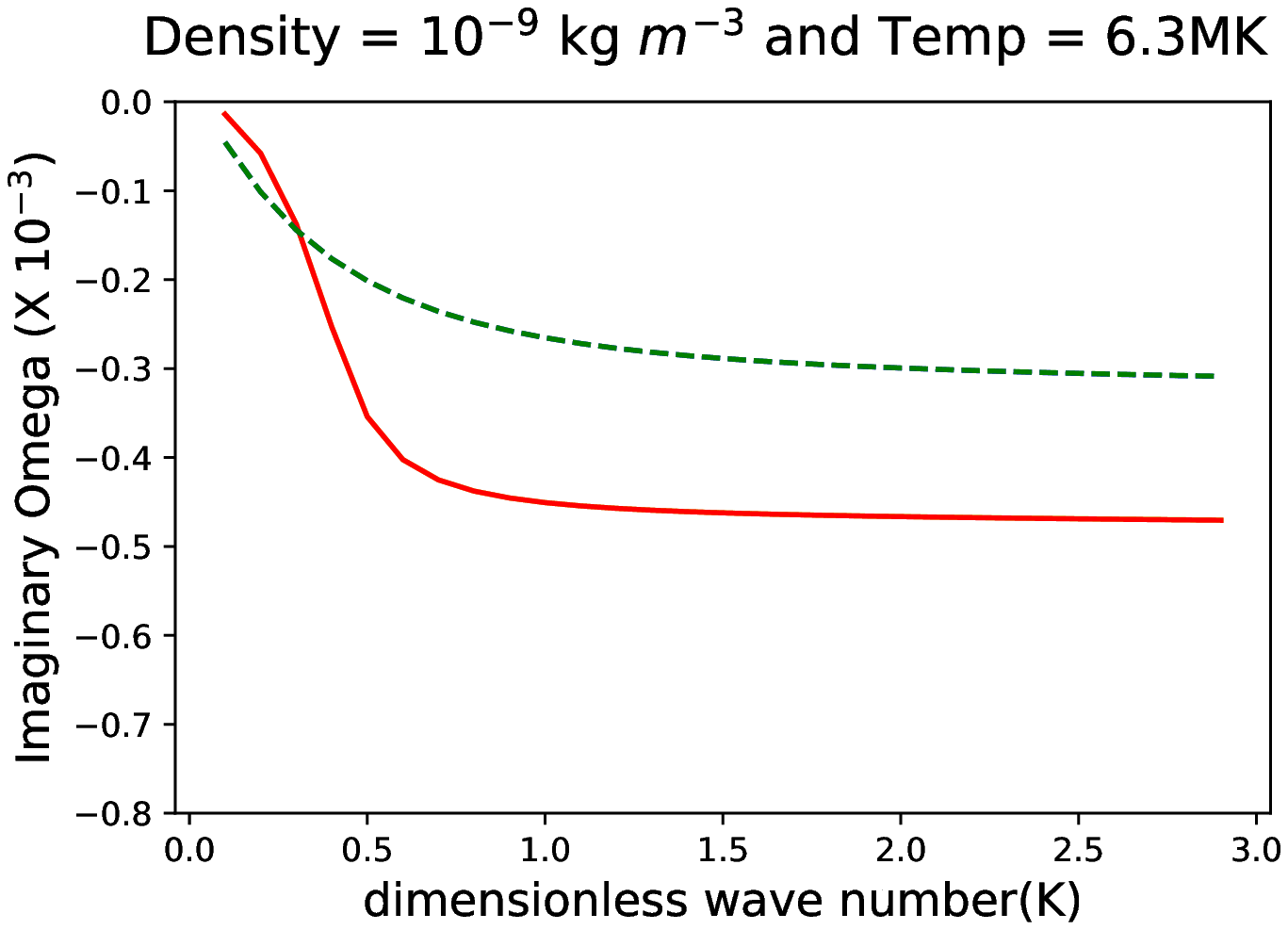}
              }
     \vspace{-0.35\textwidth}   
     \centerline{\Large \bf     
      \hspace{0.0 \textwidth} \color{white}{(c)}
      \hspace{0.415\textwidth}  \color{white}{(d)}
         \hfill}
     \vspace{0.31\textwidth}    
     
      \centerline{\hspace*{0.015\textwidth}
               \includegraphics[width=0.495\textwidth,clip=]{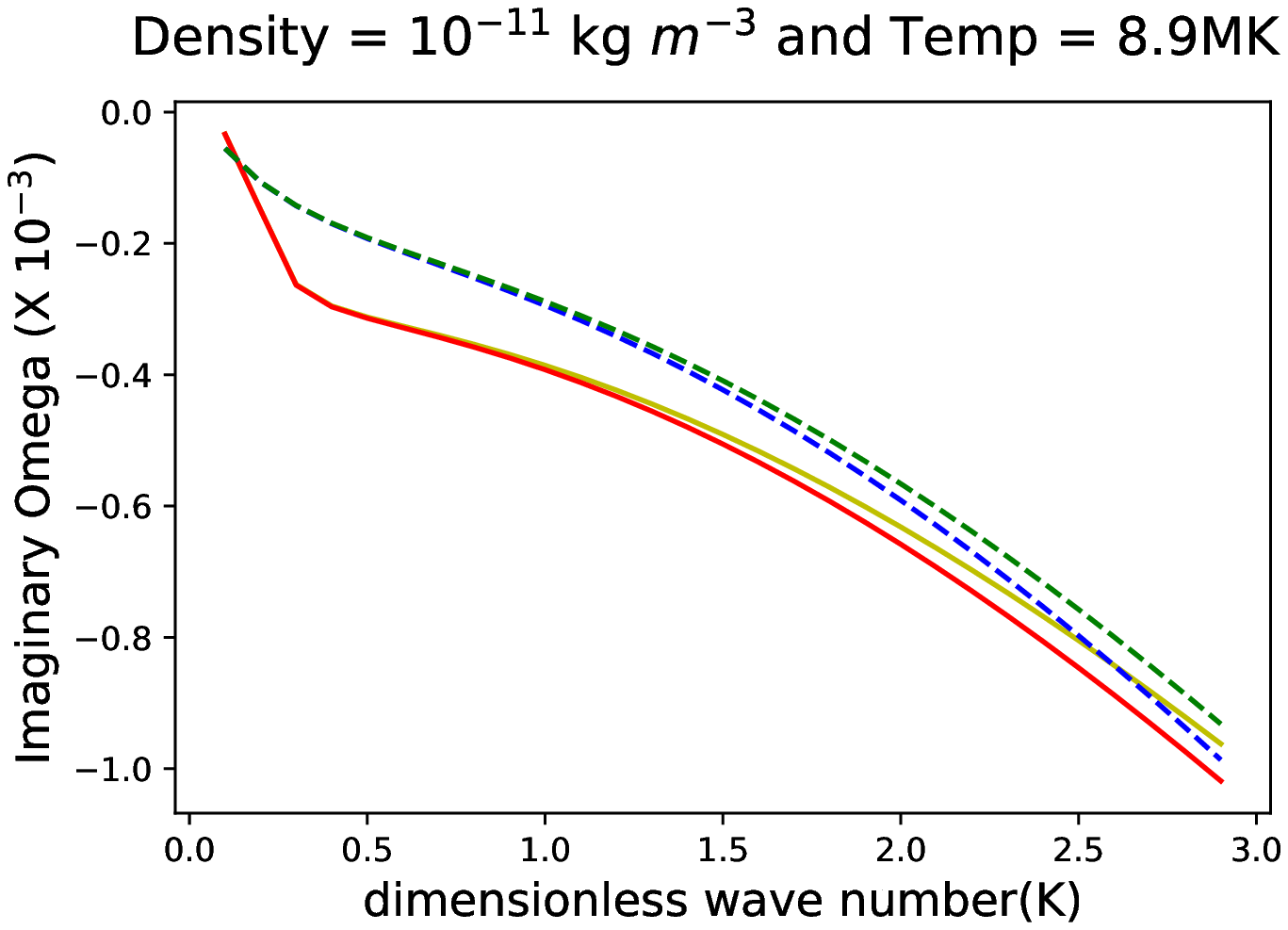}
               \hspace*{-0.015\textwidth}
               \includegraphics[width=0.495\textwidth,clip=]{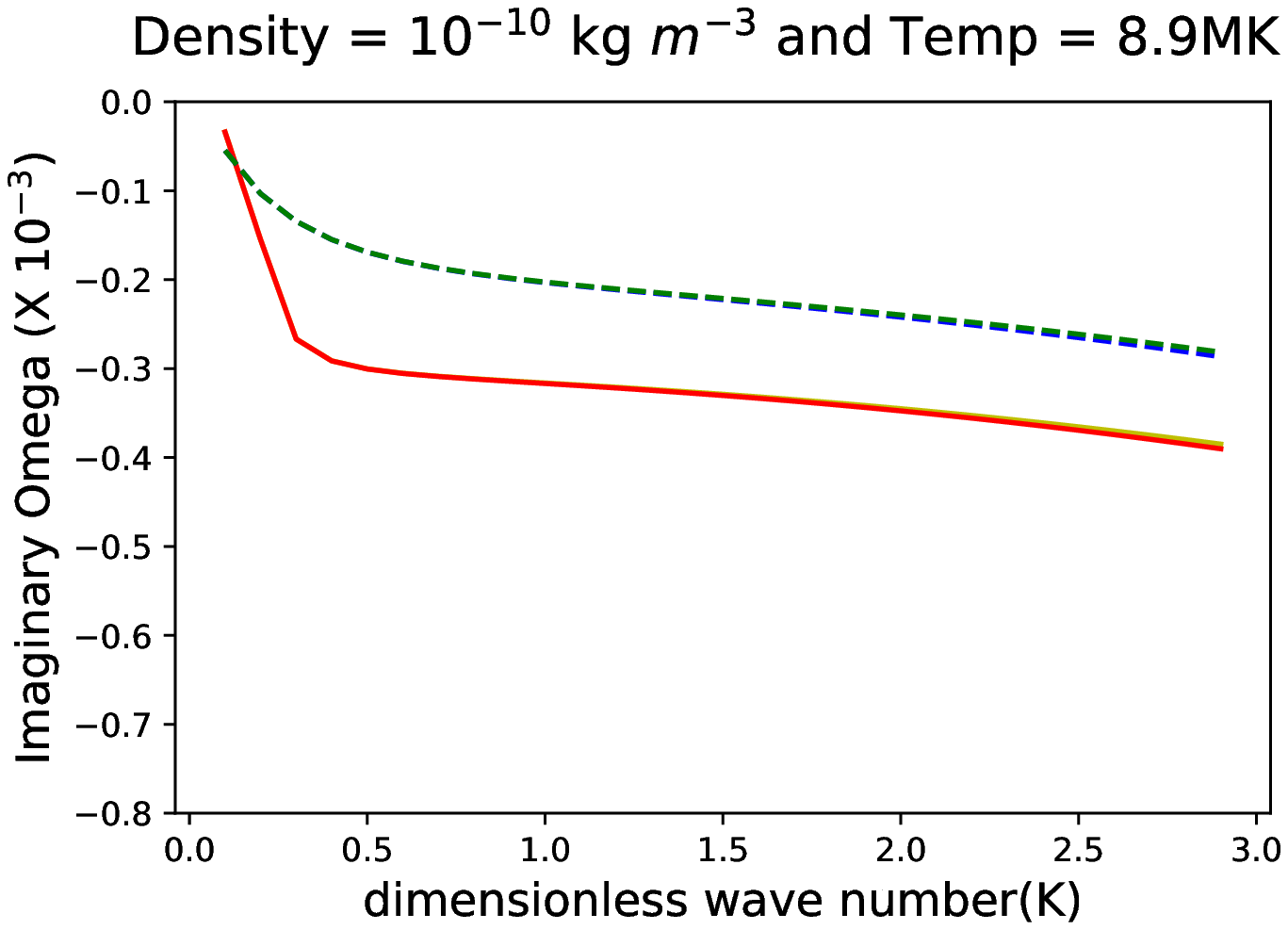}
               \hspace*{-0.015\textwidth}
               \includegraphics[width=0.495\textwidth,clip=]{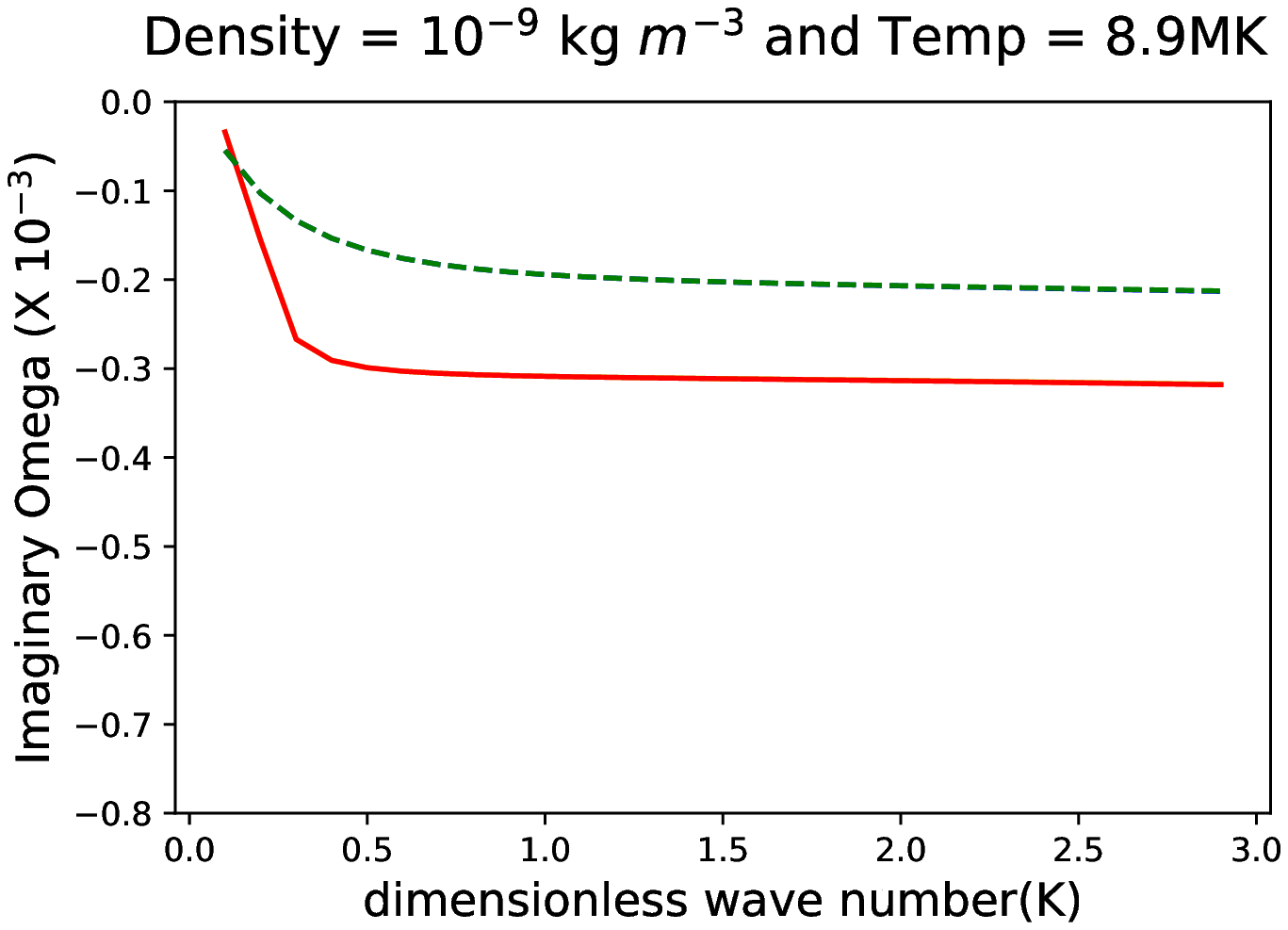}
              }
     \vspace{-0.35\textwidth}   
     \centerline{\Large \bf     
      \hspace{0.0 \textwidth} \color{white}{(c)}
      \hspace{0.415\textwidth}  \color{white}{(d)}
         \hfill}
     \vspace{0.31\textwidth}    
              
\caption{Variation of \(\omega_I\) with non dimensional wave number $K$ at 5.0, 6.3, 8.9 MK temperatures in a loop of length $L$=500 Mm. The columns from left to right show the loops with densities 10$^{-11}$ kg m$^{-3}$, 10$^{-10}$ kg m$^{-3}$, and 10$^{-9}$ kg m$^{-3}$ respectively. In each panel, the red and yellow curves correspond to the solution of the dispersion relation with and without the effect of compressive viscosity respectively when the heating-cooling imbalance is present. The blue-dotted and green-dotted curves  represent the same as the red and yellow ones but for the case when the heating-cooling imbalance is not present. Thermal conductivity is always present as a damping mechanism in these analyses.
        }
   \label{F-4panels}
   \end{figure}

  \begin{figure}    
   \centerline{\hspace*{0.015\textwidth}
               \includegraphics[width=0.695\textwidth,clip=]{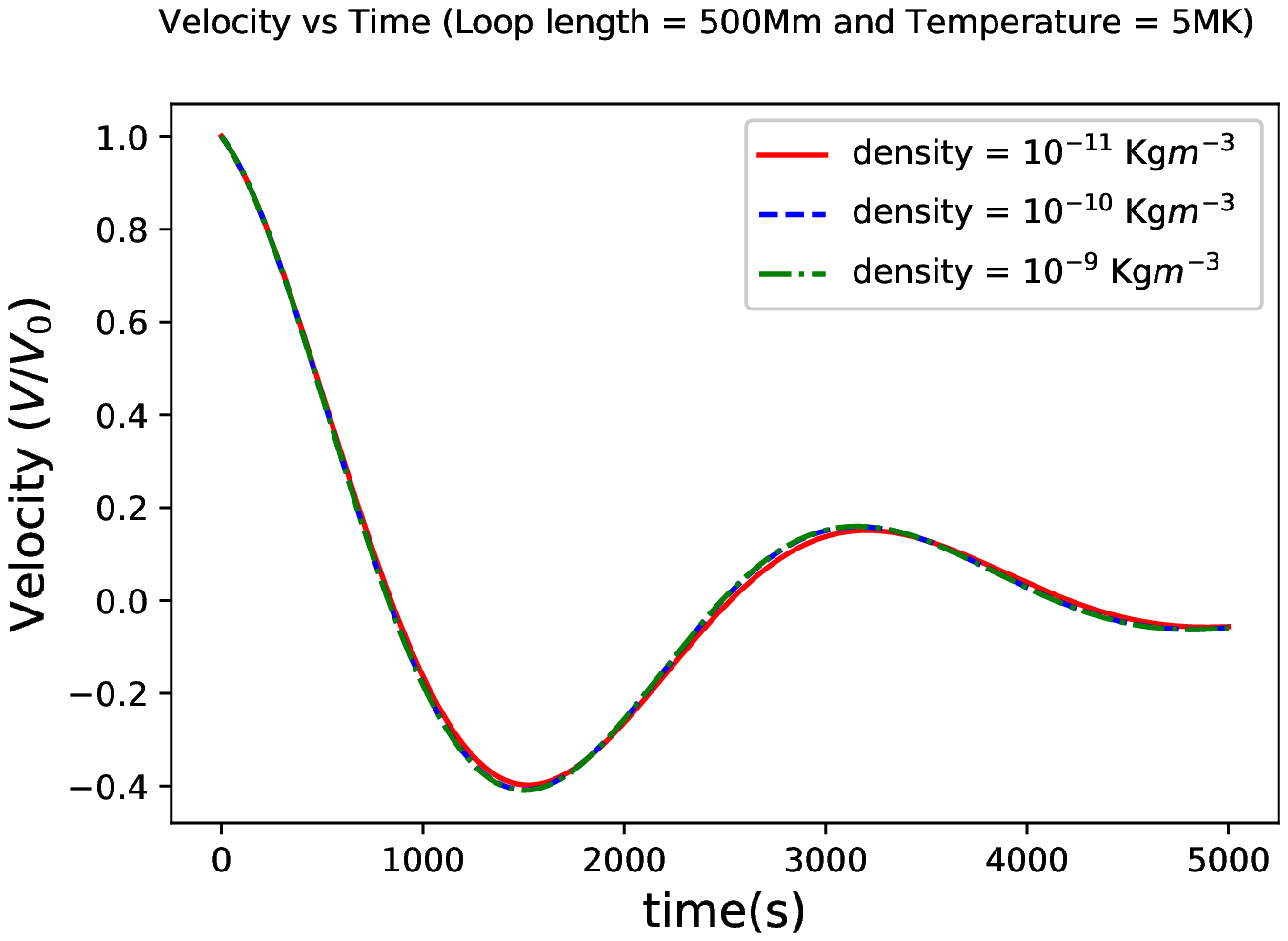}
               \hspace*{0.015\textwidth}
               \includegraphics[width=0.695\textwidth,clip=]{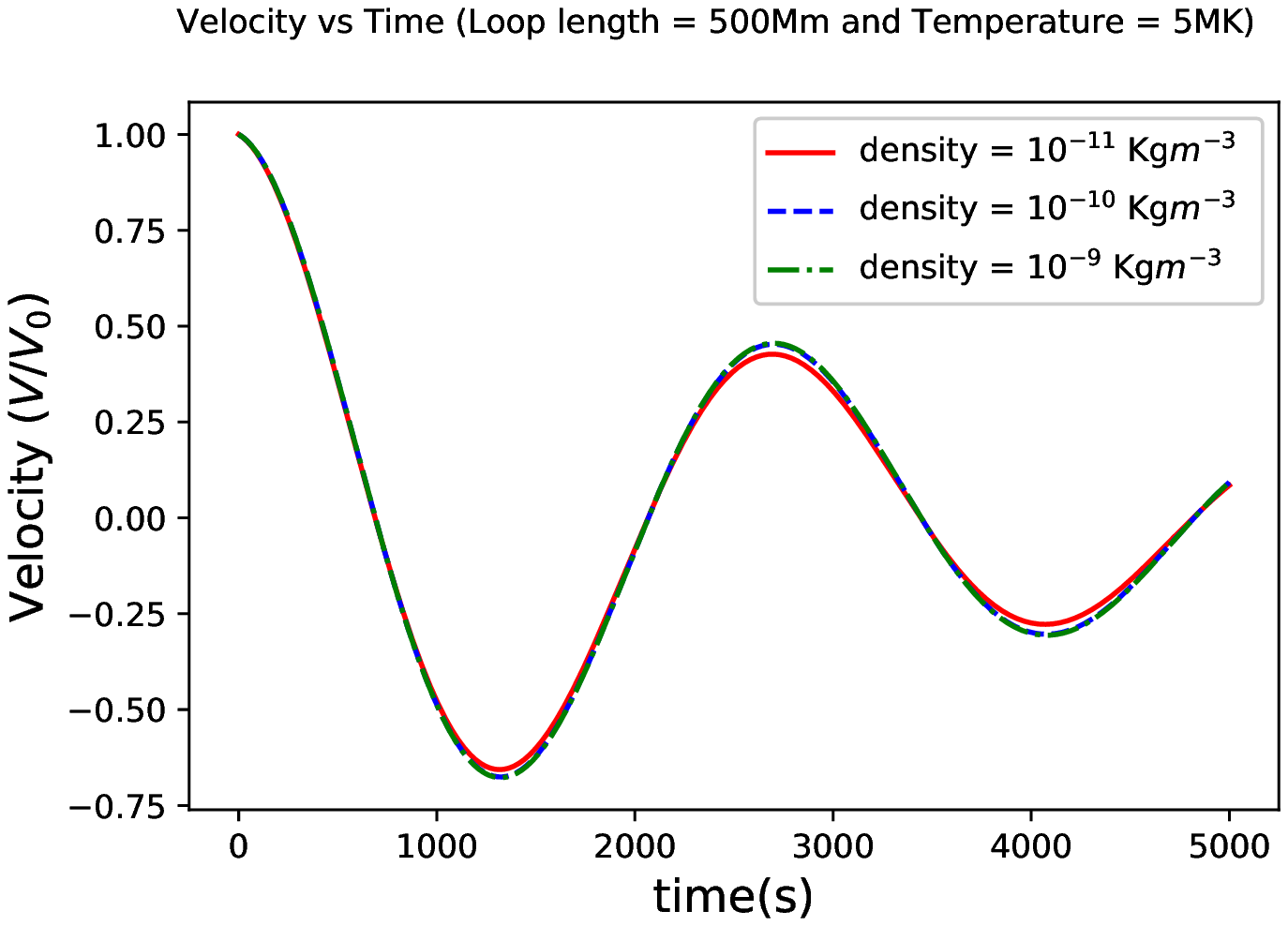}
              }
     \vspace{-0.35\textwidth}   
     \centerline{\Large \bf     
      \hspace{0.0 \textwidth}  \color{white}{(a)}
      \hspace{0.415\textwidth}  \color{white}{(b)}
         \hfill}
     \vspace{0.31\textwidth}    
   \centerline{\hspace*{0.015\textwidth}
               \includegraphics[width=0.695\textwidth,clip=]{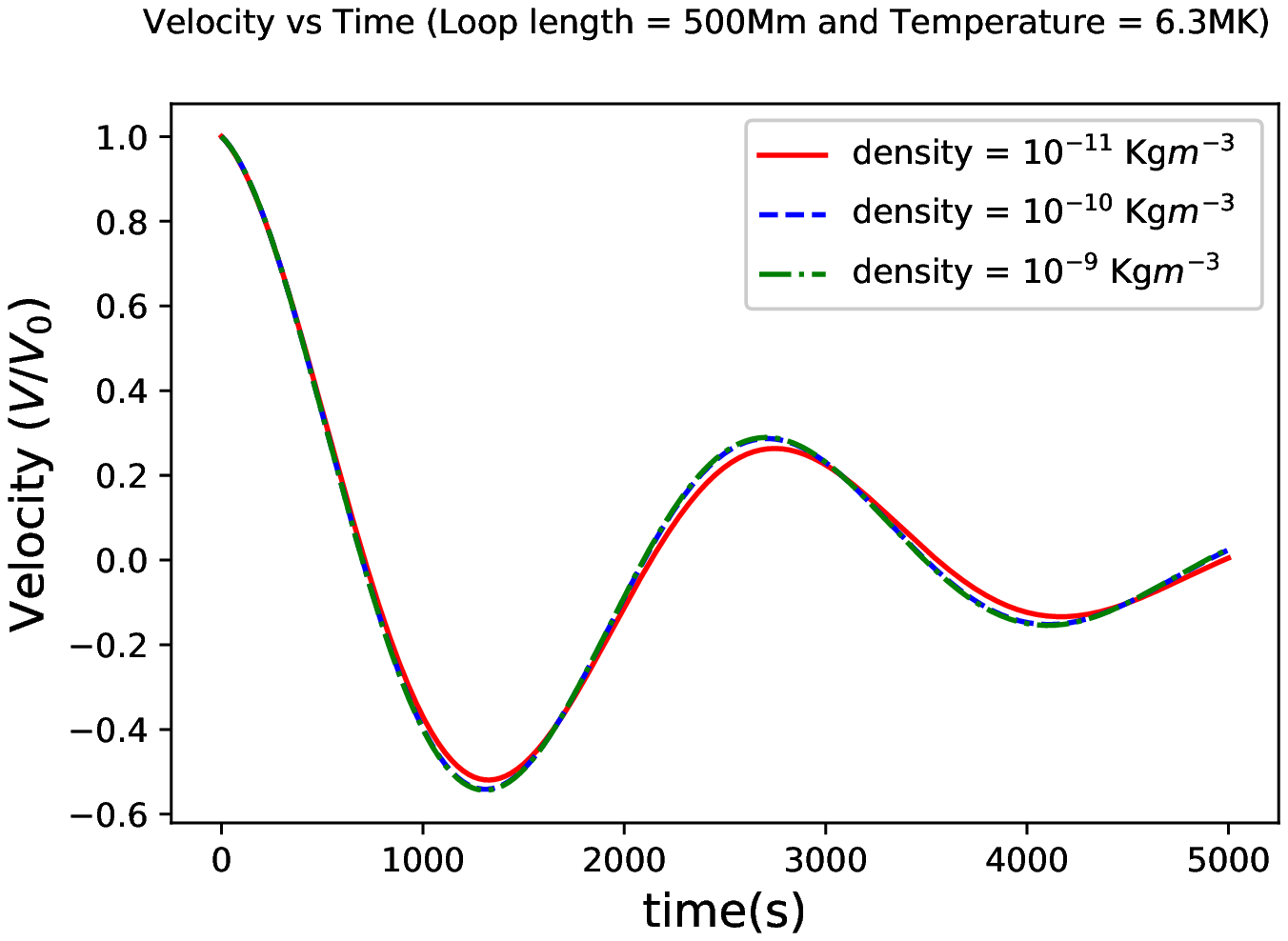}
               \hspace*{0.015\textwidth}
               \includegraphics[width=0.695\textwidth,clip=]{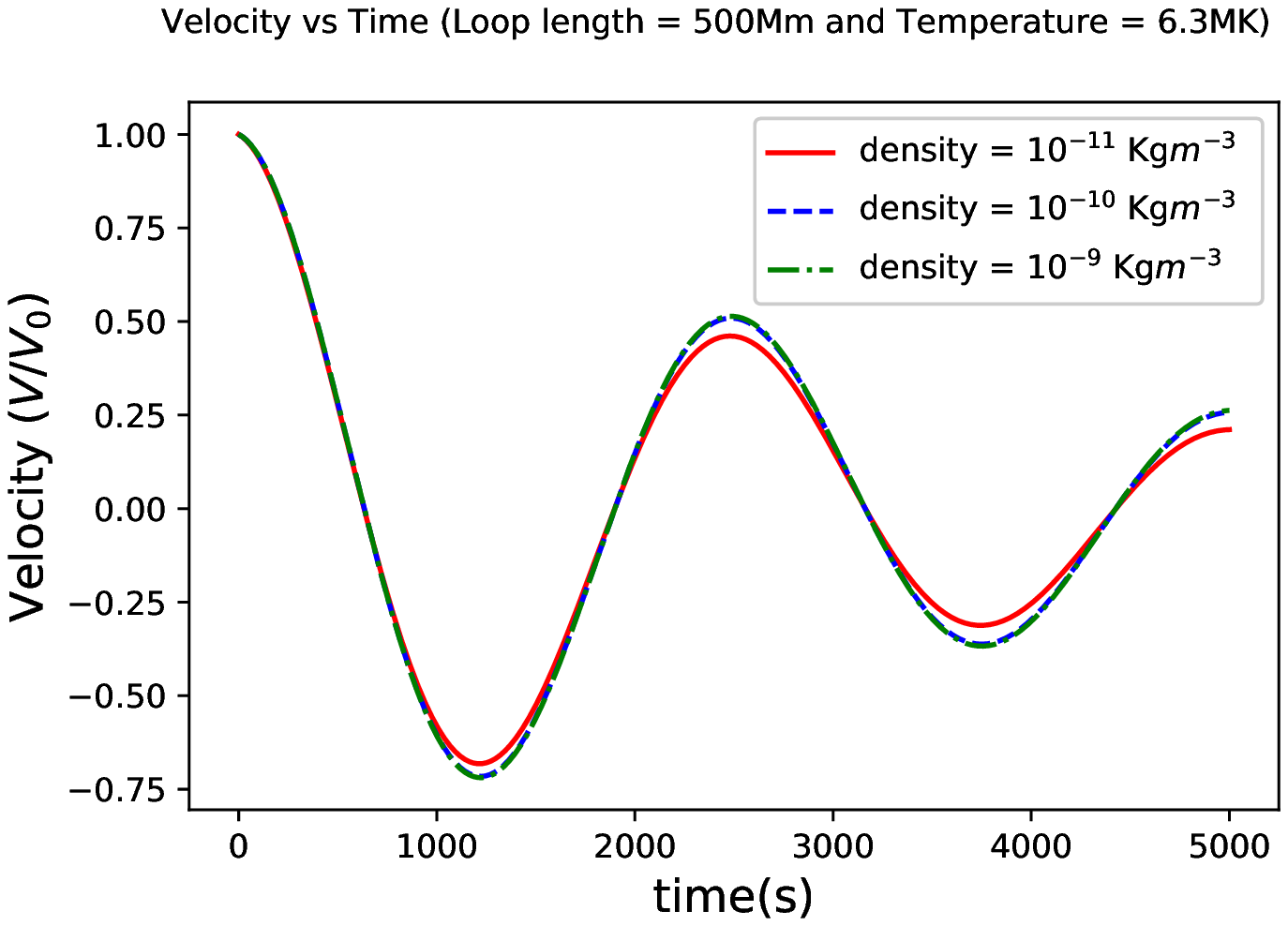}
              }
     \vspace{-0.35\textwidth}   
     \centerline{\Large \bf     
      \hspace{0.0 \textwidth} \color{white}{(c)}
      \hspace{0.415\textwidth}  \color{white}{(d)}
         \hfill}
     \vspace{0.31\textwidth}    
     
      \centerline{\hspace*{0.015\textwidth}
               \includegraphics[width=0.695\textwidth,clip=]{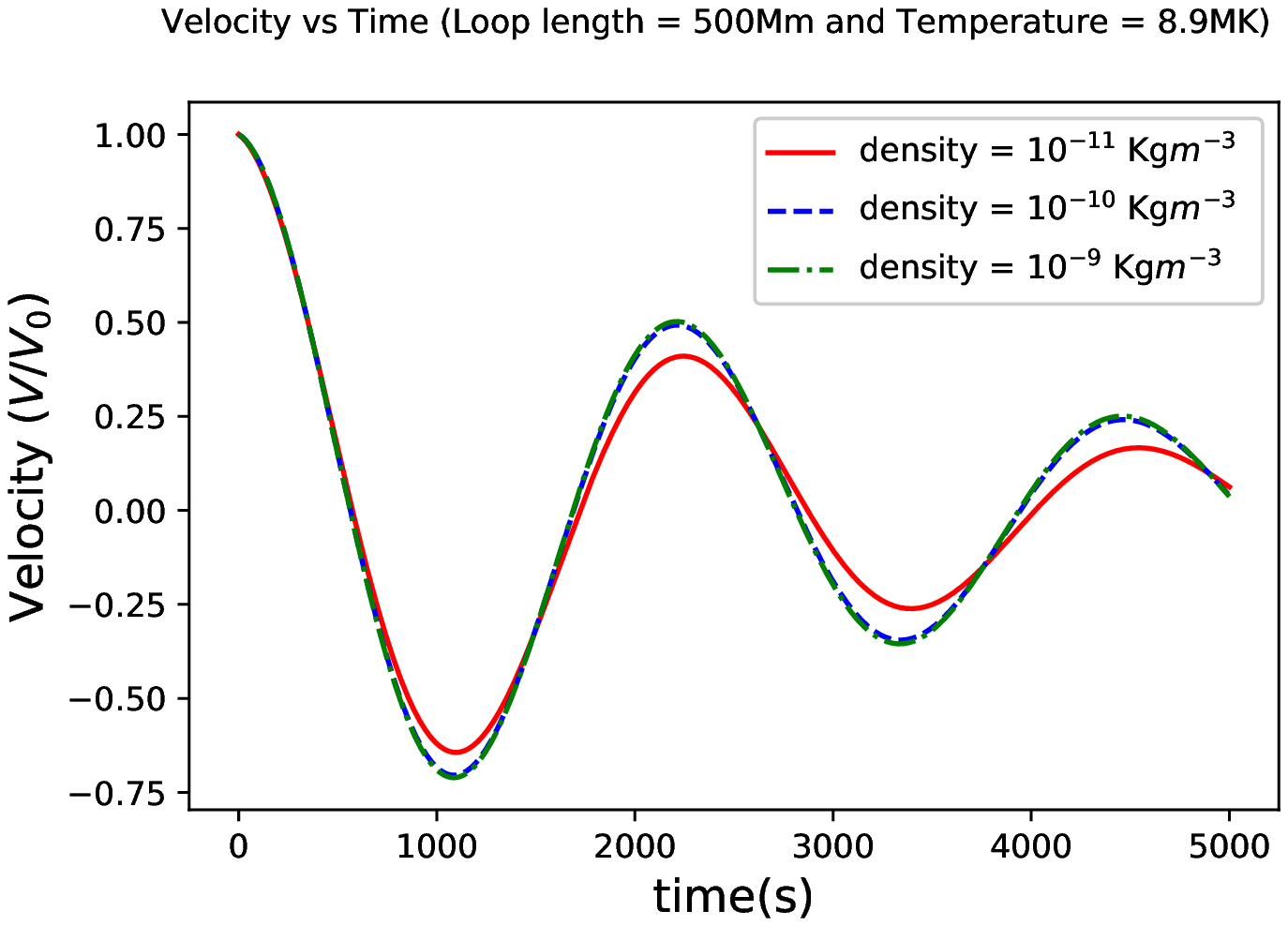}
               \hspace*{0.015\textwidth}
               \includegraphics[width=0.695\textwidth,clip=]{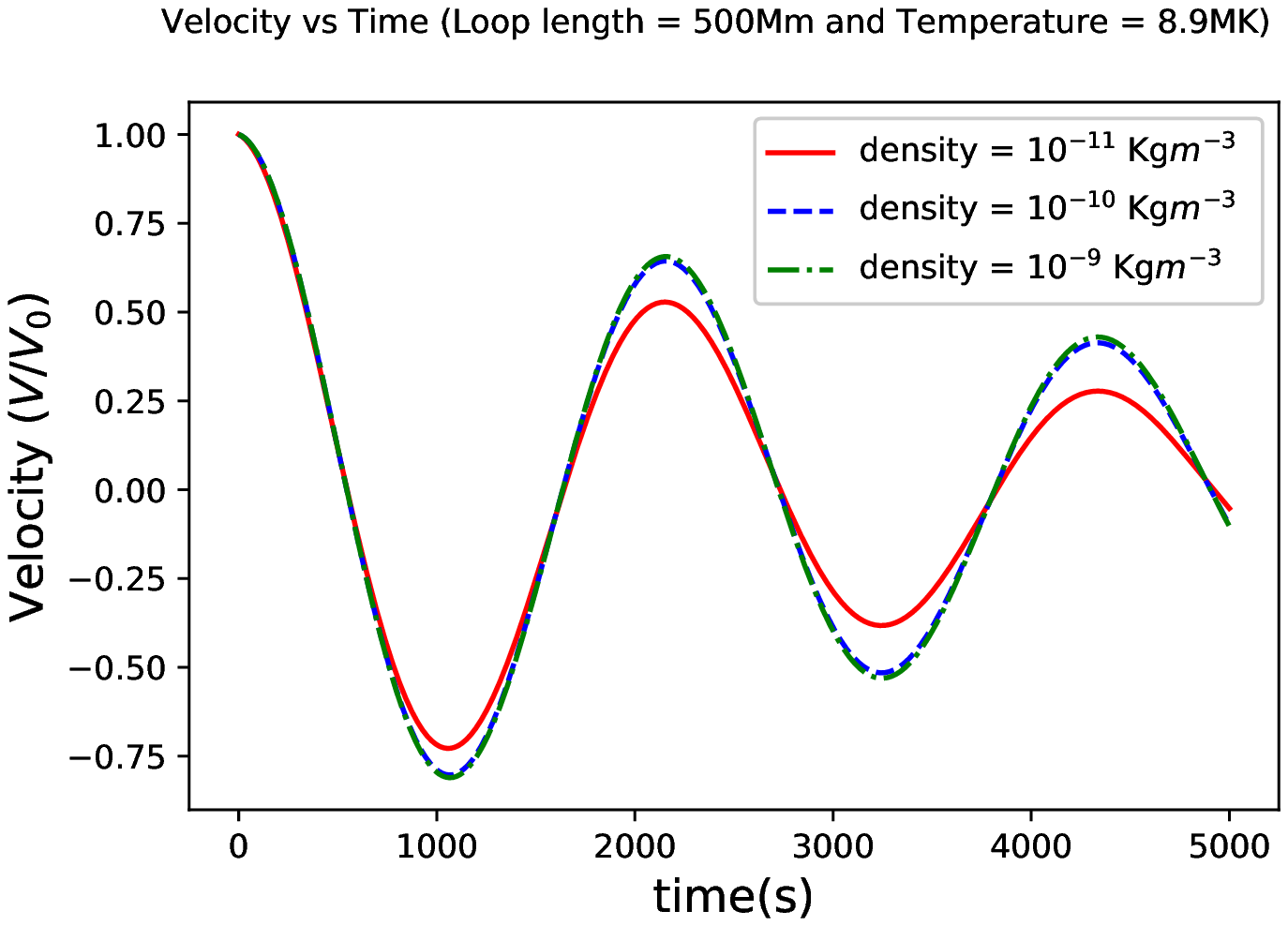}
              }
     \vspace{-0.35\textwidth}   
     \centerline{\Large \bf     
      \hspace{0.0 \textwidth} \color{white}{(c)}
      \hspace{0.415\textwidth}  \color{white}{(d)}
         \hfill}
     \vspace{0.31\textwidth}    
              
\caption{Left panels show the temporal variation of velocity $V$ for the fundamental mode of slow magnetoacoustic oscillations at $z=\frac{L}{2}$ for a loop of length $L$=500 Mm with densities $\rho$=10$^{-11}$ (red), 10$^{-10}$ (blue-dotted), 10$^{-9}$ (green-dotted) kg m$^{-3}$ at $T$=5.0, 6.3, 8.9 MK when heating-cooling imbalance is present. Right panels are similar to the ones on the left but for the case without heating-cooling imbalance.
        }
   \label{F-4panels}
   \end{figure}
  
  
In the previous section, we have found that the thermal conductivity along with the heating-cooling imbalance causes the efficient damping of the fundamental slow mode oscillations in the  hot regime ($T\leq$10 MK) of the longer loops, while compressive viscosity does not lead to any significant enhancement of the damping of this oscillation. Keeping this scenario in view, in this section we present some additional analyses about the role of loop density in the damping of slow waves in a coronal loop of length $L$=500 Mm within the hot regime. 
  
In Figure 6, the variation of \(\omega_I\) with non-dimensional wave number $K$ is presented at $T$=5.0, 6.3, 8.9 MK in the loop with densities ($\rho$) 10$^{-11}$, 10$^{-10}$, and 10$^{-9}$ kg m$^{-3}$. These panels collectively show that in the hot regime at each temperature value of 5.0, 6.3, and 8.9 MK,  the \(\omega_I\) values for the fundamental mode do not show any appreciable drop (red/yellow curves when compared to each other) as the density increases. This depicts that the damping rate due to heating-cooling imbalance is not affected much in more bulky loops (e.g. post-flare loop arcades) compared to the loops with normal coronal density. At a given temperature, the value of \(\omega_I\) shown by the red-yellow curve slightly decreases with the increase in density for the fundamental mode, which depicts that the increase in density slightly suppresses the damping due to thermal conduction and heating-cooling imbalance. The red and yellow curves are superimposed, and a similar scenario is true for the green-dotted and blue-dotted curves. This implies that the compressive viscosity does not add any enhancement to the damping of the fundamental modes, while the damping is mostly performed by thermal conductivity. A similar physical scenario is also evident for the damping of the higher order harmonics in the longest loop of 500 Mm length. The damping effects caused  by the heating-cooling imbalance and thermal conductivity on the higher order harmonics (e.g. second harmonics with $K$=2.0) do not change appreciably with the increase of the loop density. However, the curves of $\omega_I$ become flatter with the increase of the harmonic number, showing a behavior distinctly different from those where thermal conduction dominates the damping (compare the right columns of Figure 6 and Figure 8). This may suggest that the role of optically thin radiation becomes important in the damping of higher harmonic slow modes in the over-dense loops \citep{2004A&A...415..705D,2006SoPh..236..127P}. A detailed analysis of the effects of radiation is beyond the scope of this study. In the longest loop with a normal coronal density at a temperature of 6.3\,--\,8.9 MK, some weak damping effect caused by the viscosity on the higher mode harmonics can been seen (cf. middle and bottom panels in the left column in Figure 6). In all the panels, the values of \(\omega_I\) shown by the red/yellow curves are significantly lower compared to the ones represented by the blue/green-dotted curves. This implies that overall the heating-cooling imbalance has larger effects on the damping of slow waves compared to the case when it is not considered.
  
 The top-left and top-right panels in Figure 7 compare the velocity oscillations at different densities in the cases with and without heating-cooling imbalance. It is clear that the the velocity oscillations are  more damped in the case with heating-cooling imbalance (top-left panel).
As temperature increases (middle and bottom panels), this difference in the damping rate reduces. However, one can still see that the velocity oscillations with heating-cooling imbalance considered (middle and bottom-left panels) are  more damped compared to the case without it (middle and bottom-right panels). 

In summary, in the hot ($T\leq$10 MK) longer loops, both thermal conduction and heating-cooling imbalance play significant roles in the damping of fundamental mode oscillations in loops of different densities, however, the damping rate due to heating-cooling imbalance only decreases a little as the loop density increases. 
In addition, the damping effect of radiation may become more important than that of thermal conduction for the higher harmonics in over-dense long loops, 
which needs to be investigated further in the future.

  \subsubsection{Effect of Loop Density on Slow Wave Damping in a Coronal Loop of Length $L$=50 Mm within Super-Hot Regime of Temperature ($T>$10 MK)}
  
\begin{figure}    
   \centerline{\hspace*{0.015\textwidth}
               \includegraphics[width=0.495\textwidth,clip=]{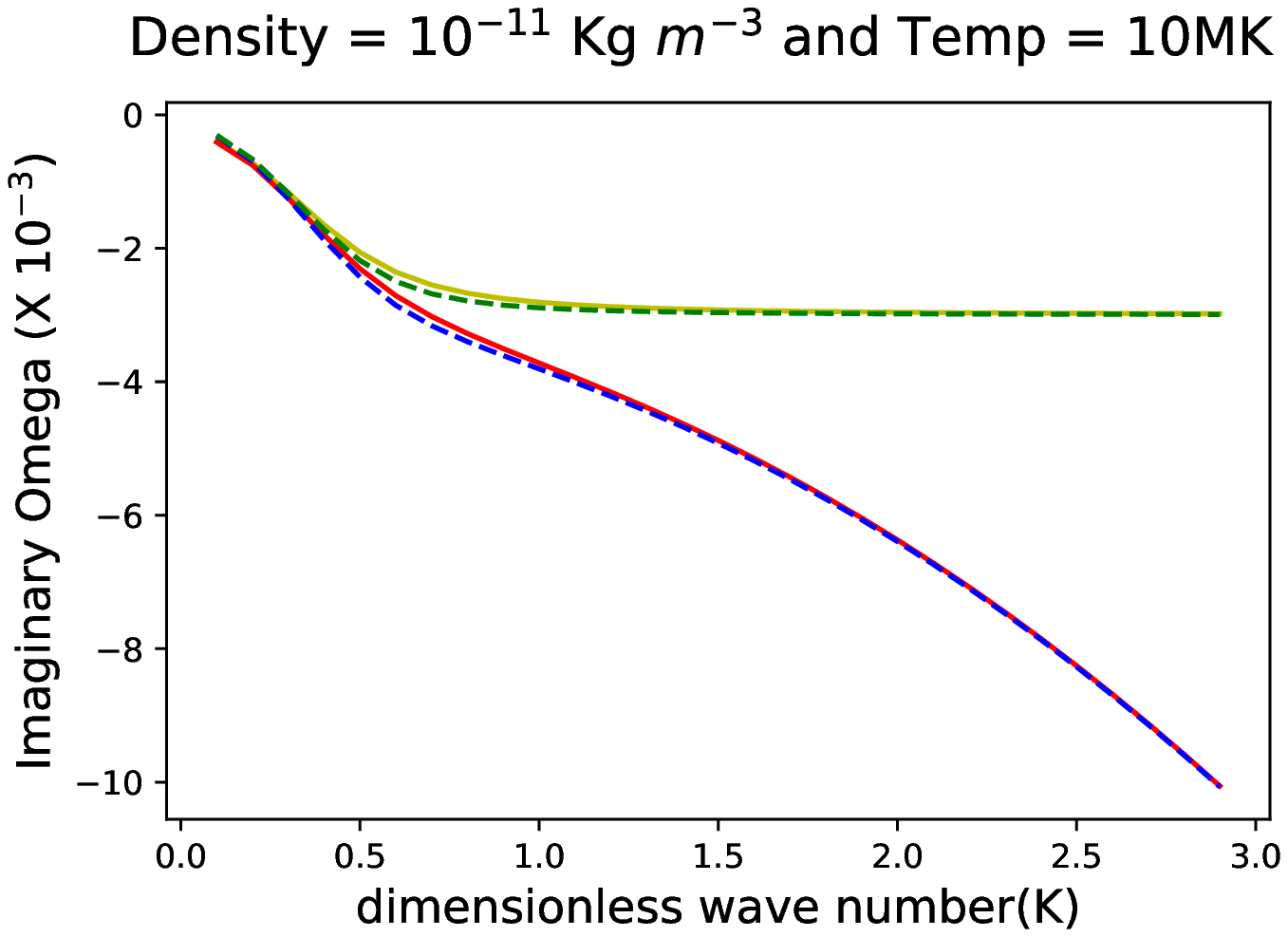}
               \hspace*{-0.015\textwidth}
               \includegraphics[width=0.495\textwidth,clip=]{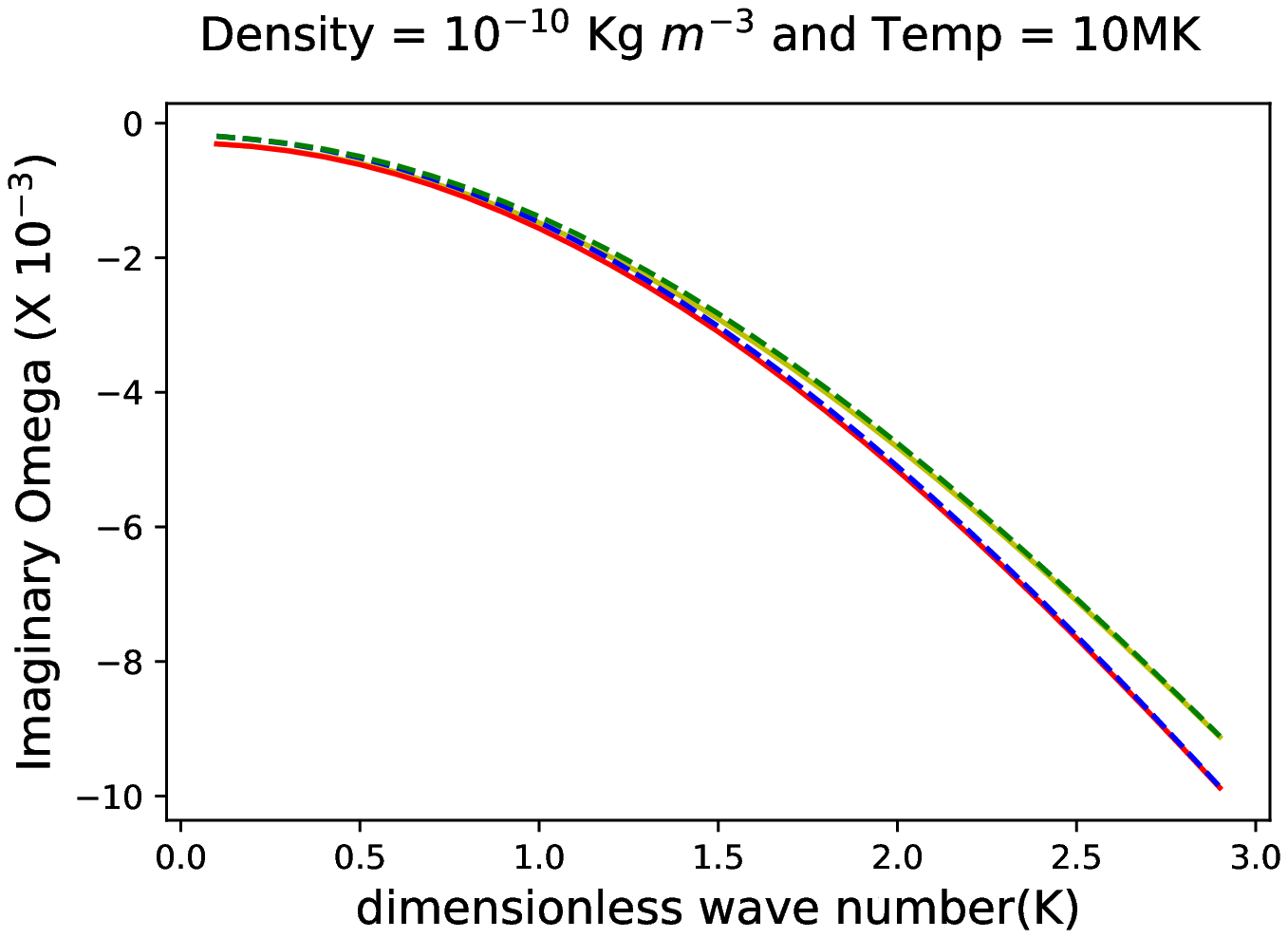}
               \hspace*{-0.015\textwidth}
               \includegraphics[width=0.495\textwidth,clip=]{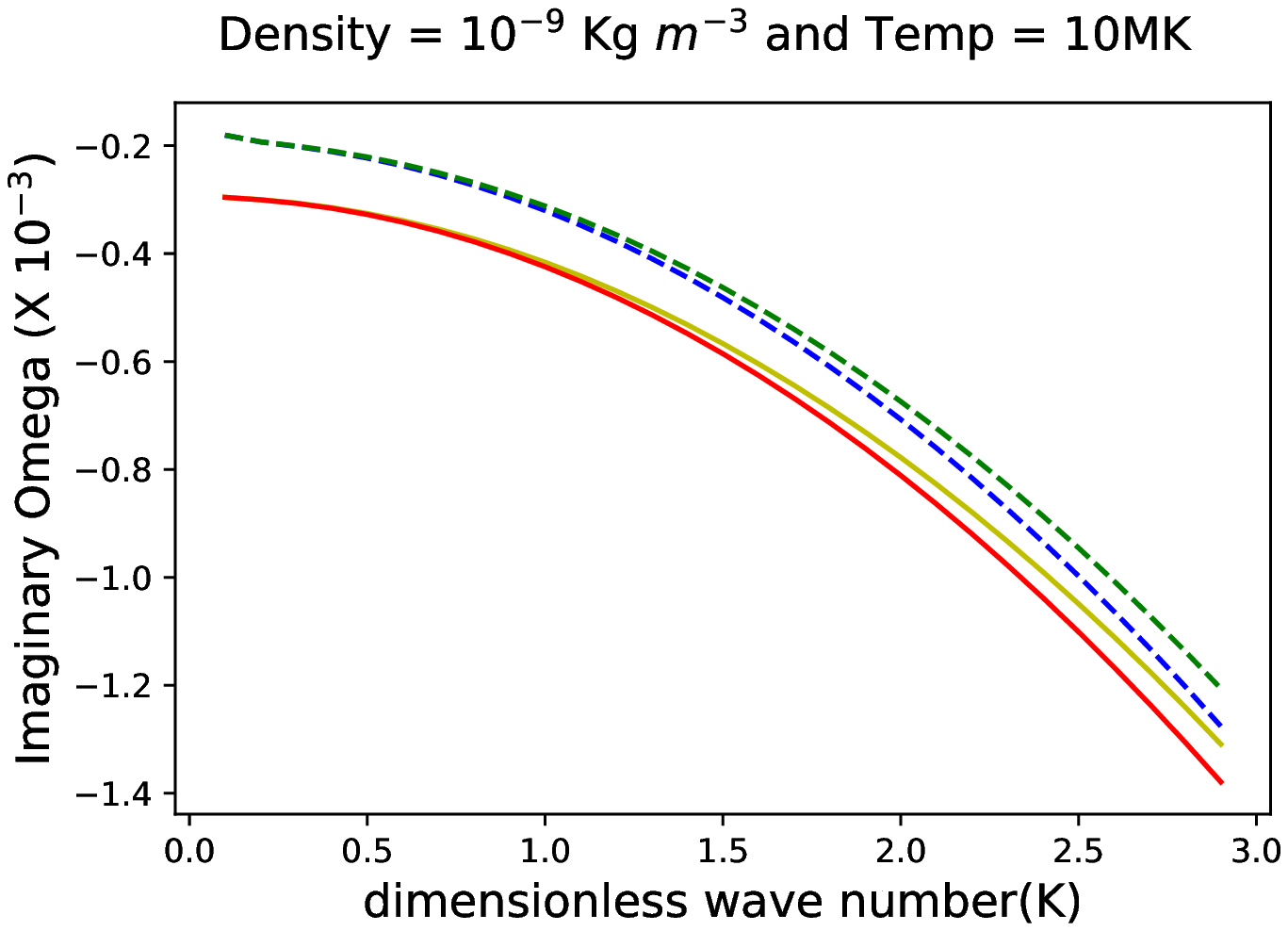}
              }
     \vspace{-0.35\textwidth}   
     \centerline{\Large \bf     
      \hspace{0.0 \textwidth}  \color{white}{(a)}
      \hspace{0.415\textwidth}  \color{white}{(b)}
         \hfill}
     \vspace{0.31\textwidth}    
   \centerline{\hspace*{0.015\textwidth}
               \includegraphics[width=0.495\textwidth,clip=]{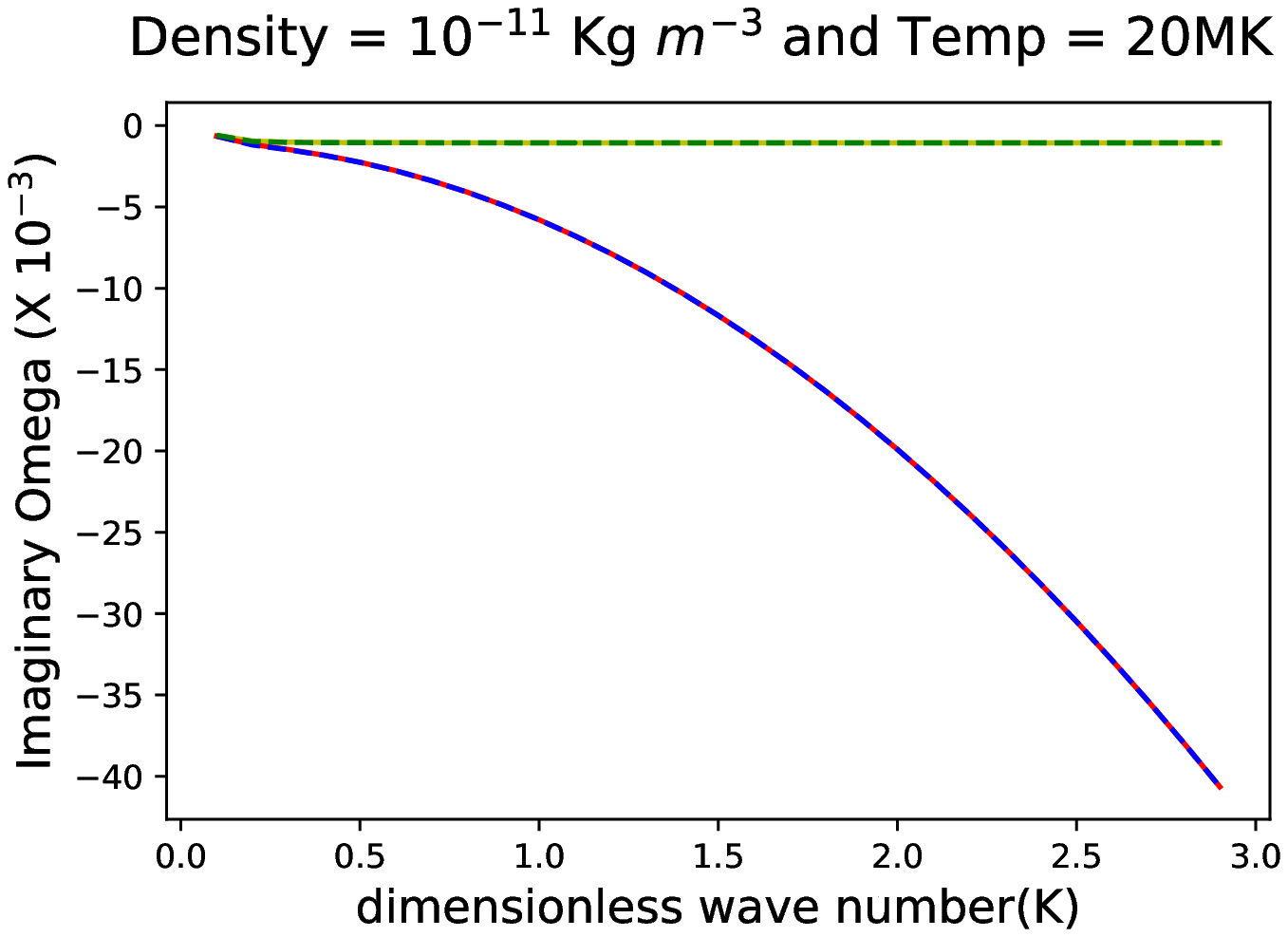}
               \hspace*{-0.015\textwidth}
               \includegraphics[width=0.495\textwidth,clip=]{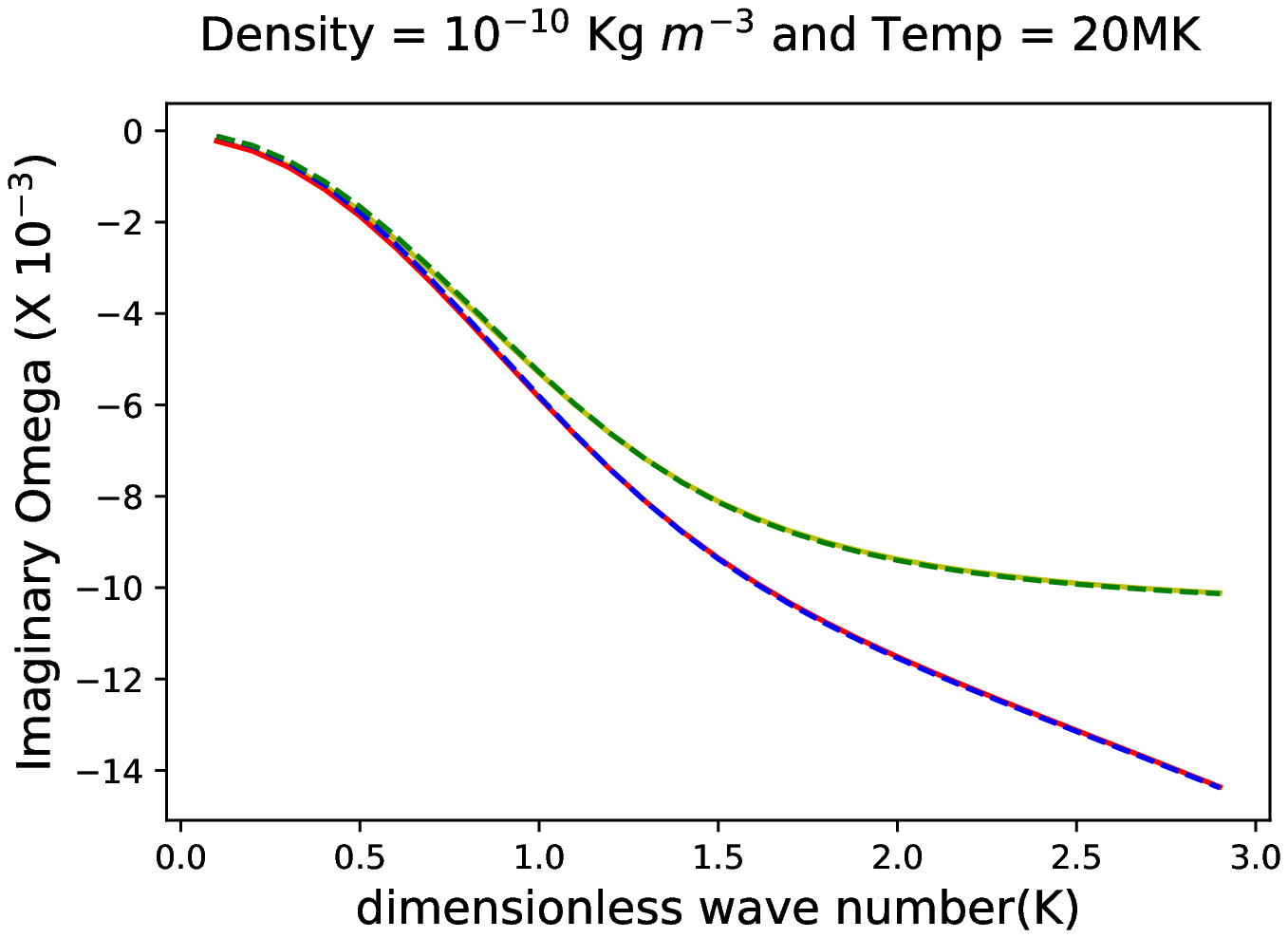}
               \hspace*{-0.015\textwidth}
               \includegraphics[width=0.495\textwidth,clip=]{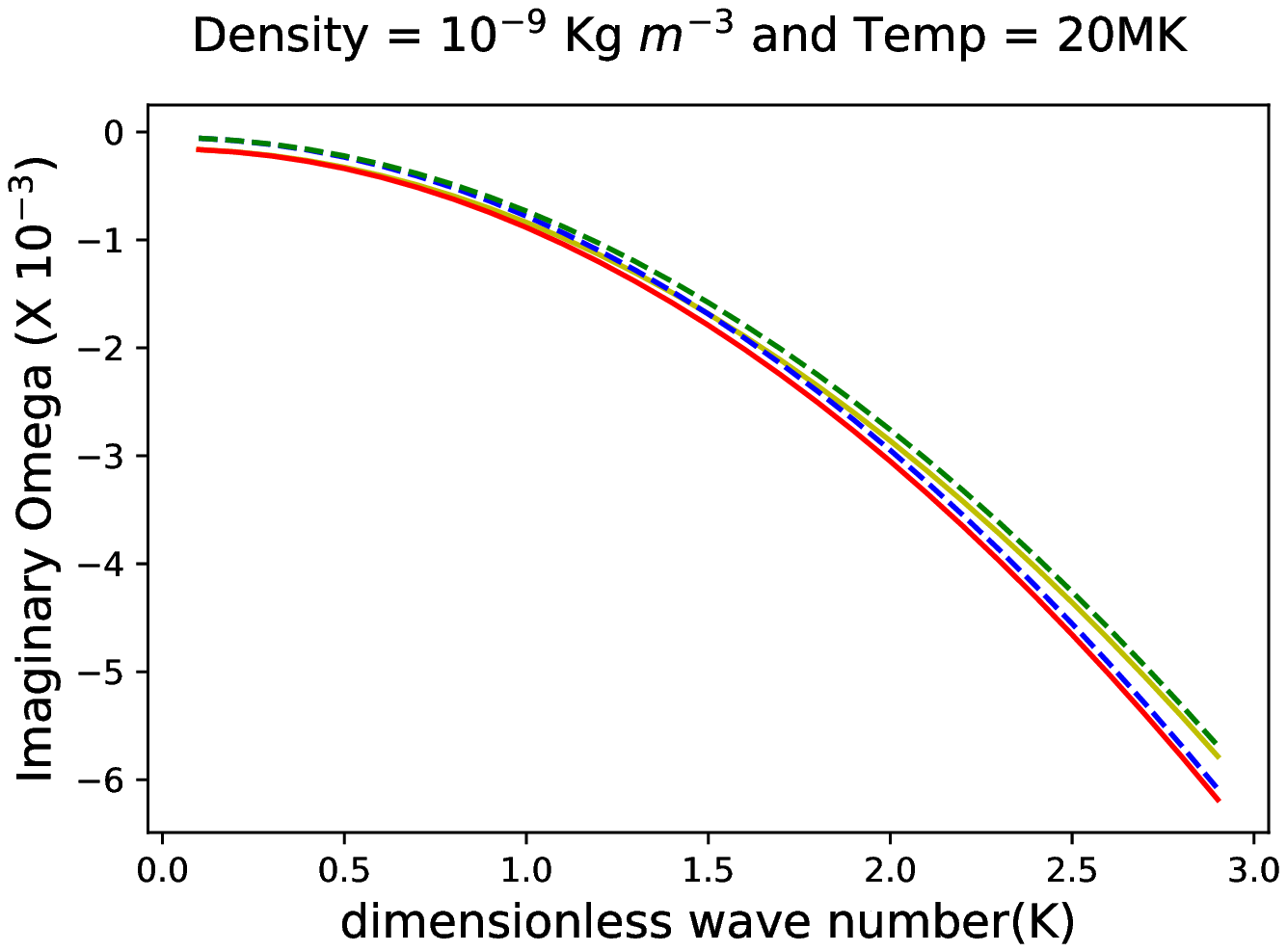}
              }
     \vspace{-0.35\textwidth}   
     \centerline{\Large \bf     
      \hspace{0.0 \textwidth} \color{white}{(c)}
      \hspace{0.415\textwidth}  \color{white}{(d)}
         \hfill}
     \vspace{0.31\textwidth}    
     
      \centerline{\hspace*{0.015\textwidth}
               \includegraphics[width=0.495\textwidth,clip=]{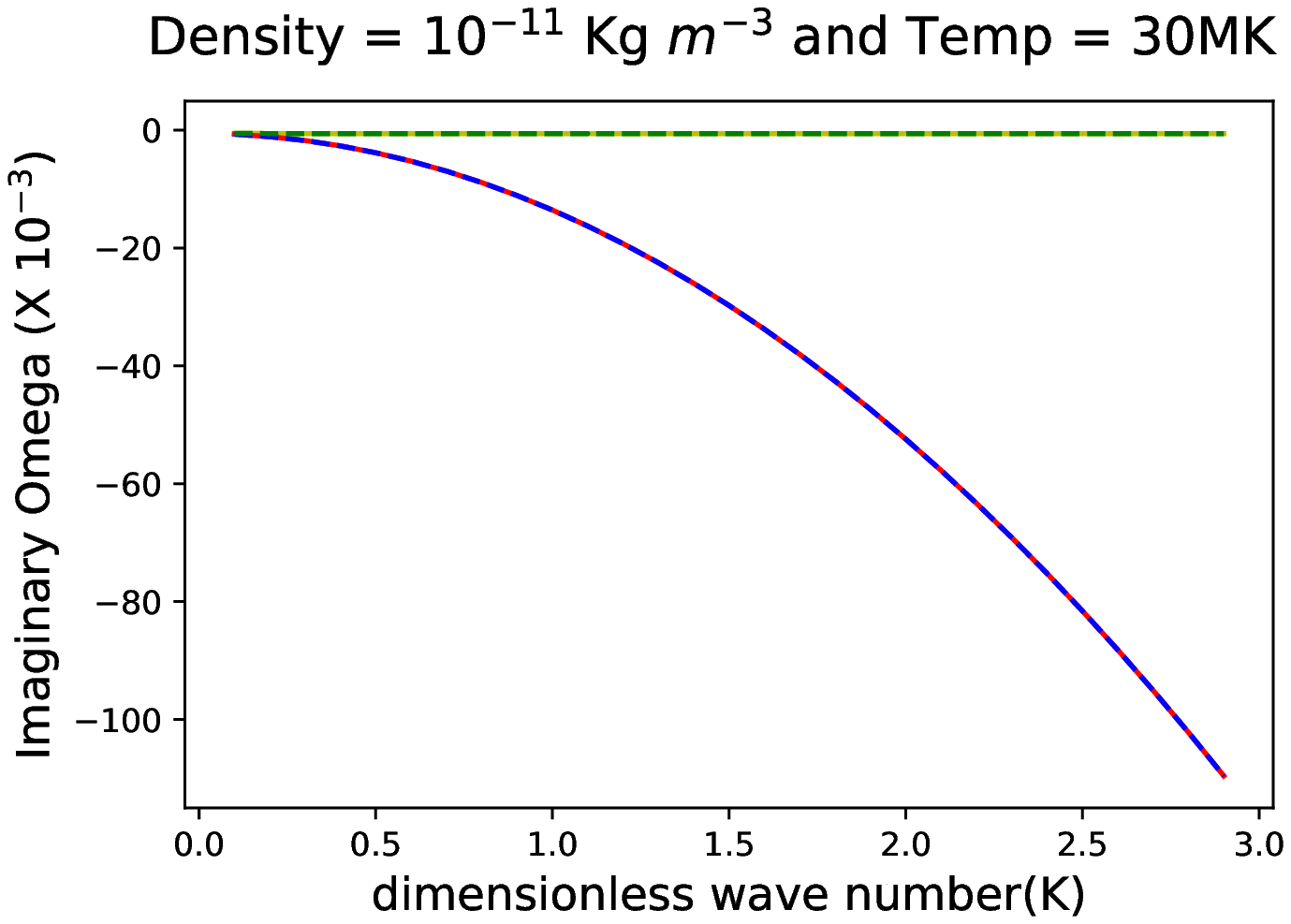}
               \hspace*{-0.015\textwidth}
               \includegraphics[width=0.495\textwidth,clip=]{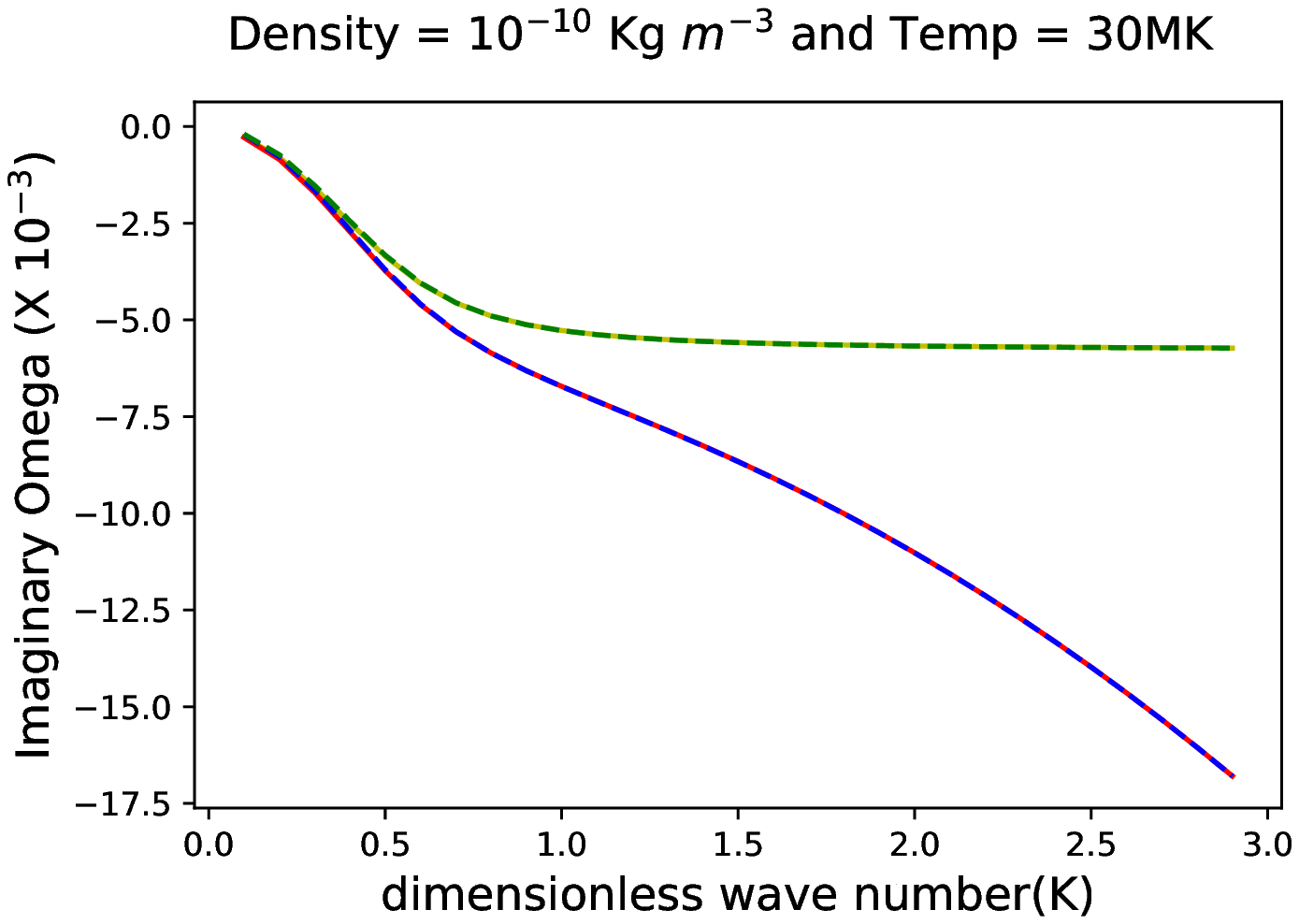}
               \hspace*{-0.015\textwidth}
               \includegraphics[width=0.495\textwidth,clip=]{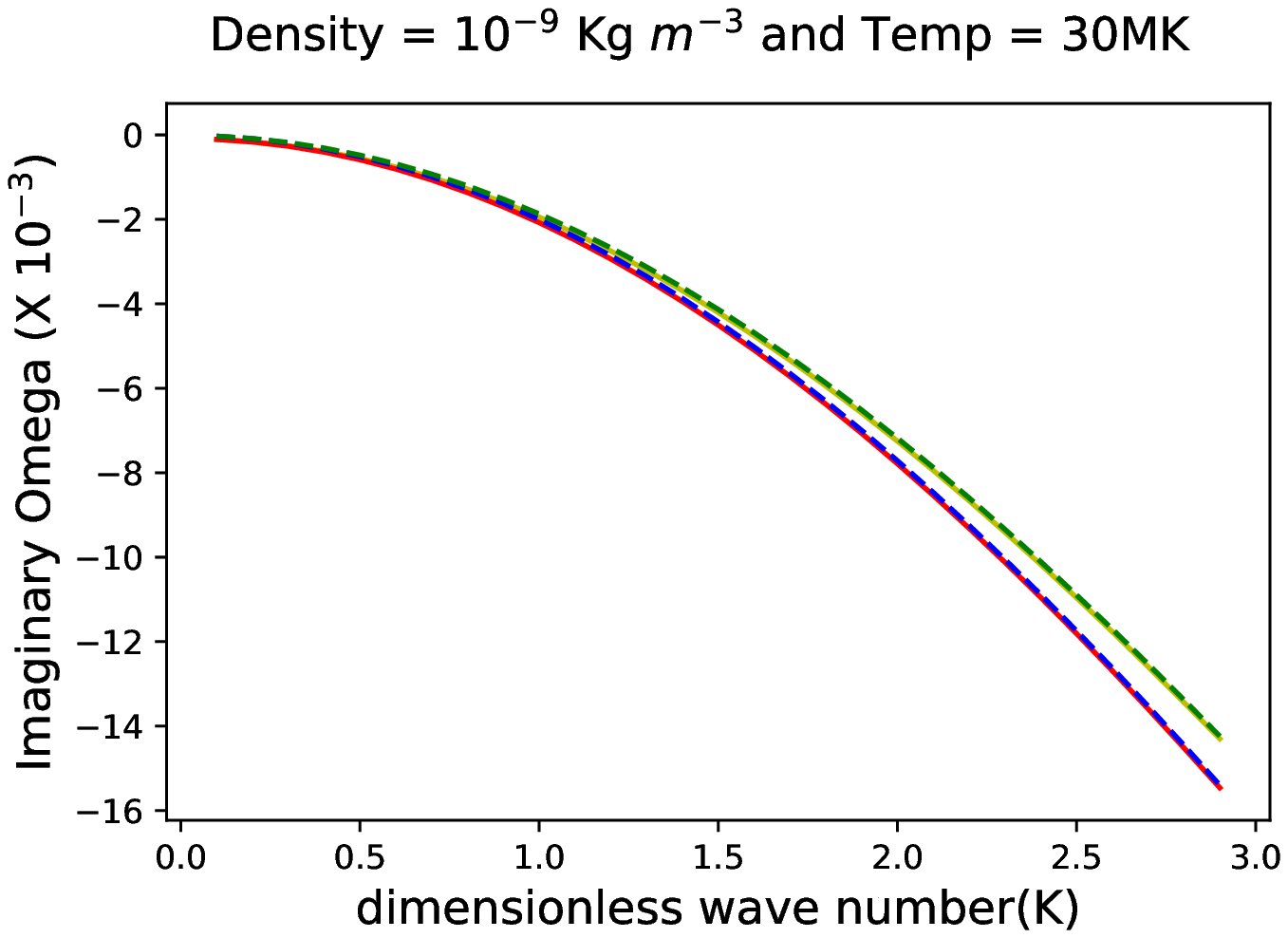}
              }
     \vspace{-0.35\textwidth}   
     \centerline{\Large \bf     
      \hspace{0.0 \textwidth} \color{white}{(c)}
      \hspace{0.415\textwidth}  \color{white}{(d)}
         \hfill}
     \vspace{0.31\textwidth}    
              
\caption{Variation of \(\omega_I\) with non-dimensional wave number K at  10, 20,30 MK temperatures in a loop of length $L$=50 Mm. The left-most column shows the variations for the loop with  normal coronal density $\rho$=10$^{-11}$ kg m$^{-3}$. The middle and right-most columns represent the estimations for the bulky loops with higher densities $\rho$=10$^{-10}$ kg m$^{-3}$ and $\rho$=10$^{-9}$ kg m$^{-3}$ respectively. In each panel, the red and yellow curves correspond to the solution of the dispersion relation with and without the effect of compressive viscosity respectively when heating-cooling imbalance is present. The blue-dotted and green-dotted curves show the same as the red and yellow ones but for the case when heating-cooling imbalance is not present. Thermal conductivity is always present as a damping mechanism in these analyses.
        }
   \label{F-4panels}
   \end{figure}
\begin{figure}    
   \centerline{\hspace*{0.015\textwidth}
               \includegraphics[width=0.695\textwidth,clip=]{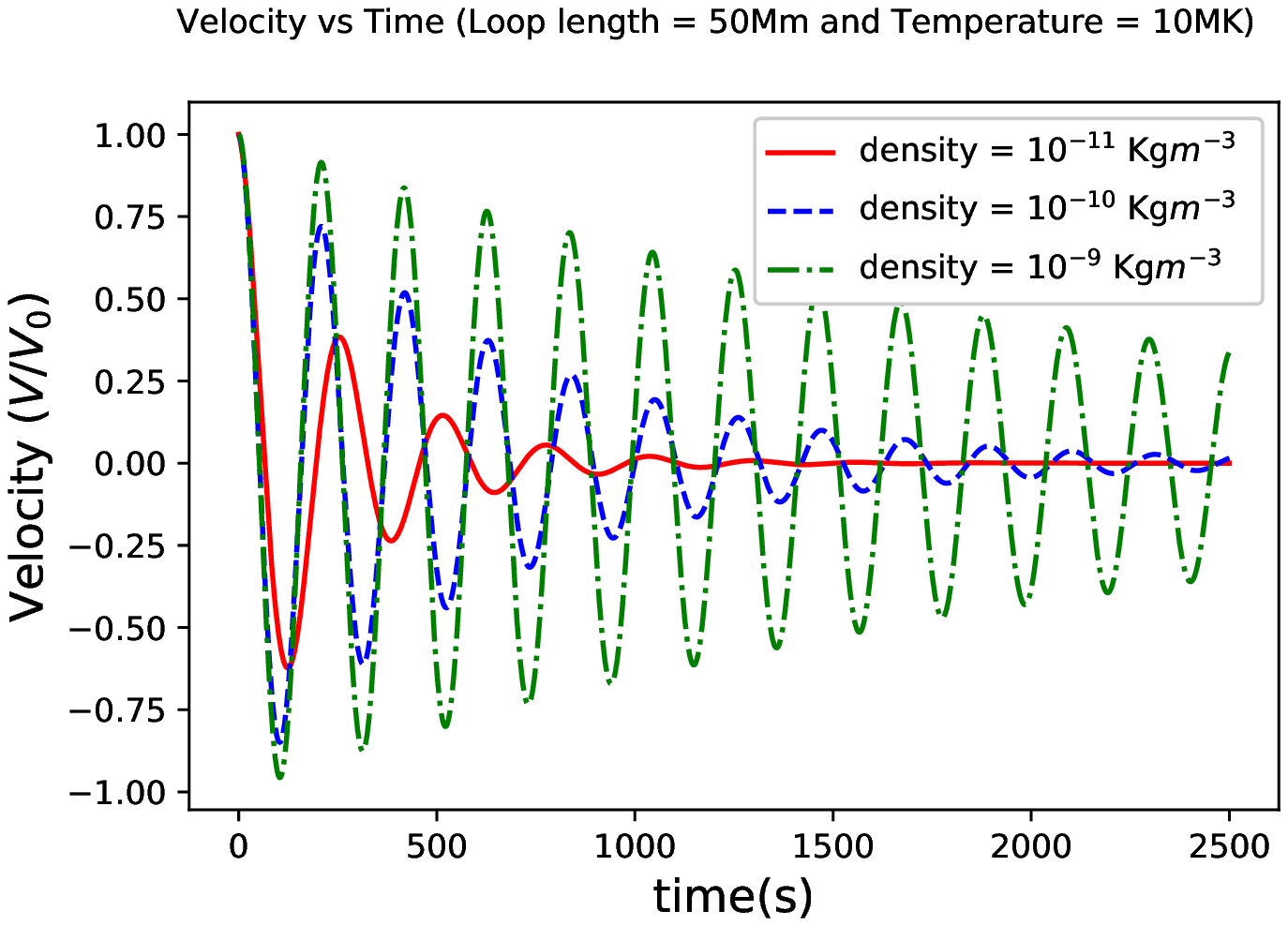}
               \hspace*{0.015\textwidth}
               \includegraphics[width=0.695\textwidth,clip=]{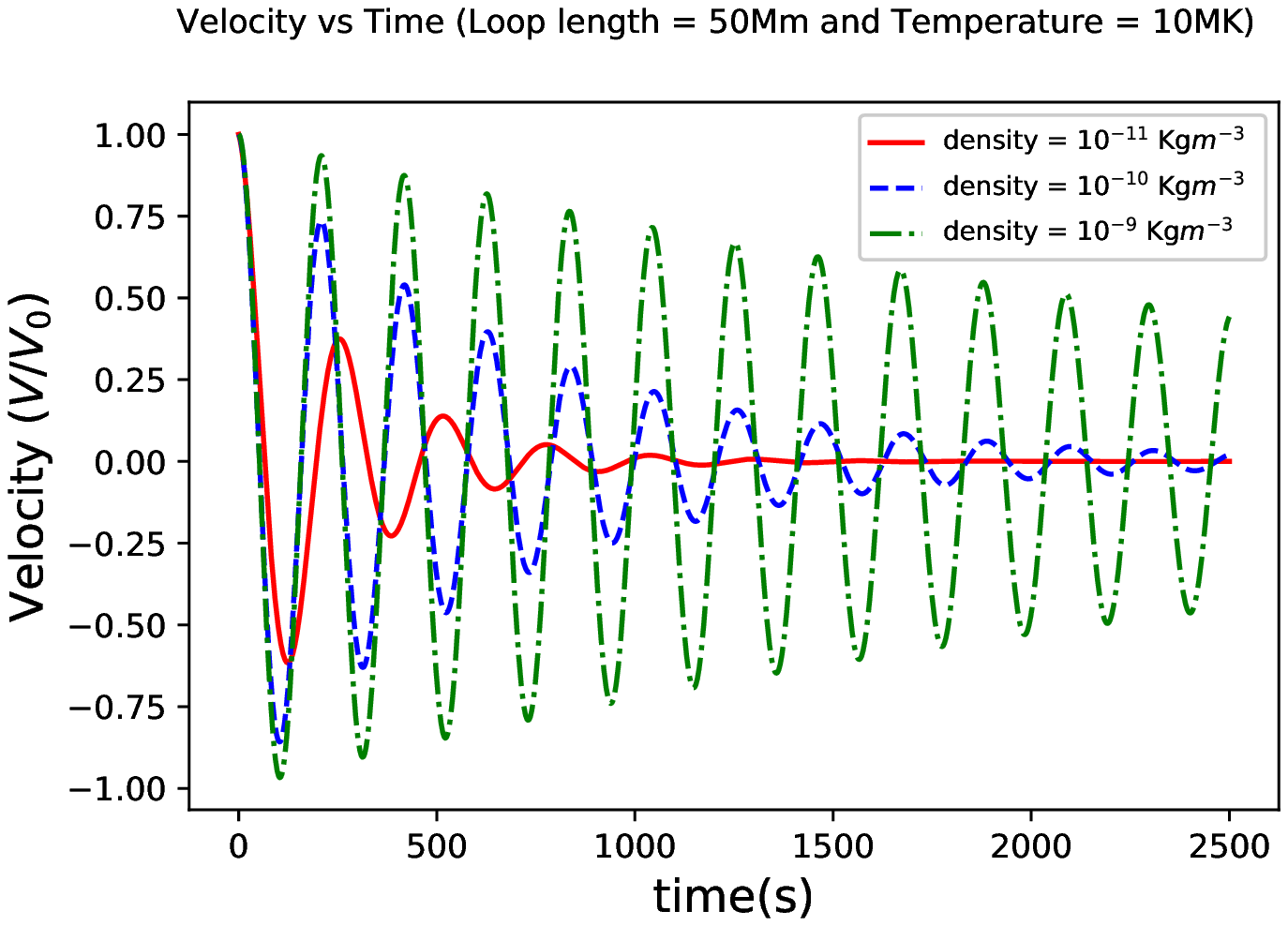}
              }
     \vspace{-0.35\textwidth}   
     \centerline{\Large \bf     
      \hspace{0.0 \textwidth}  \color{white}{(a)}
      \hspace{0.415\textwidth}  \color{white}{(b)}
         \hfill}
     \vspace{0.31\textwidth}    
   \centerline{\hspace*{0.015\textwidth}
               \includegraphics[width=0.695\textwidth,clip=]{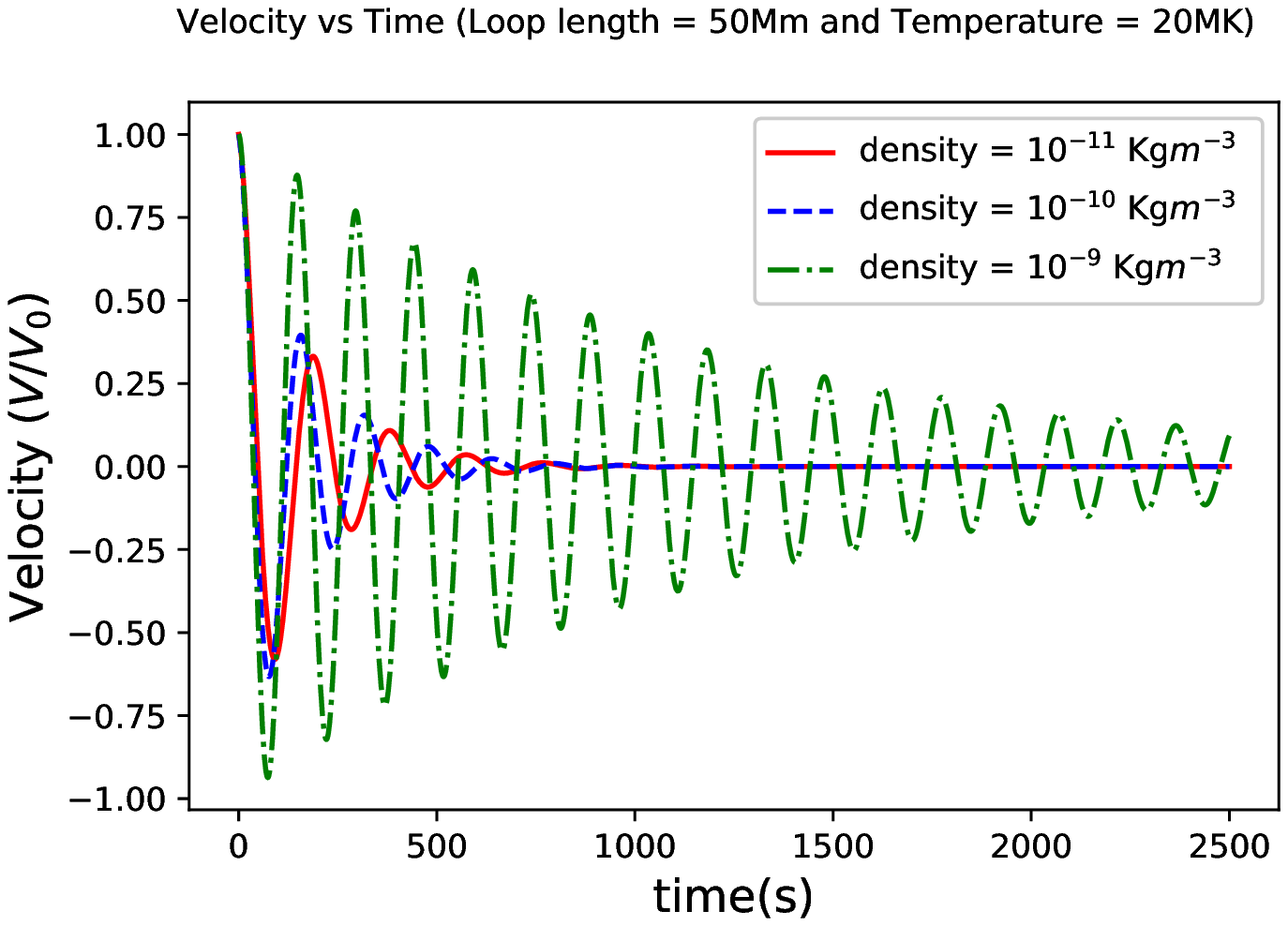}
               \hspace*{0.015\textwidth}
               \includegraphics[width=0.695\textwidth,clip=]{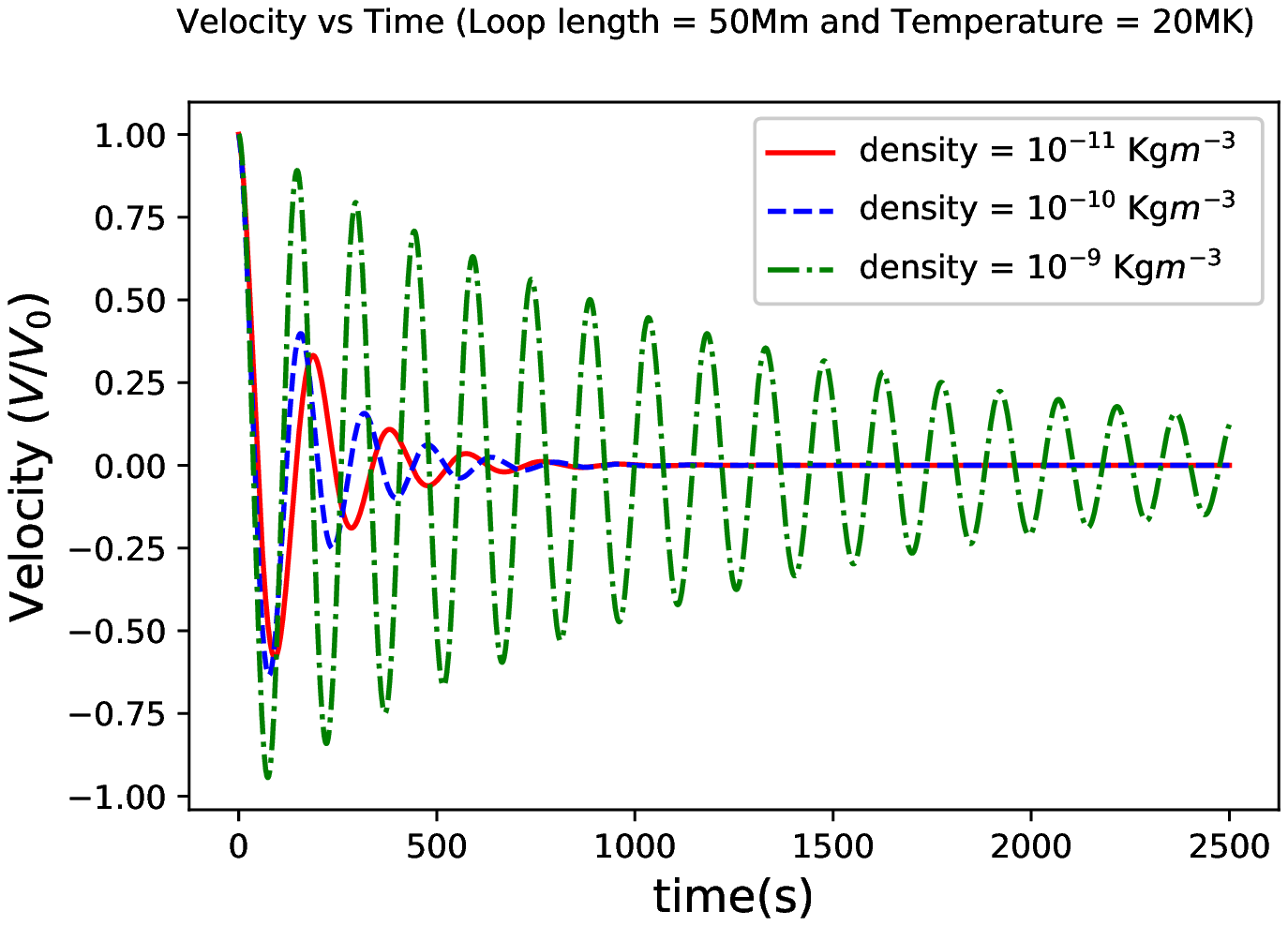}
              }
     \vspace{-0.35\textwidth}   
     \centerline{\Large \bf     
      \hspace{0.0 \textwidth} \color{white}{(c)}
      \hspace{0.415\textwidth}  \color{white}{(d)}
         \hfill}
     \vspace{0.31\textwidth}    
     
      \centerline{\hspace*{0.015\textwidth}
               \includegraphics[width=0.695\textwidth,clip=]{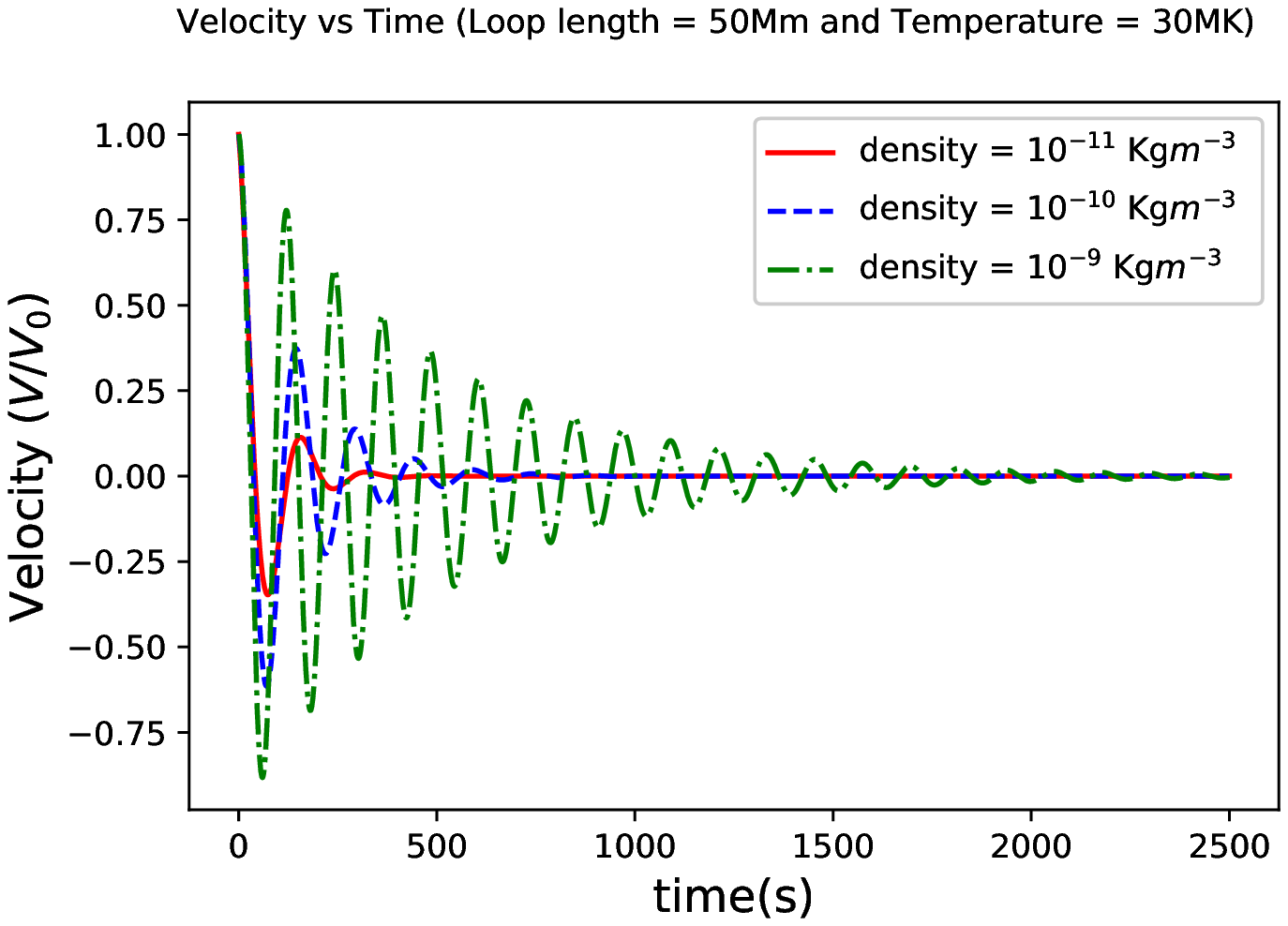}
               \hspace*{0.015\textwidth}
               \includegraphics[width=0.695\textwidth,clip=]{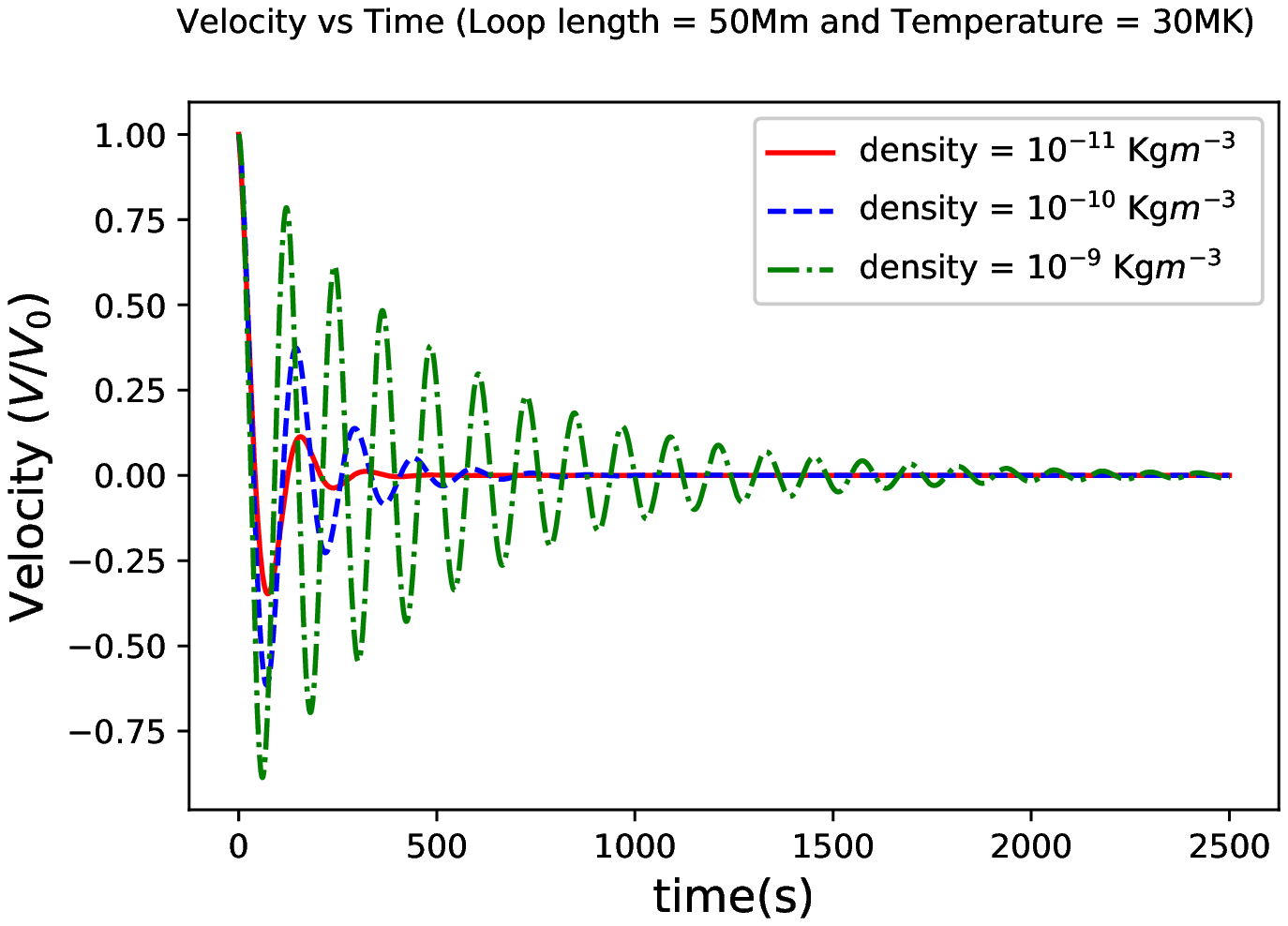}
              }
     \vspace{-0.35\textwidth}   
     \centerline{\Large \bf     
      \hspace{0.0 \textwidth} \color{white}{(c)}
      \hspace{0.415\textwidth}  \color{white}{(d)}
         \hfill}
     \vspace{0.31\textwidth}    
              
\caption{Left column: Temporal variation of the velocity amplitude $V$ of the fundamental mode of slow magnetoacoustic oscillations at $z=\frac{L}{2}$ for a loop of length $L$=50 Mm with densities $\rho$=10$^{-11}$ (red), 10$^{-10}$ (blue-dotted), 10$^{-9}$ (green-dotted) kg m$^{-3}$ at $T$=10 MK (top), 20 MK (middle), and 30 MK (bottom) when heating-cooling imbalance is present. 
Right column: Similar panels as shown in the left column but without heating-cooling imbalance.
        }
   \label{F-4panels}
   \end{figure}
   As we have previously detected a significant role of compressive viscosity and thermal conductivity jointly for the shorter loops  (e.g., $L$=50 Mm) with a typical coronal loop density at the super-hot regime of temperature ($T>$10 MK). We examine next the effect of density variations on the properties of the slow magnetoacoustic oscillations in a shorter loop ($L$=50 Mm) for different temperatures ranging between 10\,--\,30 MK. It should be noted that compressive viscosity along with thermal conductivity have significant effects on the damping of the fundamental mode for all considered temperatures ranging between 5\,--\,30 MK (Figure 1) compared to the case of thermal conductivity alone, however, the contribution of compressive viscosity becomes significant only at very high temperatures, e.g., 10\,--\,30 MK (Figure 1). We divide such loops into two categories based on their density values, namely, the normal density loops ($\rho$=10$^{-11}$ kg m$^{-3}$, and over-dense flaring loops ($\rho$=10$^{-10}$\,--\,10$^{-9}$ kg m$^{-3}$) \citep{2004psci.book.....A}, and analyze the variation of \(\omega_I\) with $K$ for the super-hot regime of $T\geq$10 MK. The results of this analysis are presented in Figure 8.
   
   In the top row of Figure 8, at $T$=10 MK and $\rho$=10$^{-11}$ kg m$^{-3}$ (normal dense loops), for the fundamental mode, the value of \(\omega_I\) represented by the red/blue-dotted curves drops compared to the same shown by the yellow/green dotted curves. As previously, it is inferred that when viscosity is added to thermal conductivity, it enhances the damping (cf. top-left panel in the first row of Figure 8). Since red and blue-dotted curves are coincident with each other this means that heating cooling imbalance does not have any significant role in enhancing the damping in the present case. A similar physical scenario also works for the higher order harmonics. However, when density is increased by an order of magnitude (mildly over-dense loops), i.e., 10$^{-10}$ kg m$^{-3}$ (cf. top-middle panel in the first row of Figure 8), for the fundamental mode, the red, yellow, green-dotted and blue-dotted curves are almost all coincident. This describes the fact that even at a higher temperature, if the density of the shorter loop is higher, the inclusion of compressive viscosity does not cause more damping than thermal conductivity alone (yellow/green dotted curves). Moreover, the effect of heating-cooling imbalance  on the damping in this condition is almost negligible. However, when examining the higher order harmonics (e.g. $K$=2.0), we find that viscosity still has some small effects in enhancing the damping along with thermal conductivity (red/blue-dotted curves still lower down) compared to thermal conductivity alone (yellow/green-dotted curves). When density is increased by one more order (over-dense loops), i.e. 10$^{-9}$ kg m$^{-3}$ (cf. top-right panel in the first row of Figure 8), then for the fundamental mode, the red and yellow (green-dotted and blue) curves coincide with each other and the values of \(\omega_I\) shown by the red/yellow curve drop compared to the ones represented by the green/blue-dotted curves. Now, the inclusion of compressive viscosity does not cause any appreciable enhancement of the damping compared to the one already caused by thermal conductivity. Therefore, thermal conductivity remains dominating the damping mechanism while viscosity does not play any significant role in the bulky loops (e.g. post flare loops) even if they are shorter and maintained at high temperature. Moreover, it is noticed that the heating-cooling imbalance enhances slightly more the damping compared to the case without its presence. For the higher order harmonics, the compressive viscosity still plays some roles in slightly enhancing the damping. 
 At temperatures $T$=20 MK (cf. middle row in Figure 8) and 30 MK (cf. bottom row in Figure 8), it is seen a similar trend as density increases, the effect of the compressive viscosity diminishes. However, it still remains effective for the second harmonics.
   In conclusion, even for the shorter and super hot loops, the damping effect of compressive viscosity may be weak or negligible if the loop density is high enough.
  
For post flare bulky loops, since the thermal ratio $d\gg1$ for super-hot and short loops, thermal conduction damping is weak but with an increase in density, $d$ decreases since $d \propto 1/\rho_0$ and comes close to the value of the peak damping rate \citep{2003A&A...408..755D} causing an increase in the damping. When viscosity is added, which is a dominant damping mechanism in the regime, the increase in density reduces the viscous ratio $(\epsilon \propto \frac{1}{\rho_0})$ thus reducing the damping effect due to viscosity. Note that the scales for Y-axis between the left and right panels are distinctly different in Figure 8.\newline
In contrast,both the thermal ratio ($d$) and viscous ratio ($\epsilon$) are $\ll1$ for hot (5\,--\,10 MK) and longer loops (500 Mm) and thermal conduction is a dominant damping mechanism while the damping due to viscosity is weak. As the value of density is increased, both $\epsilon$ and $d$ decrease even further and thus the damping rate due to thermal conductivity is reduced for over dense loops (cf. Figure 6) while the addition of viscosity does not make significant difference in the overall damping for normal as well as for post flare bulky loops. 
   
Figure 9 shows the temporal variation of the velocity $V$ of the fundamental mode for a loop of length $L$=50 Mm with densities $\rho$=10$^{-11}$ (red), 10$^{-10}$ (blue-dotted), 10$^{-9}$ (green-dotted) kg m$^{-3}$ at different temperatures with and without heating-cooling imbalance. It is seen that left and right columns are almost similar, implying that in the super-hot regime the damping of the velocity amplitude oscillations is less influenced by the heating-cooling imbalance. In the left (or right) column, when we see from the top-panel (at 10 MK) to the bottom-panel (at 30 MK), we notice that the velocity amplitude oscillations are largely damped in the loop of density  10$^{-11}$ kg m$^{-3}$ (red curve) under the inclusion of compressive viscosity together with the thermal conductivity, while it is less damped in the loops of densities 10$^{-10}$ and 10$^{-9}$ kg m$^{-3}$ (blue-dotted and green-dotted curves). The damping rate of the velocity oscillations in normal, mild, and over-dense loops increases with the increment in temperature which can be seen clearly from these plots.

\subsection{Comparison of the Role of Thermal Conductivity and Viscosity in Damping of Slow Waves in Coronal Loops with Heating-Cooling Imbalance}

\begin{figure}    
   \centerline{\hspace*{0.015\textwidth}
               \includegraphics[width=0.695\textwidth,clip=]{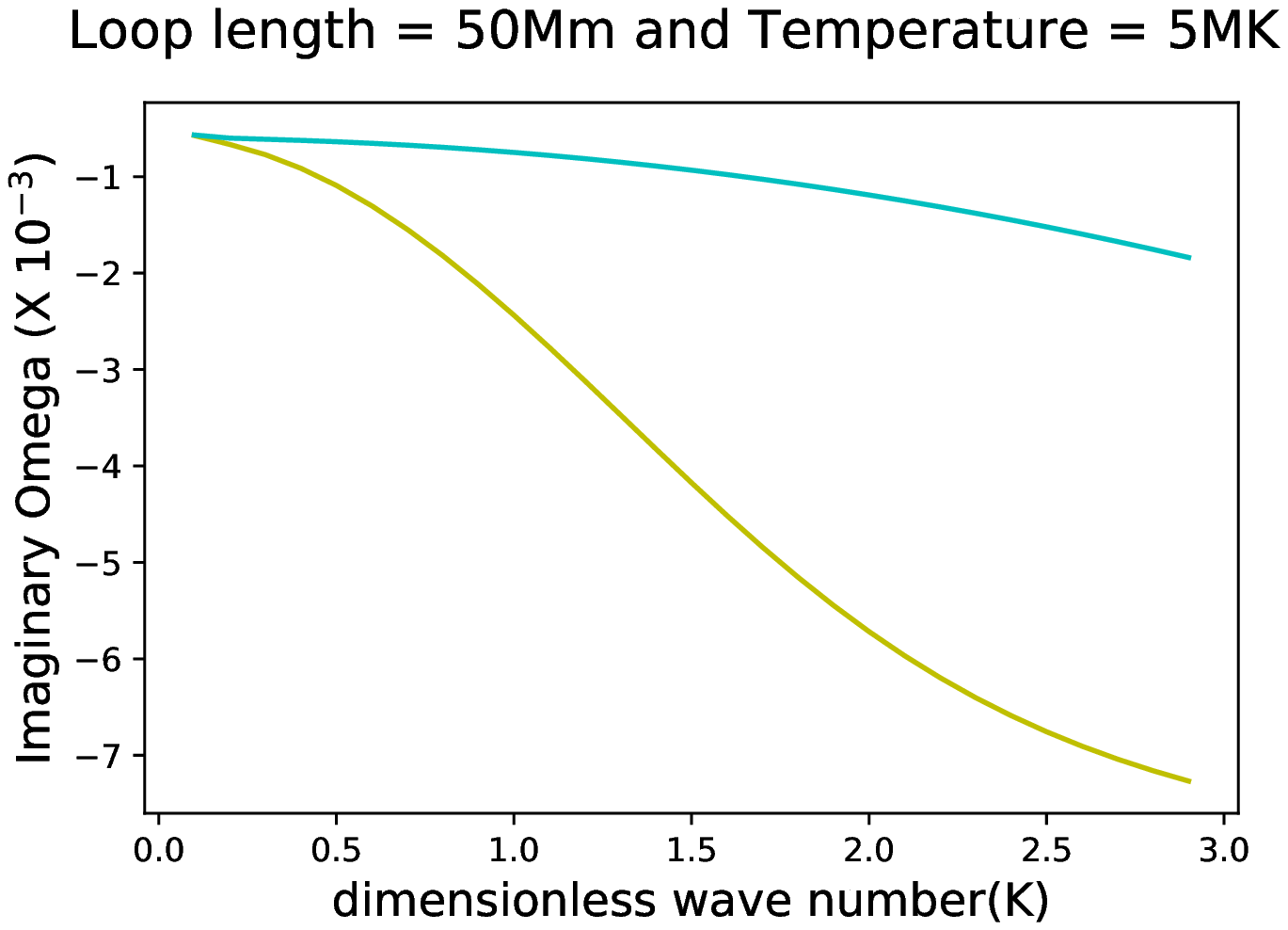}
               \hspace*{0.015\textwidth}
               \includegraphics[width=0.695\textwidth,clip=]{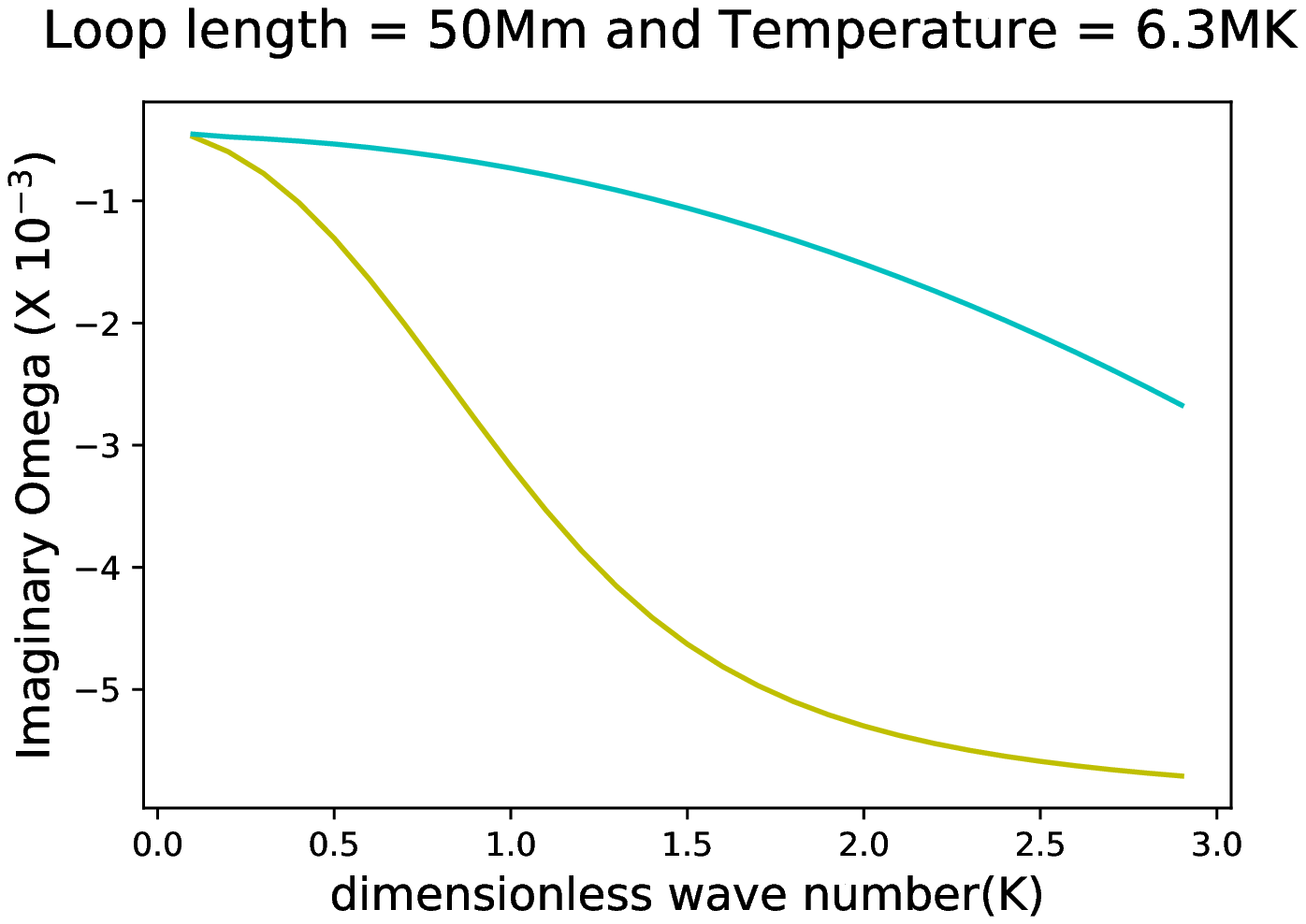}
              }
     \vspace{-0.35\textwidth}   
     \centerline{\Large \bf     
      \hspace{0.0 \textwidth}  \color{white}{(a)}
      \hspace{0.415\textwidth}  \color{white}{(b)}
         \hfill}
     \vspace{0.31\textwidth}    
   \centerline{\hspace*{0.015\textwidth}
               \includegraphics[width=0.695\textwidth,clip=]{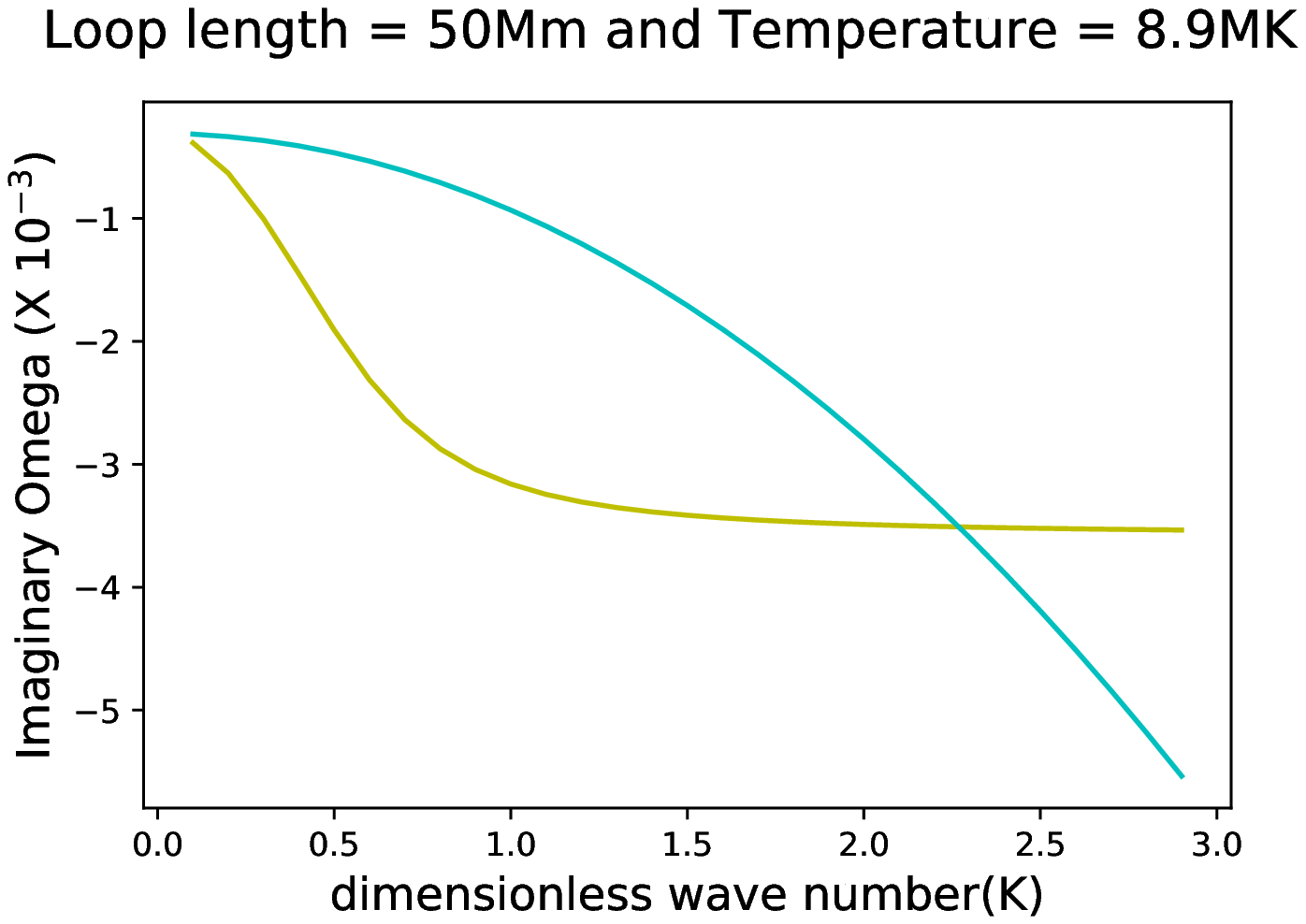}
               \hspace*{0.015\textwidth}
               \includegraphics[width=0.695\textwidth,clip=]{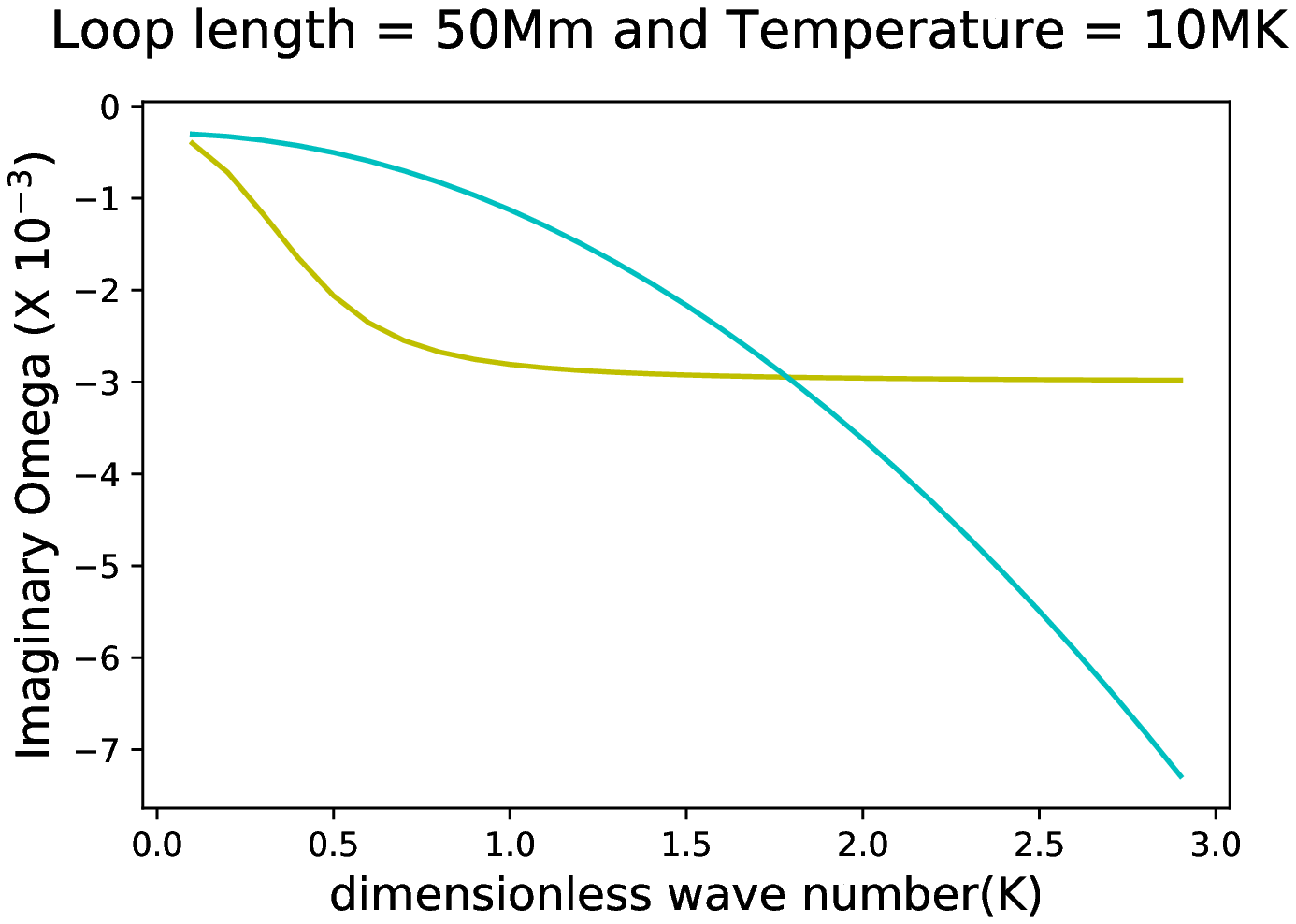}
              }
     \vspace{-0.35\textwidth}   
     \centerline{\Large \bf     
      \hspace{0.0 \textwidth} \color{white}{(c)}
      \hspace{0.415\textwidth}  \color{white}{(d)}
         \hfill}
     \vspace{0.31\textwidth}    
     
      \centerline{\hspace*{0.015\textwidth}
               \includegraphics[width=0.695\textwidth,clip=]{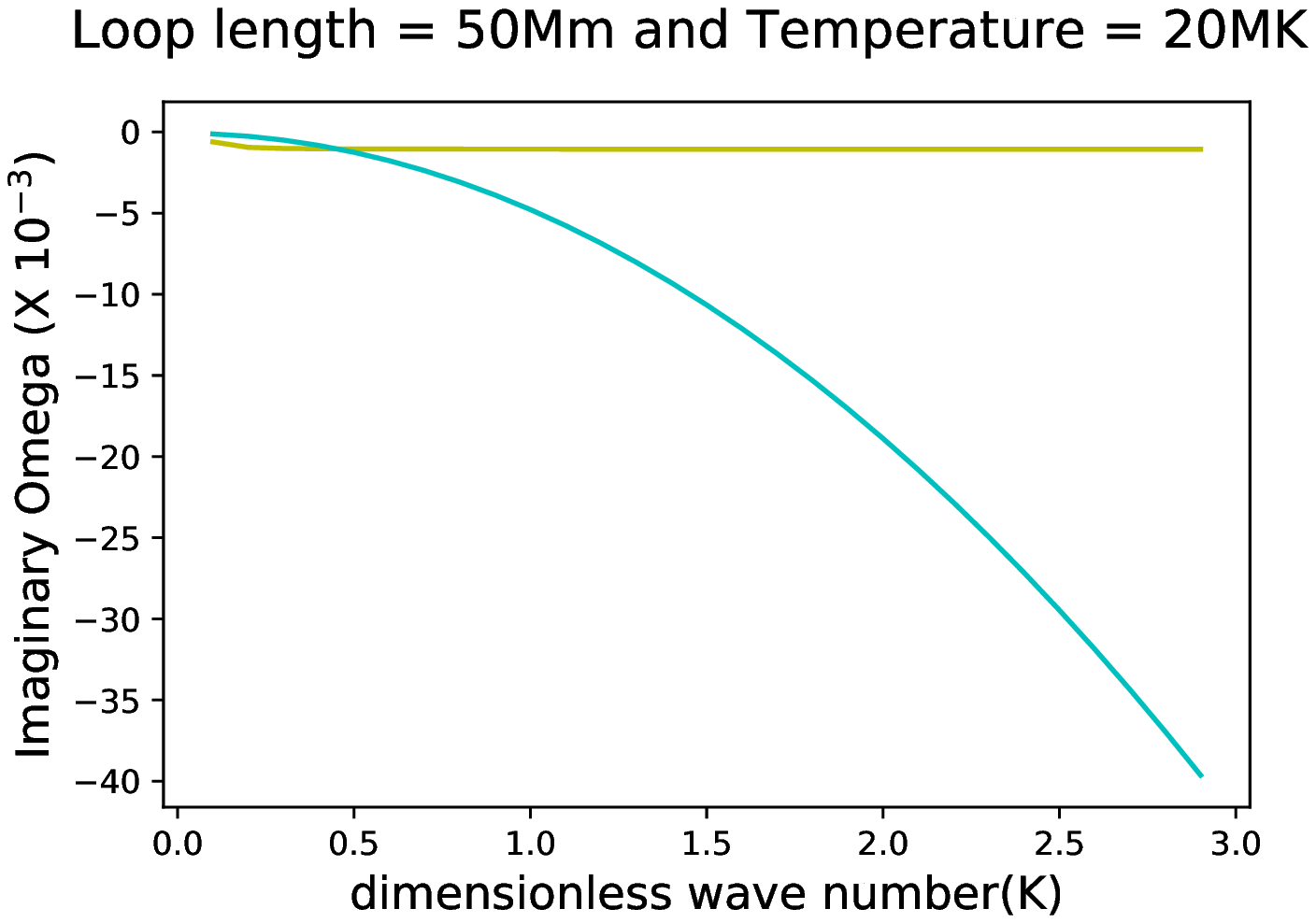}
               \hspace*{0.015\textwidth}
               \includegraphics[width=0.695\textwidth,clip=]{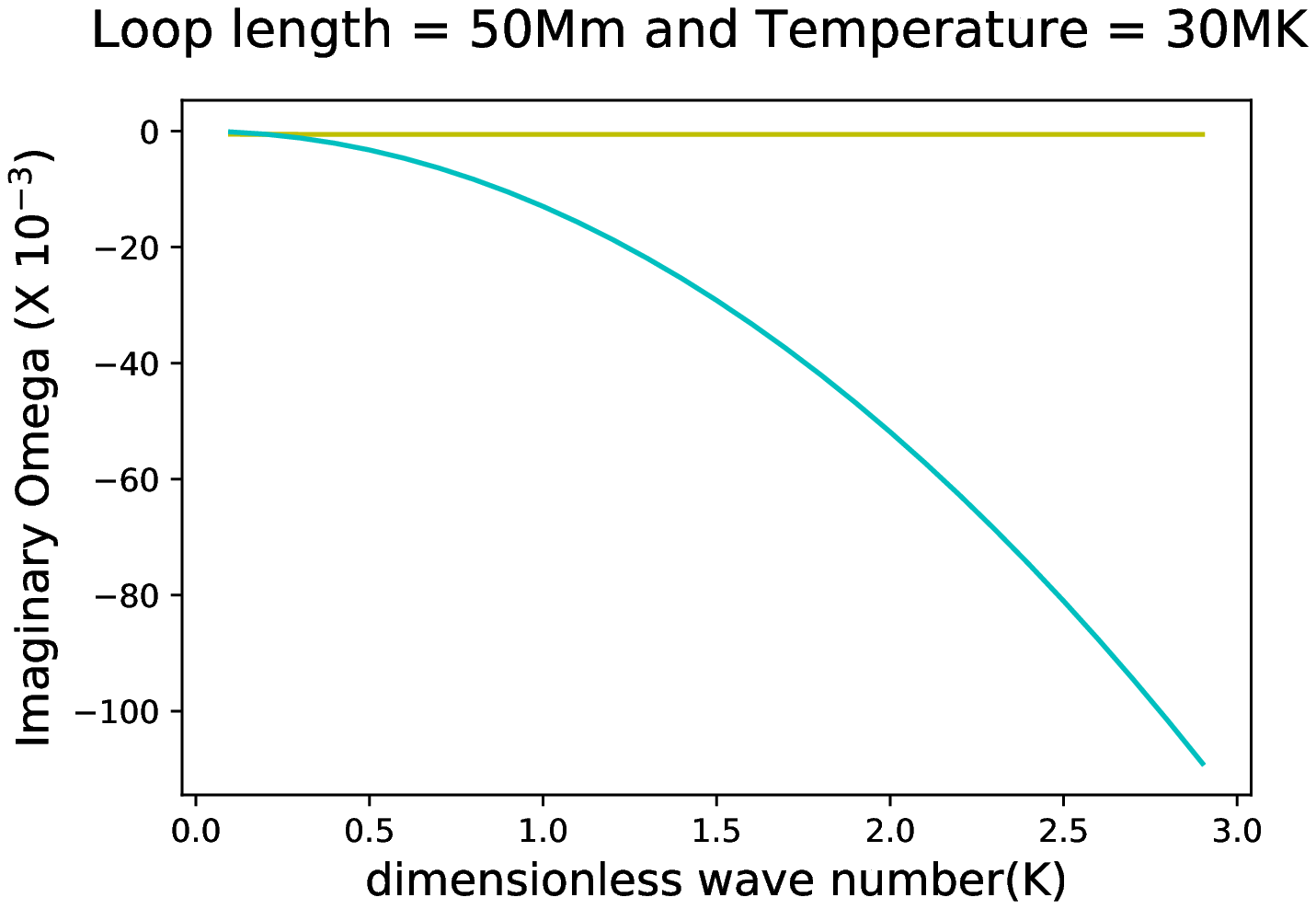}
              }
     \vspace{-0.35\textwidth}   
     \centerline{\Large \bf     
      \hspace{0.0 \textwidth} \color{white}{(c)}
      \hspace{0.415\textwidth}  \color{white}{(d)}
         \hfill}
     \vspace{0.31\textwidth}    
              
\caption{The panels show the variation of \(\omega_I\) with  dimensionless wave number $K$ at a fixed loop-length of 50 Mm for different temperatures ranging from 5 MK to 30 MK. In each panel, the cyan and yellow curves correspond to the solution of the dispersion relation for compressive viscosity and thermal conductivity respectively when heating-cooling imbalance is present.
        }
   \label{F-4panels}
   \end{figure}
\begin{figure}    
   \centerline{\hspace*{0.015\textwidth}
               \includegraphics[width=0.695\textwidth,clip=]{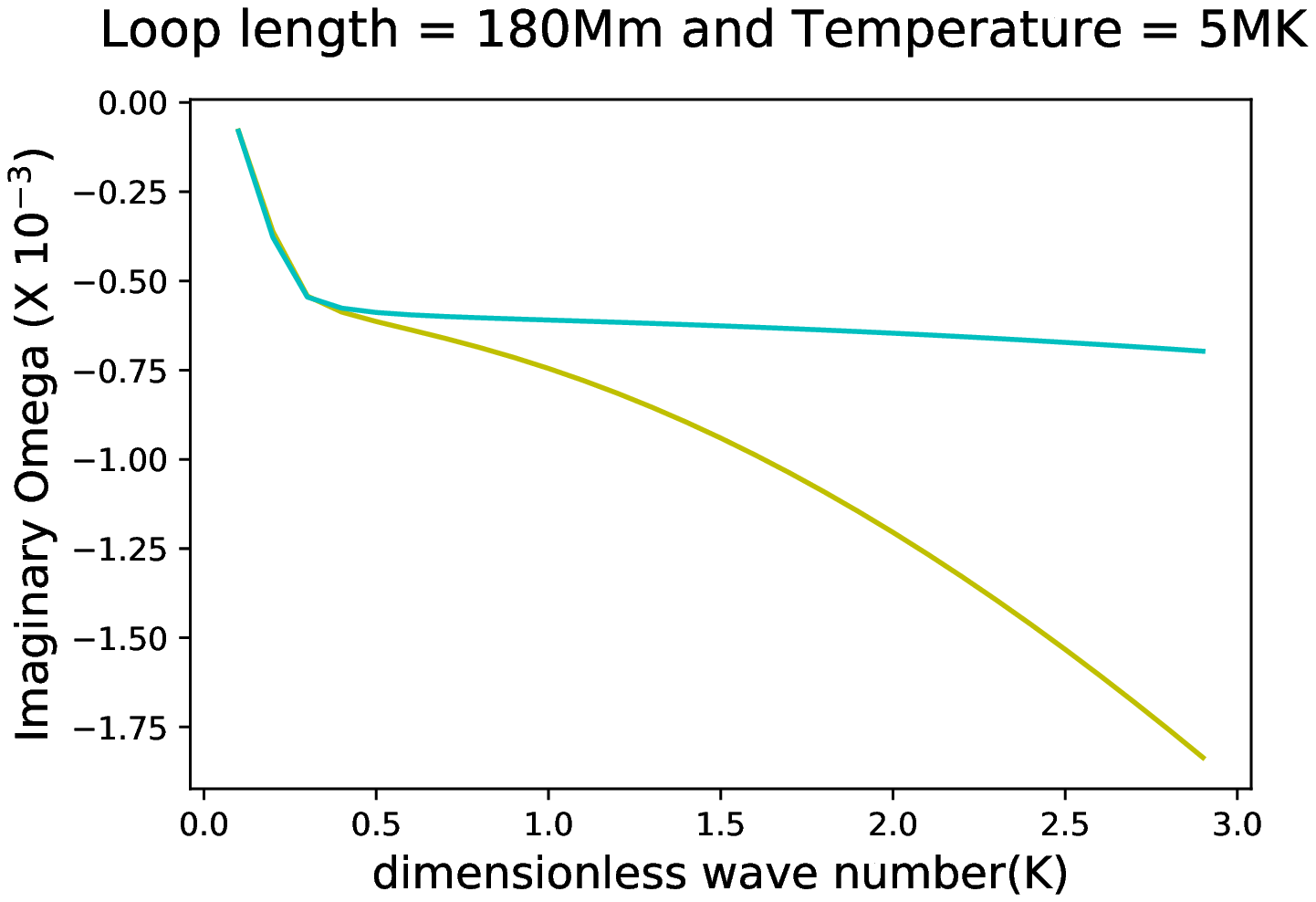}
               \hspace*{0.015\textwidth}
               \includegraphics[width=0.695\textwidth,clip=]{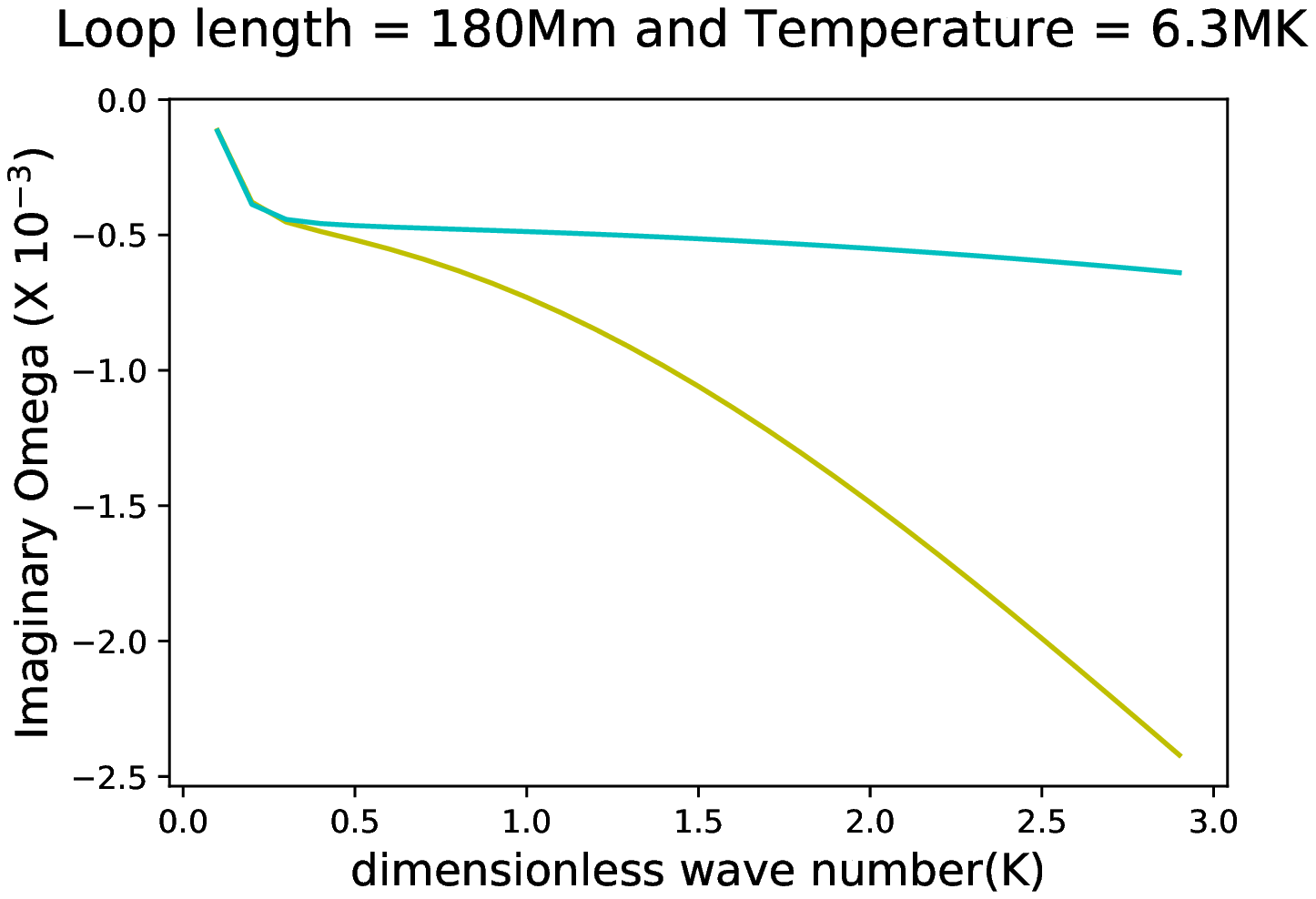}
              }
     \vspace{-0.35\textwidth}   
     \centerline{\Large \bf     
      \hspace{0.0 \textwidth}  \color{white}{(a)}
      \hspace{0.415\textwidth}  \color{white}{(b)}
         \hfill}
     \vspace{0.31\textwidth}    
   \centerline{\hspace*{0.015\textwidth}
               \includegraphics[width=0.695\textwidth,clip=]{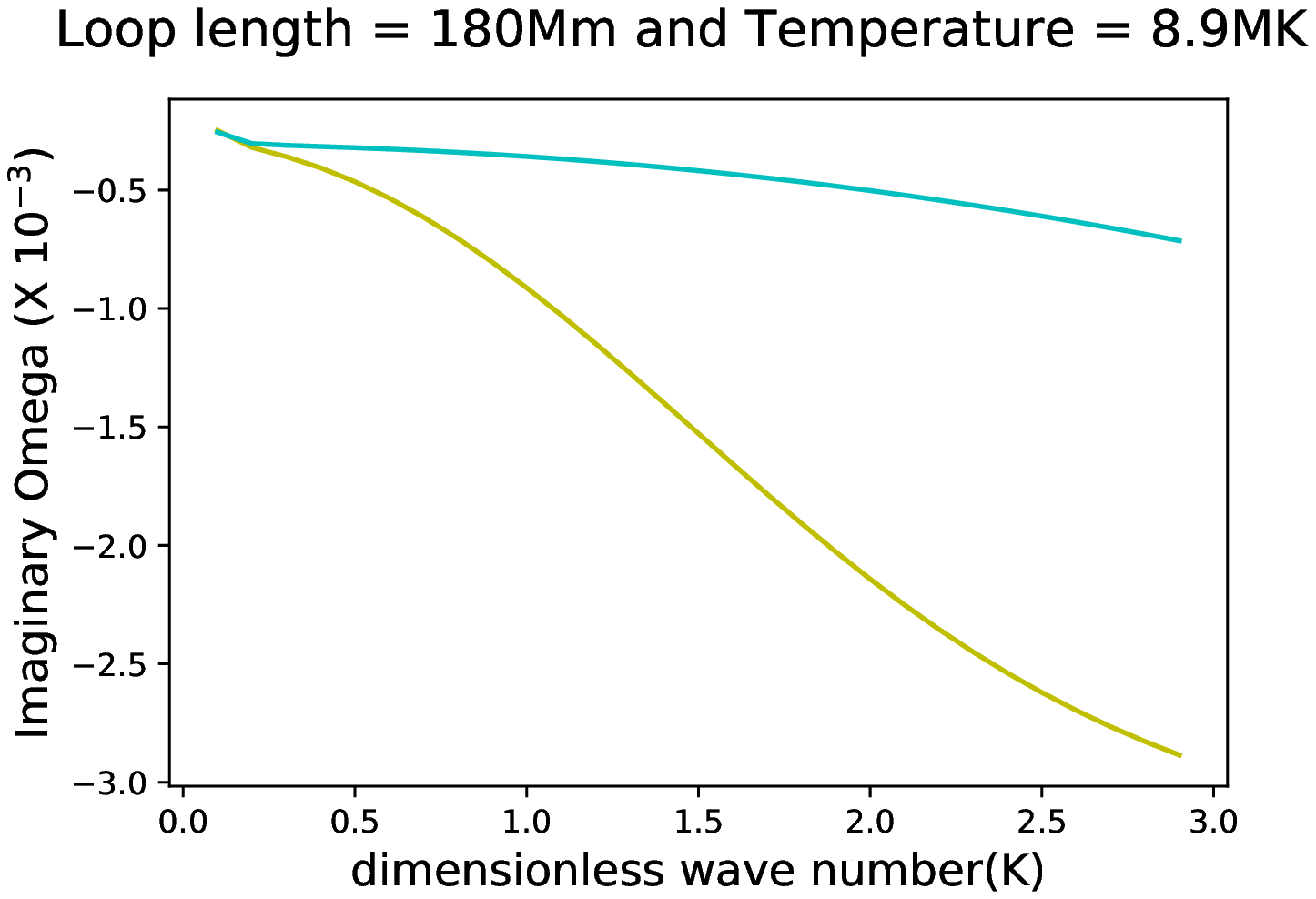}
               \hspace*{0.015\textwidth}
               \includegraphics[width=0.695\textwidth,clip=]{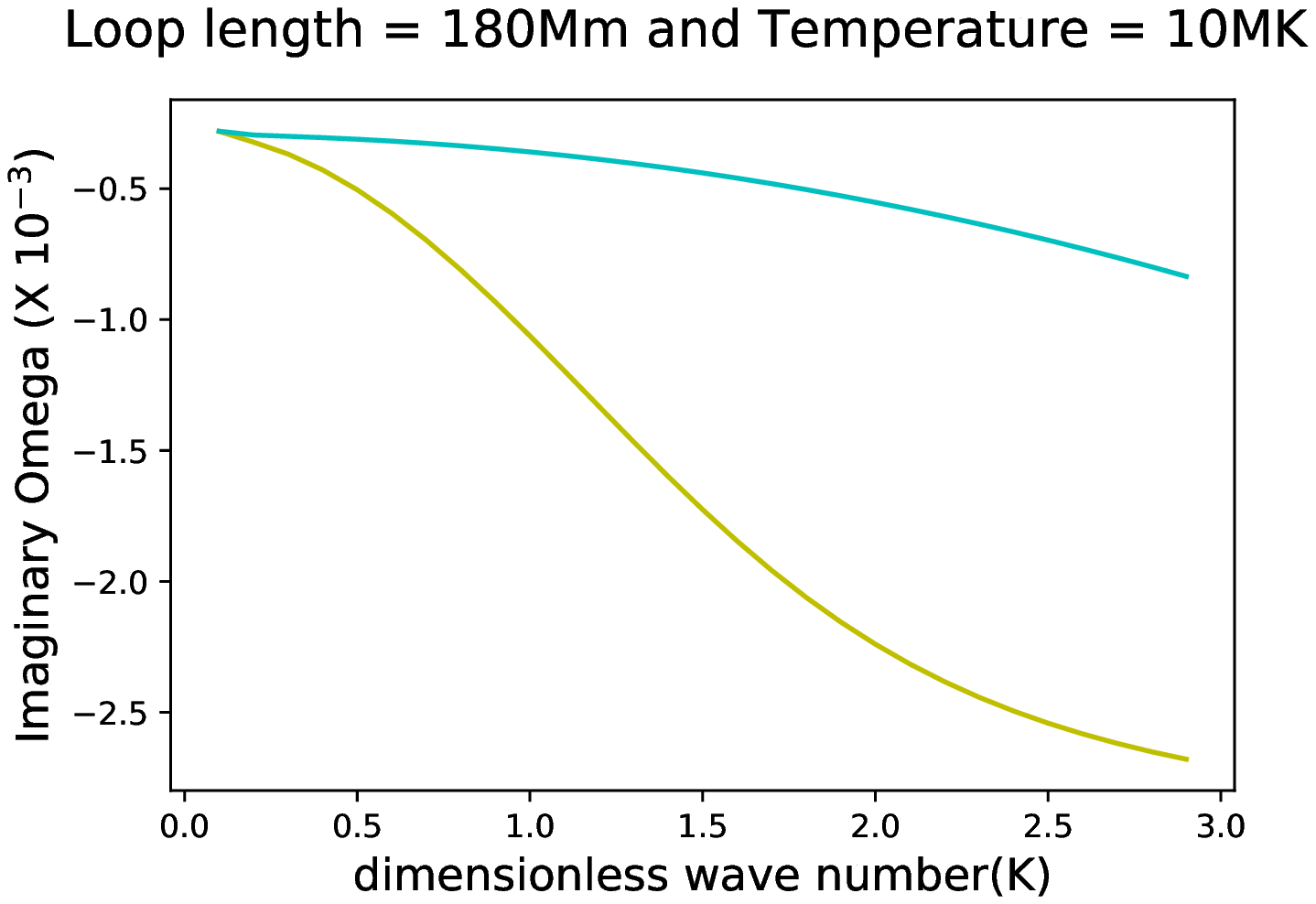}
              }
     \vspace{-0.35\textwidth}   
     \centerline{\Large \bf     
      \hspace{0.0 \textwidth} \color{white}{(c)}
      \hspace{0.415\textwidth}  \color{white}{(d)}
         \hfill}
     \vspace{0.31\textwidth}    
     
      \centerline{\hspace*{0.015\textwidth}
               \includegraphics[width=0.695\textwidth,clip=]{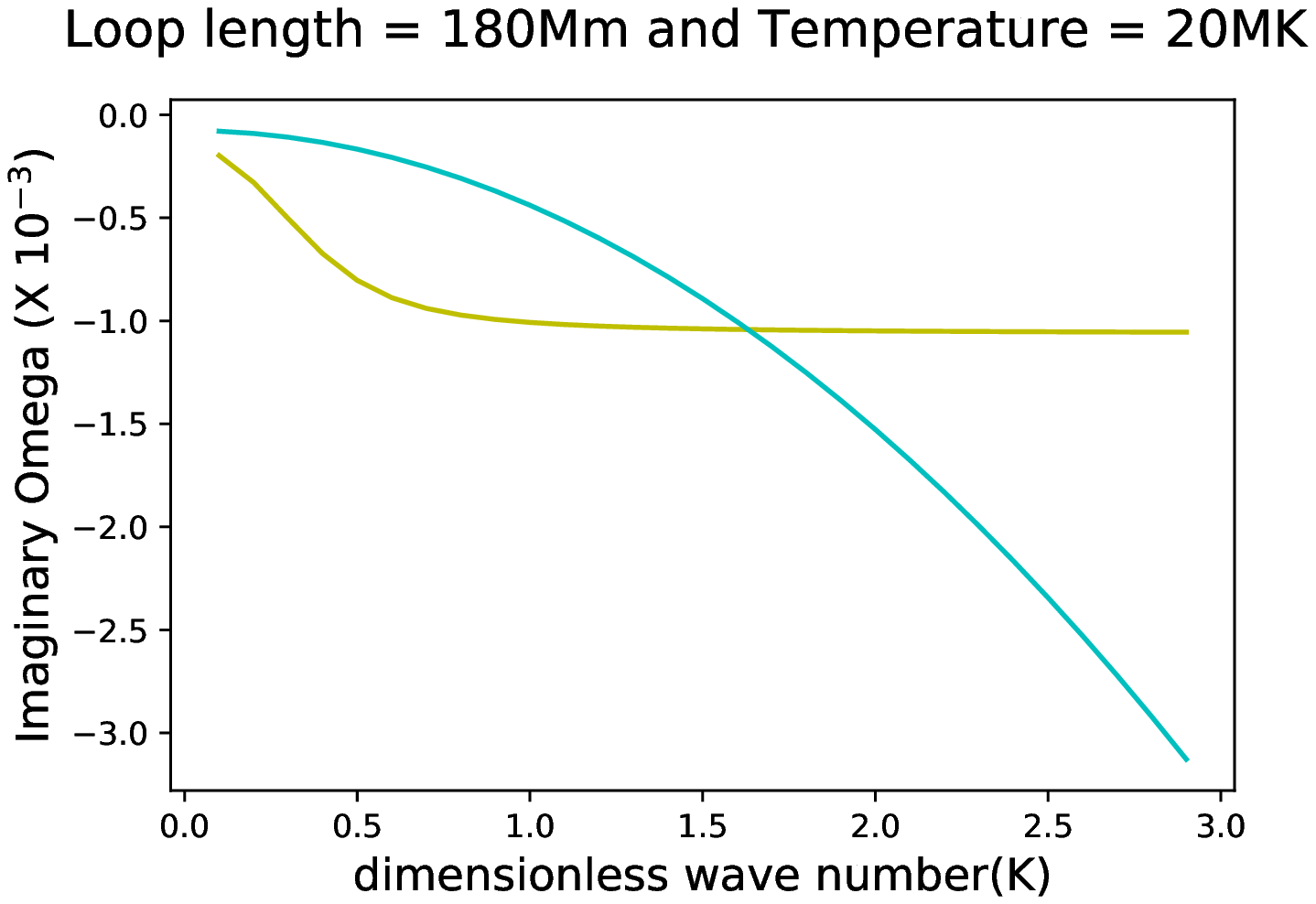}
               \hspace*{0.015\textwidth}
               \includegraphics[width=0.695\textwidth,clip=]{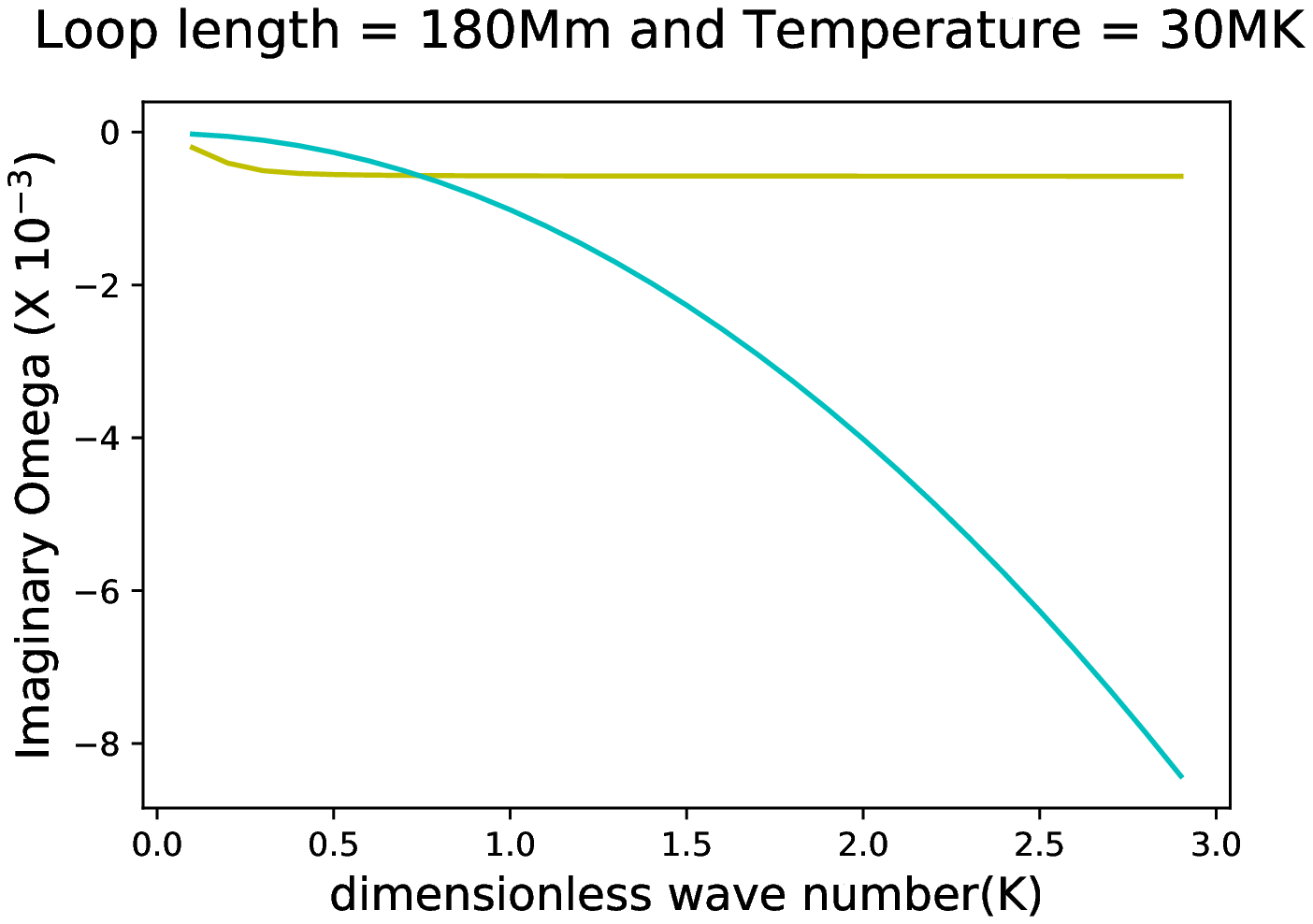}
              }
     \vspace{-0.35\textwidth}   
     \centerline{\Large \bf     
      \hspace{0.0 \textwidth} \color{white}{(c)}
      \hspace{0.415\textwidth}  \color{white}{(d)}
         \hfill}
     \vspace{0.31\textwidth}    
              
\caption{The panels show the variation of \(\omega_I\) with  dimensionless wave number $K$ at a fixed loop-length of 180 Mm for different temperatures ranging from 5 MK to 30 MK. In each panel, the cyan and yellow curves correspond to the solution of the dispersion relation for compressive viscosity and thermal conductivity respectively when heating-cooling imbalance is present.
        }
   \label{F-4panels}
   \end{figure}
\begin{figure}    
   \centerline{\hspace*{0.015\textwidth}
               \includegraphics[width=0.695\textwidth,clip=]{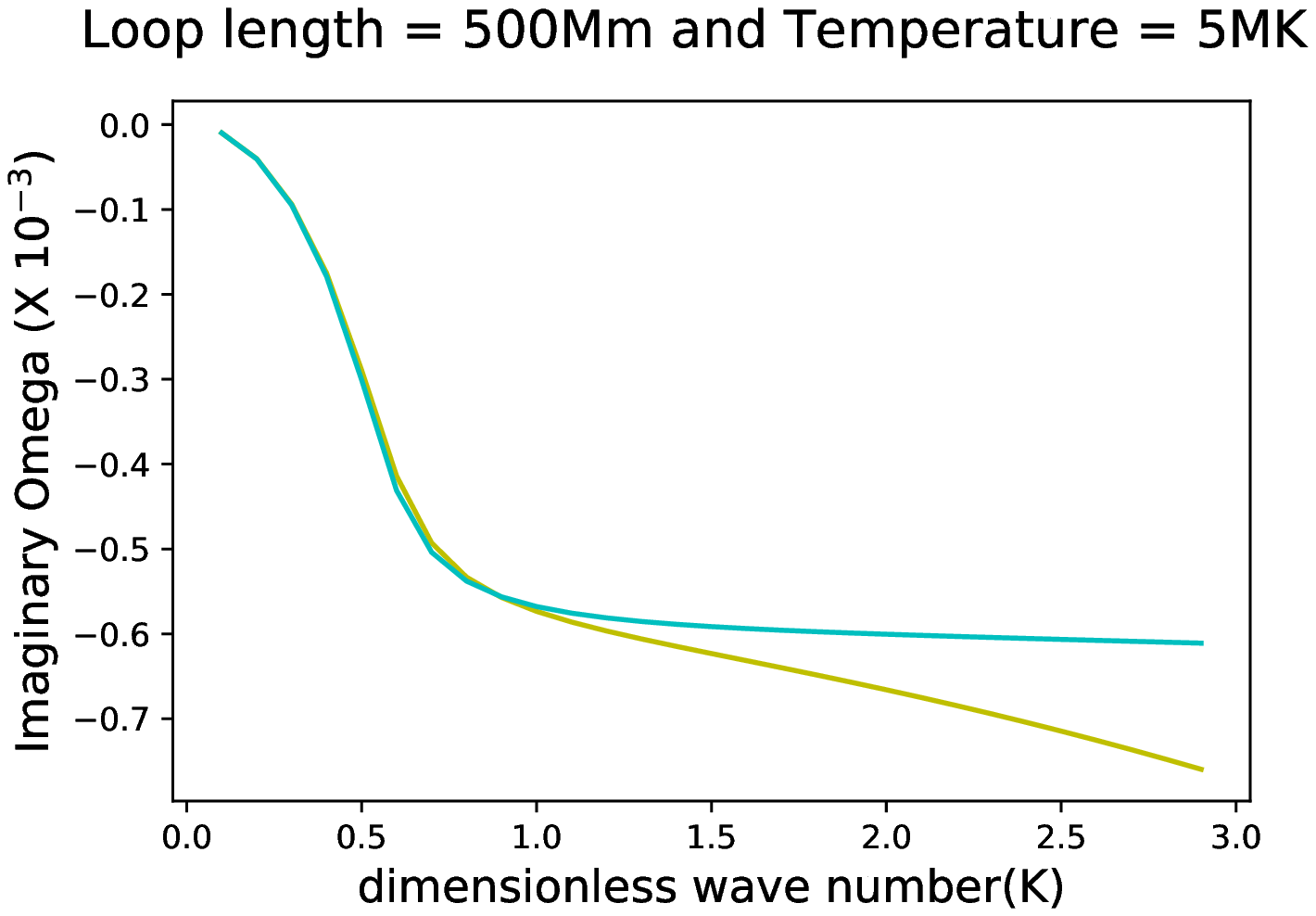}
               \hspace*{0.015\textwidth}
               \includegraphics[width=0.695\textwidth,clip=]{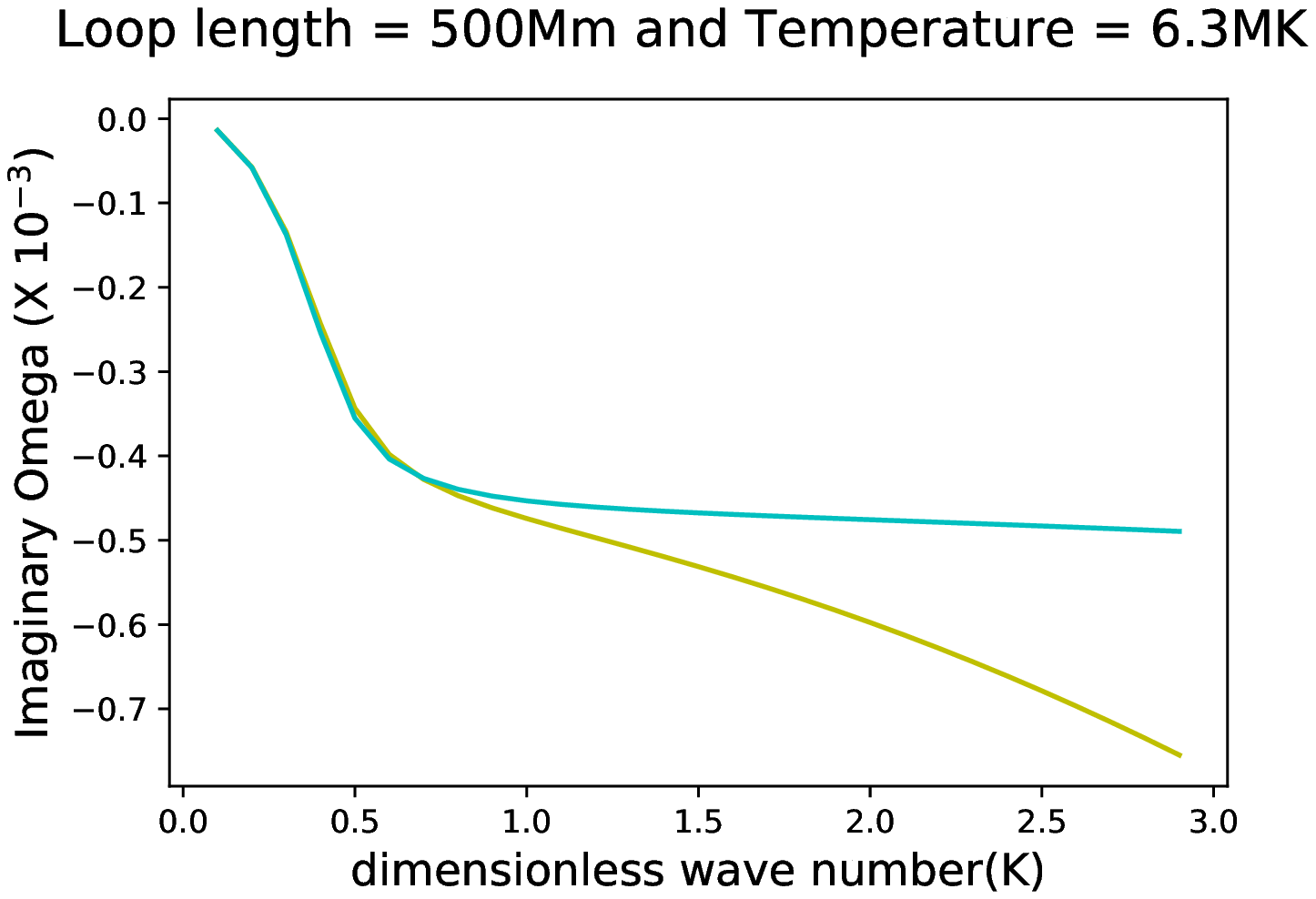}
              }
     \vspace{-0.35\textwidth}   
     \centerline{\Large \bf     
      \hspace{0.0 \textwidth}  \color{white}{(a)}
      \hspace{0.415\textwidth}  \color{white}{(b)}
         \hfill}
     \vspace{0.31\textwidth}    
   \centerline{\hspace*{0.015\textwidth}
               \includegraphics[width=0.695\textwidth,clip=]{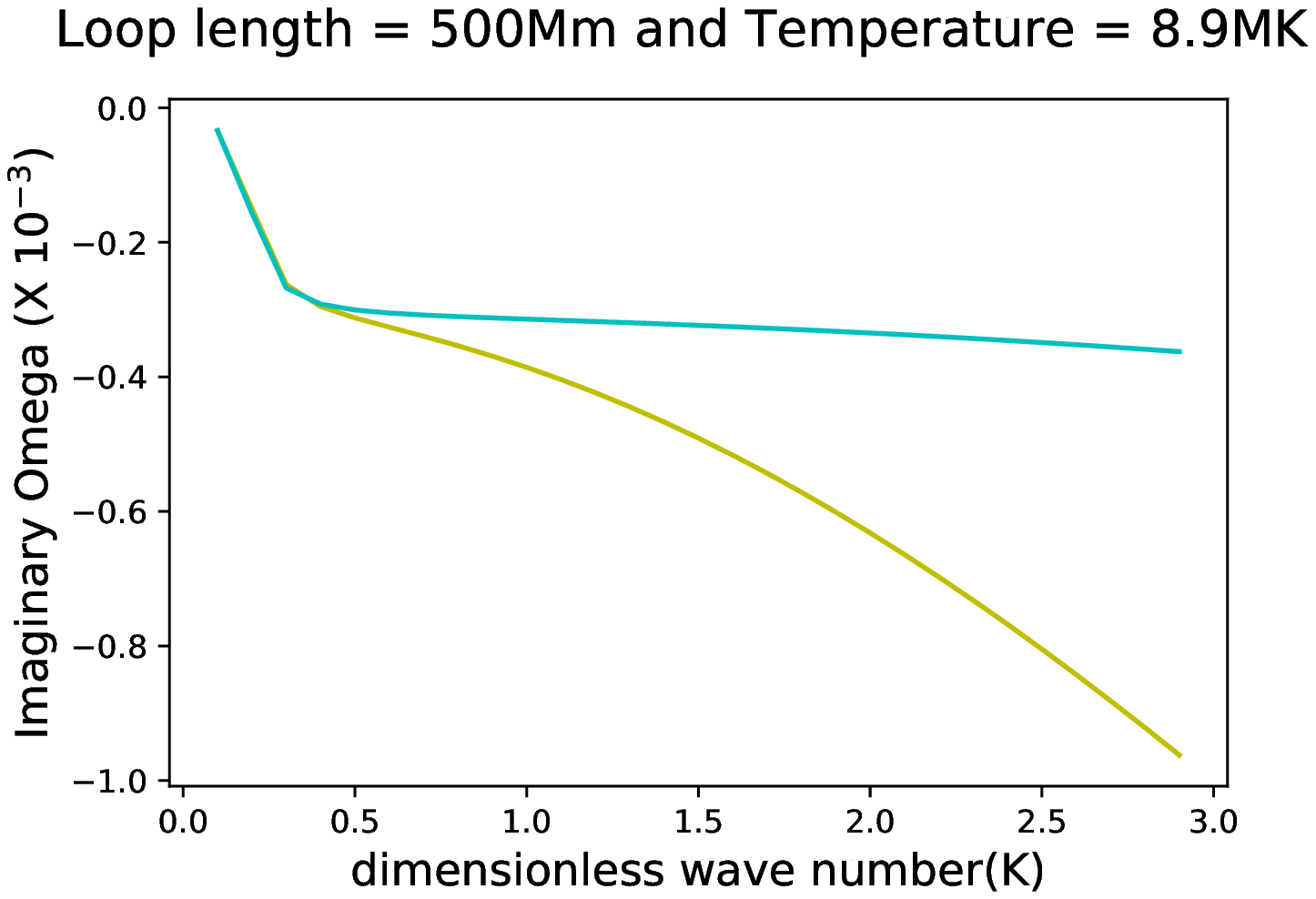}
               \hspace*{0.015\textwidth}
               \includegraphics[width=0.695\textwidth,clip=]{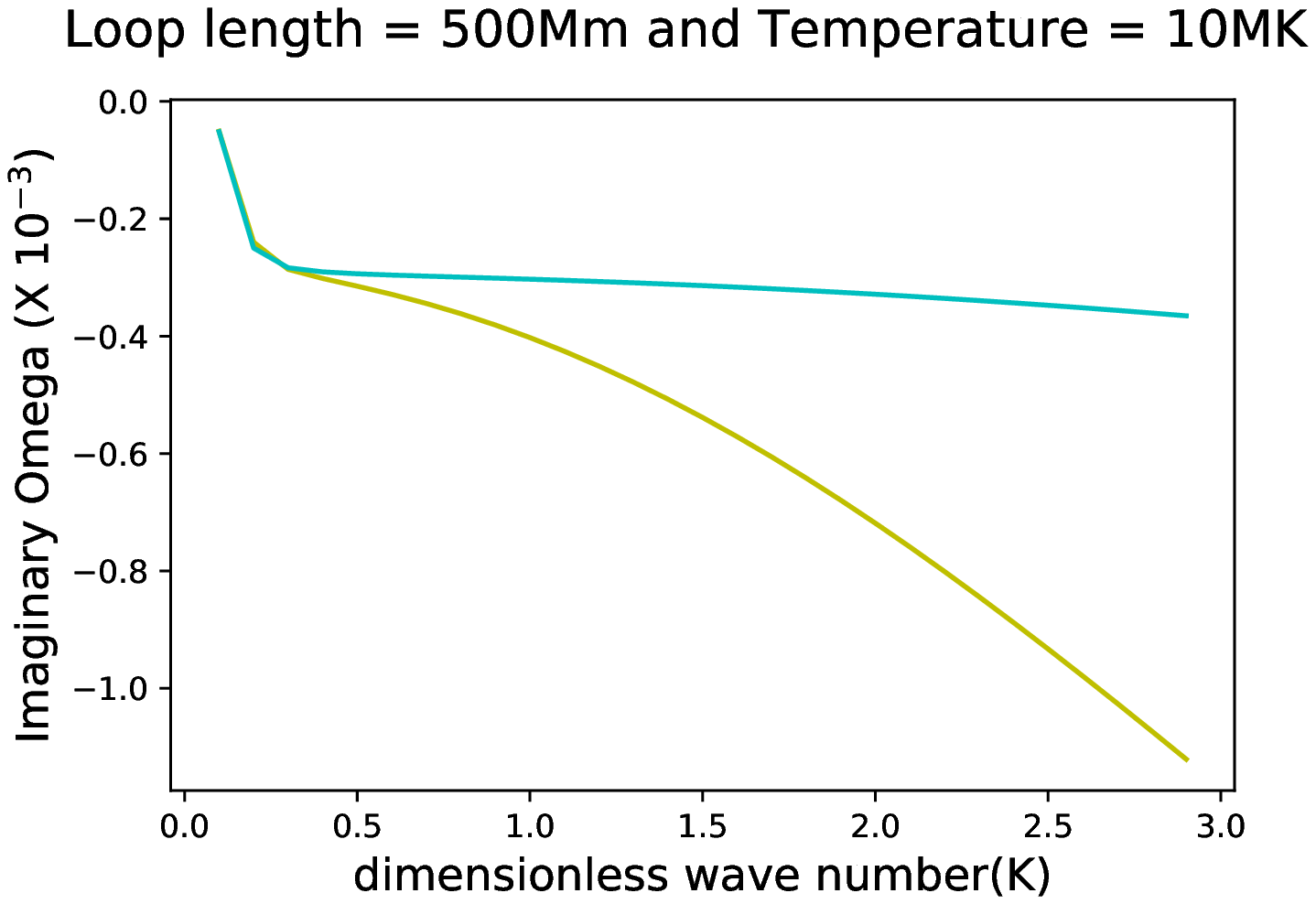}
              }
     \vspace{-0.35\textwidth}   
     \centerline{\Large \bf     
      \hspace{0.0 \textwidth} \color{white}{(c)}
      \hspace{0.415\textwidth}  \color{white}{(d)}
         \hfill}
     \vspace{0.31\textwidth}    
     
      \centerline{\hspace*{0.015\textwidth}
               \includegraphics[width=0.695\textwidth,clip=]{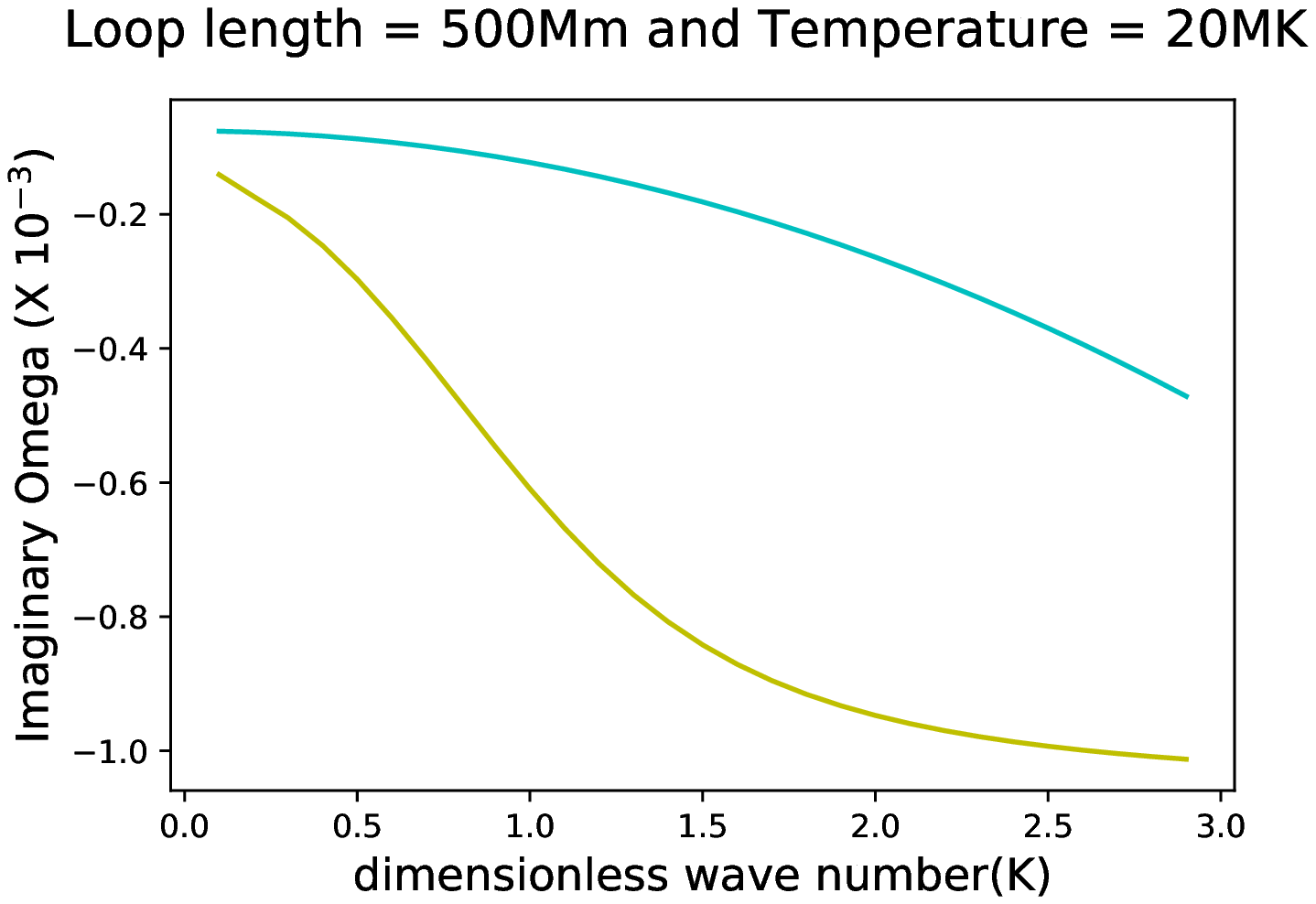}
               \hspace*{0.015\textwidth}
               \includegraphics[width=0.695\textwidth,clip=]{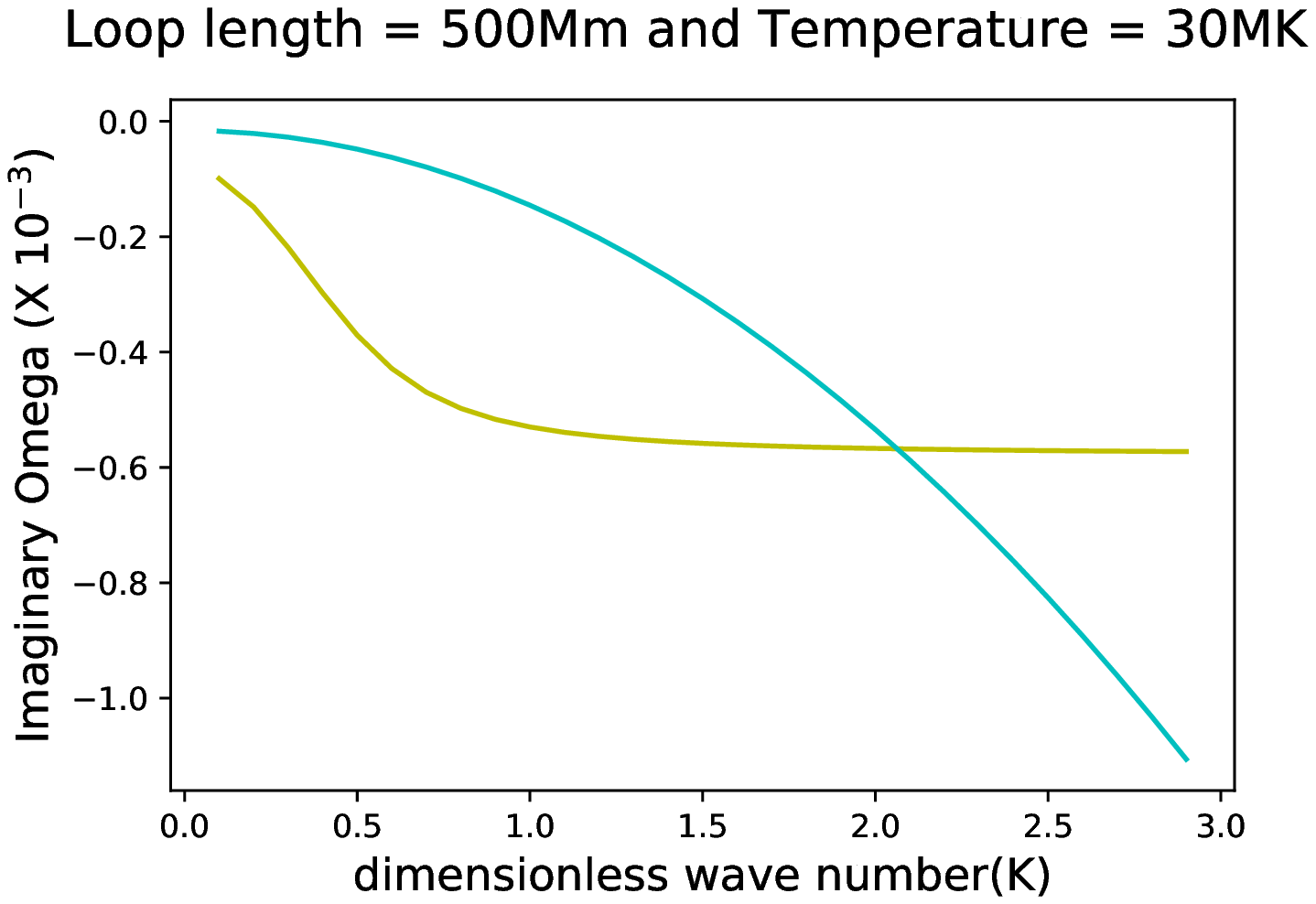}
              }
     \vspace{-0.35\textwidth}   
     \centerline{\Large \bf     
      \hspace{0.0 \textwidth} \color{white}{(c)}
      \hspace{0.415\textwidth}  \color{white}{(d)}
         \hfill}
     \vspace{0.31\textwidth}    
              
\caption{The panels show the variation of \(\omega_I\) with  dimensionless wave number $K$ at a fixed loop-length of 500 Mm for different temperatures ranging from 5 MK to 30 MK. In each panel, the cyan and yellow curves correspond to the solution of the dispersion relation for compressive viscosity and thermal conductivity respectively when heating-cooling imbalance is present.
        }
   \label{F-4panels}
   \end{figure}
   
   In the previous sections, we have performed comprehensive analyses on the damping of slow magnetoacoustic oscillations. We find that at a hot regime of the temperature ($T\leq$10 MK), thermal conductivity along with heating-cooling imbalance plays a significant role in enhancing the damping of the fundamental mode. While, in the super-hot regime of temperature ($T>$10 MK), the inclusion of compressive viscosity along with thermal conductivity causes an enhanced wave damping. 
   Therefore, in the present section, we estimate and analyze the individual roles of compressive viscosity and thermal conductivity in the wave damping under the effect of heating-cooling imbalance.
   
In Figure 10, we find that the damping of the fundamental mode is much higher due to thermal conductivity (yellow) compared to compressive viscosity (cyan) in the hot regime $T$=5\,--\,10 MK in the shortest considered loop of length 50 Mm.  However, their role is getting reversed in the super-hot regime $T>$10 MK as the compressive viscosity dominates the damping of the fundamental mode at 20 and 30 MK temperatures. This result also supports the conclusions of \citet{2004ApJ...605..493M}, \citet{2007SoPh..246..187S}, and \citet{2012SoPh..280..137A}. A similar physical scenario is also valid for the higher order harmonics with $K\geq$2.0. In Figure 11, for a loop of 180 Mm length, thermal conductivity dominates over the temperature range of 5\,--\,20 MK for the dissipation of the fundamental mode oscillations. This result also supports the conclusion derived in \citet{2002ApJ...580L..85O} that the thermal conductivity is a dominant damping mechanism in typical hot coronal loops. The compressive viscosity begins to play some roles in the damping at 30 MK, and its effect is slightly stronger than that of thermal conduction.
At $T>$20 MK, the damping of the higher harmonics is dominated by compressive viscosity, while at $T<$20 MK, it is dominated by thermal conduction. In Figure 12, for the longest loop of 500 Mm length, thermal conductivity at all temperatures dominates over compressive viscosity in the damping of the fundamental mode.
   
 \subsection{Comparison of Various Scaling Laws between $\tau$ and $P$}

 \begin{figure}[ht]
    \includegraphics[width = 12.0cm, angle=0]{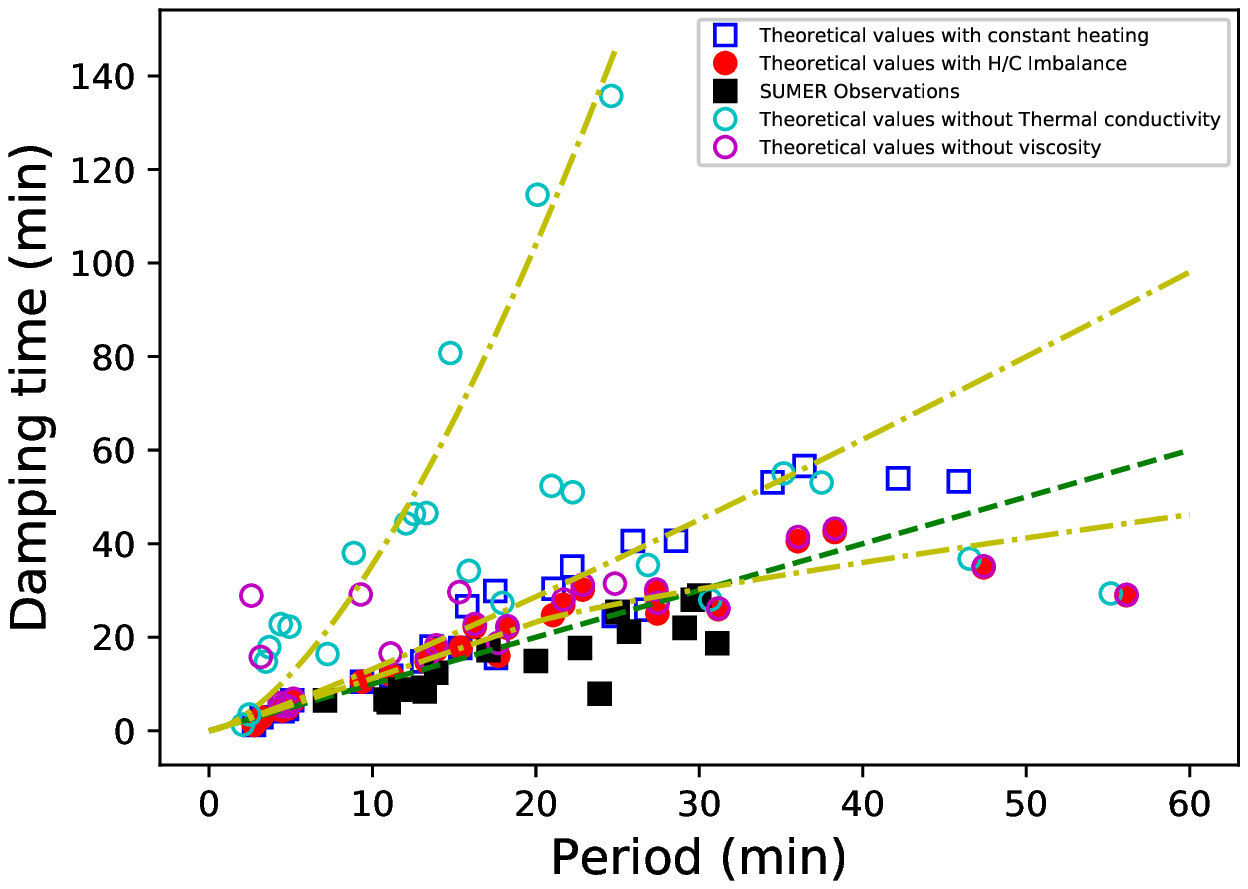}
    \caption{Variation of damping time ($\tau$) vs period ($P$): (i) filled-red circles (with thermal conductivity plus compressive viscosity plus heating-cooling imbalance); (ii) cyan circles (only with compressive viscosity plus heating-cooling imbalance; (iii) pink circles (only with thermal conductivity plus heating-cooling imbalance (iv) blue rectangles (without imbalance plus compressive viscosity plus thermal conductivity; (v) filled-black rectangles (observed SUMER oscillations). Various dashed-dot yellow lines show the fittings on  (i)-(iii) theoretical data points. The dark green-dashed line is $\tau$=$P$ line.
    }
    \label{fig:PropProf}
\end{figure}    
Figure 13 displays the damping time vs period for all kinds of theoretically estimations. Filled red circles are the measurements related to the theoretically estimated $\tau$ and $P$ for the damped fundamental mode when we consider thermal conductivity, compressive viscosity, and heating-cooling imbalance. Cyan circles are the theoretical data points when we consider only compressive viscosity and heating-cooling imbalance. While, pink circles are the data points when we consider only thermal conductivity and heating-cooling imbalance.  Blue rectangles are the points under the effect of compressive viscosity and thermal conductivity without heating-cooling imbalance. Finally, filled black rectangles  are data related to the observed SUMER oscillations \citep{2003A&A...406.1105W} that are overplotted. Various dashed-dot yellow lines show the fittings on  various sets of theoretical data points while dark green-dashed line is $\tau$=$P$ line. It should be noted that we have taken all the estimated data ($\tau$, $P$) corresponding to the fundamental mode oscillations derived from coronal loops with a normal density of 10$^{-11}$ kg m$^{-3}$.

These damped oscillations and related $\tau$ and $P$ are being studied and plotted in Figure 13 without any information regarding loop length and temperature. Keep in mind that SUMER oscillations were measured in a variety of loops with different length and temperature \citep{2011SSRv..158..397W}. It is clear that the observed data points (black rectangles) from SUMER oscillations match quite closely the theoretically estimated data of the slow-mode oscillations when we consider thermal conductivity, compressive viscosity, and heating-cooling imbalance (red circles). However, a close inspection of Figure 13 provides some more interesting scientific facts. For the period $P\leq$25 min, an almost linear trend is observed between $\tau$ and $P$ with a scaling  $\tau\propto$P$^{1.05}$. Note that dark green-dotted line is $\tau$=$P$ line. These first 14 theoretically estimated data points are basically associated with the oscillations for the shorter loops, and they are closely matching the observed SUMER oscillations (black rectangles) which were shown to follow a scaling law of $\tau \propto P^{0.96}$ as per the improved measurements taking into account the flow effects \citep{2005A&A...435..753W} and also the more recent oscillations detected by RHESSI \citep{2016ApJ...830..110C} have shown a similar scaling law for hotter and shorter loops. For  $P\geq$25 min,
the next 10 theoretically estimated data points are more scattered and  basically related to the various oscillations detected in the longer loops. The lower dot-dashed yellow line that fits the red filled circles is indeed composed of two power-law scalings. For the period $P\leq$25 min the scaling between $\tau$ and $P$ is 
$\tau\propto$P$^{1.05}$. While beyond this period, it is found to be $\tau\propto$P$^{0.95}$. A break point is detected at $P=25$ min, and the next linear trend/fit of the data has a lower slope with a scaling  $\tau\propto$P$^{0.95}$ close to the one estimated by  \citet{2008ApJ...681L..41M} and \citet{2019ApJ...874L...1N}. It should be noted that this break point might occur due to the fact that the heating-cooling imbalance becomes effective in the loops of longer length and help to account for the observed excessive damping. Therefore, the analytical data and corresponding scaling law falls more towards the observations. 
The data points (blue rectangles) estimated under the consideration of a constant heating rate (i.e., without heating-cooling imbalance), has a scaling law (middle yellow-dotted line in Figure 13) 
expressed by $\tau\propto$P$^{1.12}$. This is slightly deviated from the SUMER observations, and damping is underestimated when heating-cooling imbalance is not considered.
The other set of data points (cyan circles) estimated under the consideration of only compressive viscosity (i.e., without considering thermal conductivity), has a scaling law (upper yellow-dashed line on these points especially below $P$=35 min) 
 expressed by $\tau\propto$P$^{1.55}$. This is far beyond the scaling related to SUMER observations, and damping is heavily underestimated in the case when thermal conduction is not considered. Note that in the period regime of $P>$35 min, which is mostly related to longer loops, the heating-cooling imbalance significantly enhances the damping and can cause the deviation of some data points (cyan circles) towards the $\tau$=$P$ line.
 In conclusion, the consideration of the joint effect of thermal conductivity, compressive viscosity, and heating-cooling imbalance predicts a theoretical scaling law in between $\tau$ and $P$ better matching the SUMER observations. Even if we remove the effect of viscosity, and only consider thermal conductivity and heating-cooling imbalance, most of the data points in Figure 13 (empty pink circles) still fall close to the filled red circles except for a few ones related to shorter loops associated with the lowest periods. This suggests that thermal conductivity along with heating-cooling imbalance may play a vital role in explaining the strongly damped slow-mode oscillations observed with SUMER for most of the events, while the role of compressive viscosity may be essential in interpreting the damping of short loops in the super-hot temperature regime such as in postflare loops in solar and stellar flares \citep{2016ApJ...830..110C}.

\section{Discussion and Conclusions} 
      \label{S-Conclusion}

\begin{table}
\begin{tabular}{ |m{2.5cm}|m{1.2cm}|m{1.2cm}|m{1.8cm}|m{1.8cm}| }

\hline
{\bf Loop Length} $\rightarrow$ & \multicolumn{2}{|c|}{ \bf Short Loops (50Mm)} & \multicolumn{2}{|c|}{\bf Long Loops (500Mm)} \\
\hline
{\bf Loop Density} $\rightarrow$ & {\bf Normal Loops} & {\bf Bulky Loops} & {\bf Normal loops} & {\bf Bulky Loops} \\
\hline
{\bf Hot Loops \newline (5-10MK)} & T.C & T.C & T.C $+$ H/C & T.C $+$ H/C \\
 \hline
{\bf Super-Hot Loops \newline (20-30MK)} & Viscosity & T.C & T.C $+$ Vis. & T.C\\
\hline
\end{tabular}
\caption{Summary of the dominant damping mechanisms of the fundamental slow modes in the various regimes of loop parameters considered throughout our study. T.C $:$ Thermal conductivity, H$/$C $:$ heating-cooling imbalance, Vis $:$ Viscosity }
\end{table}

To the best of our knowledge, the present article firstly provides a comprehensive overview on the physical scenario of the damping of the fundamental mode of slow magnetoacoustic oscillations and its higher harmonics in coronal loops with diverse length (50\,--\,500 Mm), temperature(5\,--\,30 MK), and density (10$^{-11}$\,--\,10$^{-9}$ kg m$^{-3}$) under the consideration of thermal conductivity, compressive viscosity, and heating-cooling imbalance. 
In the present work, the temperature range chosen for the theoretical analysis is based on actual observations. Similar damped harmonic oscillations found with RHESSI observations showed average periods of 0.9 min and average decay times of 1.5 min, much shorter compared to those of SUMER oscillations \citep{2016ApJ...830..110C}, while their oscillation quality (i.e. ratio of damping time to period) is similar. Therefore, the assumption of quasi-steady equilibrium may still hold if the wave period is much smaller than the cooling timescale at T=20\,--\,30 MK. The quasi-static assumption here has two-fold meanings: 1) linear theory of different damping mechanisms is applicable on a time scale comparable to the wave period corresponding to the "equilibrium" background condition ($T_0$, $\rho_0$, and $L$) at a certain instant; 2) the WKB (Wentzel-Kramers-Brillouin) approximation, i.e. $\frac{dP}{dt}$ $\ll$ 1, the change of the wave period is very slow during a certain period (e.g. the decay time considered). This allows us to analyze the decay of relative perturbations, $\frac{\Delta T}{T_0}$ and $\frac{\Delta \rho}{\rho_0}$ where $T_0$ and $\rho_0$ are the averaged parameters over this period. For example, \citet{2015ApJ...808...72S} showed observations of super-hot flare loop with GOES that can reach T$>$30 MK for a few minutes. Their damping behavior can be understood based on our parametric study in the regime of super-hot short loops. It is known that the RHESSI energy band (3\,--\,25 keV) used to detect the oscillations (or QPPs) is associated with super-hot loops with T=20\,--\,30 MK \citep[e.g.,][and references  therein]{2014ApJ...781...43C,2014SoPh..289.2547R}. The QPPs are commonly detected in solar flares \citep[e.g.][]{2014ApJ...781...43C,2014SoPh..289.2547R,2015ApJ...808...72S}  which most likely correspond to short loops, however, parametric studies of long loops with super-hot temperatures may help understand the case of stellar flares which have a wide length scale range \citep[e.g.][]{2005A&A...436.1041M,2013ApJ...778L..28S}, therefore, T=20\,--\,30 MK is well relevant to the solar atmosphere and related dynamics in the confined heated loops. On the other hand, as we know the slow-mode oscillations in cooler loops (1 MK$<$T$<$5 MK), particularly related to impulsive heating, are rarely observed compared to the temperature range considered in this study.

In general, the role of heating-cooling imbalance highly depends on the form of the heating function assumed which is however unknown, so one can only say this result  is correct based on our assumption (i.e., $a=-\frac{1}{2}$, $b=-3$), which suits the damped oscillatory regime of the slow modes as described by \citet{2019A&A...628A.133K}. On the other hand, as the result in Figure 13 suggested, the inclusion of heating-cooling imbalance can produce a theoretical prediction better matching the observations. Moreover, the results are compared with the case where no heating-cooling imbalance is present, or heating rate is assumed to be constant. The present model also adds in background the  radiative  cooling term. However, the radiative effect is insignificant compared to other dissipation mechanisms, actually we did not study its effect separately \citep{2012SoPh..280..137A}. In the present work, we found that compressive viscosity along with thermal conductivity causes strong damping of the fundamental mode oscillations in the shorter (e.g., $L$=50 Mm) and super-hot ($T>$10 MK) loops, and compressive viscosity plays a dominant role in this regime.  Nevertheless, the effect of viscosity is insignificant in the damping of these modes in longer (e.g. $L$=500 Mm) and hot loops (T$\leq$10 MK), instead thermal conductivity along with the presence of heating cooling imbalance plays an important role in this condition. Moreover, for longer loops at the hot regime of temperature, the increase in density slightly decreases the damping due to thermal conduction and heating-cooling imbalance (cf. Figure 6 and left bottom panels of Figure 7). Whereas, for shorter loops at the super-hot regime of temperature, the increment in the loop density substantially enhances the damping of the fundamental modes due to  the thermal conductivity when viscosity is absent (cf. bottom panels of Figure 8). Individual role of thermal conductivity is found to be dominant in longer loops at lower temperatures (T$\leq$10 MK), while compressive viscosity dominates the damping at super-hot temperatures ($T>$10 MK) in the shorter loops only. We have summarized our results for the fundamental mode in various physical conditions of loops in Table 1. The scaling law between $\tau$ and $P$ obtained by fitting the theoretical data is found to be closer to the observed SUMER oscillations when we add the effect of heating-cooling imbalance to the case with thermal conduction and viscosity for the damping of fundamental slow mode oscillations.

\citet{2006SoPh..236..127P} have reported that by varying the loop density from 10$^{8}$ to 10$^{10}$ cm$^{-3}$ (or $\rho$=2$\times$10$^{-13}$ to 2$\times$10$^{-11}$ kg m$^{-3}$) at a fixed temperature in the range 6\,--\,10 MK (we consider it as a hot regime, i.e. $T\leq$10 MK in our present work), they get two sets of damping of fundamental slow mode oscillations, in which one was for $\tau$/$P$$\sim$1 (dark-green dotted line in Figure 13) related to the strong damping  occurred at a lower density (10$^{8}$ cm$^{-3}$), while the other was for $\tau$/$P\geq$2 corresponding to a weak damping occurring at higher density (10$^{10}$ cm$^{-3}$). Note that $\rho$=10$^{-11}$ kg m$^{-3}$ corresponds to $N_{e}$=5$\times$10$^{9}$ cm$^{-3}$ or log$_{10}$($N_{e}$)=9.6, which gives $\tau$/$P$$\sim$1.5 from Figure 1 of \citet{2006SoPh..236..127P}, while all theoretical data in Figure 13 of our article are for $\rho=$10$^{-11}$ kg m$^{-3}$. Therefore, all the predicted data here are below $\tau$=2$P$ line as given in \citet{2006SoPh..236..127P} which is consistent with their prediction. The results presented here are thus consistent with the findings of \citet{2006SoPh..236..127P}. In the present article, a two-part power-law scaling $\tau\propto$P$^{1.05}$ (at $P\leq$25 min) and $\tau\propto$P$^{0.95}$ (at $P\geq$25 min) is achieved for the fundamental mode oscillations for the range of loop-length (50\,--\,500 Mm) and temperature (5\,--\,30 MK) under the effect of thermal conductivity, compressive viscosity, and heating-cooling imbalance, which is very close to the strong damping of $\tau$/$P\approx$1.0 and also better satisfy the observed SUMER oscillations. 
Joint effect of thermal conductivity, compressive viscosity, along with heating-cooling imbalance produces strong damping of the fundamental mode oscillations in the normal coronal loops themselves, which match well with the SUMER observations. 

\citet{2007SoPh..246..187S} have numerically studied the damped oscillations observed by SUMER in the linear and nonlinear regimes, and concluded that the damping times of the oscillations are mostly shaped by compressive viscosity rather than thermal conduction, and the damping due to optically thin radiation is negligible when considering a constant heating rate. They showed that thermal conduction alone results in slower damping of the density and velocity oscillations than the observed, while it is required to add the compressive viscosity so that the waves can be damped quickly enough to match the SUMER observations. However, their conclusion for viscosity dominating over thermal conduction in damping was deduced in the loops with very low density of 10$^{8}$ cm$^{-3}$. Instead, here we find that in coronal loops of typical density ($\rho$=10$^{-11}$ kg m$^{-3}$ or $N_{e}$=5$\times$10$^{9}$ cm$^{-3}$), the effect of compressive viscosity on the damping of the slow waves is significant only in very short loops at a super-hot temperature regime. Thus, thermal conductivity along with heating-cooling imbalance dominates in the damping of slow waves. Moreover, to better match the observed damped SUMER oscillations (figure 13), the joint effect of the thermal conductivity, compressive viscosity, and heating-cooling imbalance is required (red circles;  $\tau\propto$P$^{1.05}$ and $\tau\propto$P$^{0.95}$). On the other hand, if we consider compressive viscosity alone (cyan circles), in the observed range of SUMER oscillations (P$\leq$35 min), its variation ($\tau\propto$P$^{1.55}$) goes out of even the weak damping regime ($\tau\geq$$2P$ line) as shown in Figure 13. This suggests that compressive viscosity alone can not explain the observed damping of the SUMER oscillations as pointed by many previous reports \citep{2007SoPh..246..187S,2012SoPh..280..137A}. If we exclude the effect of the compressive viscosity, and only consider thermal conductivity supported by heating-cooling imbalance, then most of the data points in Figure 13 (empty pink circles) lie very close to the filled red circles (thermal conductivity$+$viscosity$+$imbalance) except of a few points for the oscillations of very short periods associated with very short/super-hot loops. Therefore, we conclude that thermal conductivity along with heating-cooling imbalance may be the dominant in the damping mechanism for interpreting the strongly-damped slow-mode oscillations observed with SUMER (filled black rectangles in Figure 13).

There were also several attempts in the past that investigated the damping of slow waves in nonisothermal, hot, gravitationally stratified coronal loops \citep[e.g.][]{2008SoPh..252..305E}. In hot and super-hot regime, the non-uniformity in $T_0$ and $\rho_0$
along the loop is expected to be negligible because of highly efficient thermal conduction and very large density scale height.
 For example, the density scale height is H$\approx$500 Mm for T=10 MK plasma, which implies that the effect of gravitational stratification is very small on the damping even for the longest loop of L=500 Mm that has a height h=160 Mm if the loop is semi-circular.\newline
 We would like to mention that the background heating function which is described as a function of density and temperature throughout our study has been taken just as a typical physical scenario based on the model of \citet{2019A&A...628A.133K}. Many previous works in the area have studied different functional dependencies of this unknown heating function such as its dependence on magnetic field and loop length \citep[e.g.][]{2009ApJ...690..902, 2017ApJ...849...62N} as well as on time \citep[e.g.,][]{2007ApJ...659L.173T, ApJL....826...L20} since the specific heating mechanism in the solar corona is still unknown and remains an open research problem to date.\newline
We conclude finally that although thermal conduction along with heating-cooling imbalance may be the dominant damping mechanism for interpreting the strongly damped slow-mode oscillations observed with SUMER, the role of compressive viscosity is essential to explain the damping of SUMER oscillations in coronal loops at short length and super-hot temperature regimes such as the slow modes excited in postflare loops in solar and stellar flares.

\begin{acks}
We thank the reviewer for his/her constructive comments that improved our manuscript.
AP thanks IIT (BHU) for the computational facility, and AKS acknowledges the support of UKIERI (Indo-UK) research grant for the present research. 
The work of TW was supported by NASA grants 80NSSC18K1131 and 80NSSC18K0668 as well as the NASA Cooperative Agreement NNG11PL10A to CUA.
AKS also acknowledges the ISSI-BJ regarding the science team project on "Oscillatory Processes in Solar and Stellar Coronae".
\end{acks}
\newline \newline
{\footnotesize {\bf Disclosure of Potential Conflicts of Interest:}\newline The authors declare that there are no conflicts of interest.}

\end{article} 

\end{document}